\documentclass[global]{svjour}
\usepackage{graphics}
\usepackage{natbib}
\journalname{Light Scattering Reviews}
\begin{document}
\title{Light Scattering and Thermal Emission by Primitive Dust Particles in Planetary Systems}
\author{Hiroshi Kimura\inst{1} \and Ludmilla Kolokolova\inst{2} \and Aigen Li\inst{3} \and J\'{e}r\'{e}my Lebreton\inst{4,5}
}                     
\offprints{H. Kimura}          
\institute{Graduate School of Science, Kobe University, C/O CPS (Center for Planetary Science), Chuo-Ku Minatojima Minamimachi, 7-1-48, Kobe 650-0047, Japan, \email{hiroshi\_kimura@cps-jp.org} \and Planetary Data System Group, Department of Astronomy, University of Maryland, Rm. 2337, Computer and Space Science Bldg, College Park, MD 20742, USA \and Department of Physics and Astronomy, University of Missouri, 314 Physics Building, Columbia, MO 65211, USA \and Infrared Processing and Analysis Center, California Institute of Technology, Pasadena, CA 91125, USA \and NASA Exoplanet Science Institute, California Institute of Technology, 770 S. Wilson Ave., Pasadena, CA 91125, USA}
%
%
\titlerunning{Primitive Dust Particles in Planetary Systems}
\maketitle
%
%
\section{Introduction}
\label{intro}

Stardust grains are tiny solid samples of stars, newly condensed in an expanding atmosphere of a dying star and injected into interstellar space by stellar wind and radiation pressure. 
Such a dust grain in the interstellar medium goes back and forth between diffuse clouds and dense clouds before it experiences star formation at the core of a dense cloud \citep{greenberg1984}. 
The majority of prestellar interstellar grains evaporates during star formation, but circumstellar gas condenses as pristine dust grains when a gaseous envelope of a newly born star cools down to the melting temperatures of solids \citep{keller-messenger2011}.
In consequence, it is no wonder that interstellar grains are chemically similar to primitive dust particles in planetary systems, if the particles are least processed \citep{kimura2013}.
Pristine dust grains coagulate together and form fluffy agglomerates of the grains in a protoplanetary disk around a young star \citep{weidenschilling-et-al1989}. 
In 1970s, a collection of interplanetary dust particles, which originate from comets and/or asteroids, revealed that the particles are indeed fluffy agglomerates of submicron constituent grains \citep{brownlee-et-al1976}.
Collisional growth of fluffy agglomerates ends up with planetesimals, which are at present observed as comets and asteroids in planetary systems around main-sequence stars \citep{weidenschilling1997}. 
Therefore, planetesimals (i.e., comets and asteroids) are time capsules of primitive dust particles, which are reservoirs of information on the time of planetary system formation. 
At the present time, primitive dust particles released from comets and asteroids have been observed in the Solar System and debris disks\footnote{A debris disk is defined as a dust disk that surrounds a main-sequence star with its age exceeding the lifetime of the disk. Therefore, the disk is not primordial, but must be replenished with dust particles released from their parent bodies such as asteroids and comets.} around other main-sequence stars through stellar radiation scattered or reradiated by the particles. 
The interpretation of observational data is, however, often not straightforward, unless light-scattering and thermal-emission properties of the particles are well known a priori. 

Numerical simulation is a flexible and powerful approach to deducing light-scattering and thermal-emission properties of primitive dust particles in planetary systems from astronomical observations. 
In this review, we focus on numerical approaches to light scattering and thermal emission of fluffy agglomerates that simulate primitive dust particles in the Solar System and debris disks.
In consequence, it must be emphasized that dust particles in protoplanetary disks around pre-main-sequence stars go beyond the scope of this review.
This paper is organized as follows: we review and comment on popular models of dust agglomerates in Sect.~\ref{sec:2}, light-scattering techniques applicable to agglomerates in Sect.~\ref{sec:3}, light scattering by agglomerates in Sect~\ref{sec:4}, thermal emission from agglomerates in Sect~\ref{sec:5}, and integral optical quantities of agglomerates such as bolometric albedo, radiation pressure, and equilibrium temperature in Sect~\ref{sec:6}.
Furthermore, we pass our concluding remarks in Sect~\ref{sec:7} and finally provide a summary in Sect~\ref{sec:8}.

\section{Models of dust agglomerates in planetary systems}
\label{sec:2}

\subsection{Artificial configuration of constituent grains}

\begin{figure*}
\center
\resizebox{1.0\textwidth}{!}{%
  \includegraphics{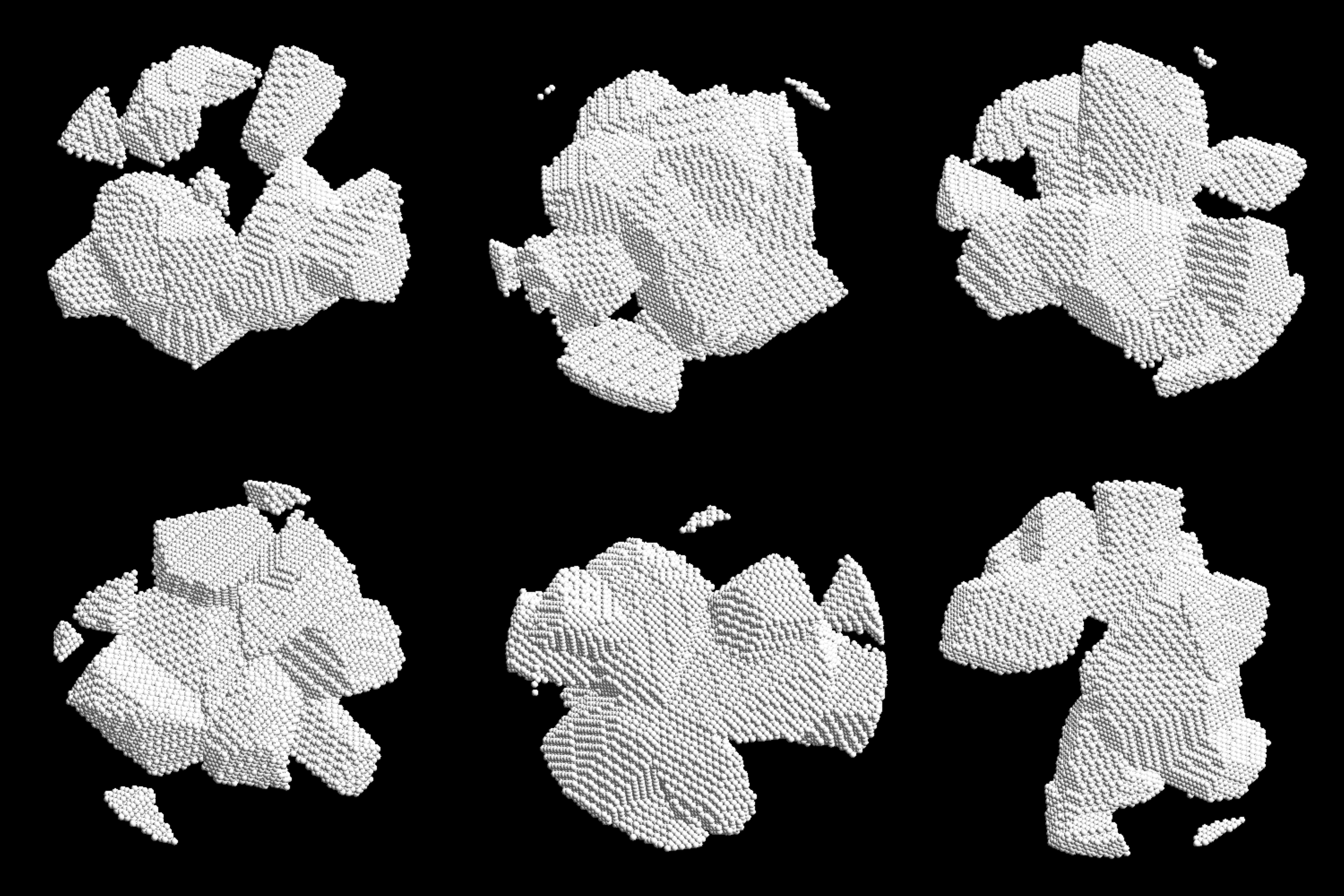}
}
\caption{Six realizations of dust agglomerates that formed by the removal of subvolumes consecutively from an initially spherical volume. 
The same algorithm was applied to generate the agglomerates, while different random numbers were used for their respective operations. 
From \citet{zubko-et-al2009}}
\label{zubko-et-al2009:f2}       
\end{figure*}

There are a variety of ways to numerically configure constituent grains in an agglomerate, if the formation process of the agglomerate is not taken into account.
The most simple algorithm for creating a fluffy agglomerate in control of its porosity would be to first form a cubic or a sphere with subvolumes and then remove a portion of the subvolumes according to the desired porosity \citep[e.g.,][]{hage-greenberg1990,zubko-et-al2009,petrova-tishkovets2011,kirchschlager-wolf2013}.
This algorithm has the advantage of easily producing irregularly shaped constituent grains, but the disadvantage is in producing equi-dimensional agglomerates.
Moreover, there is a risk that an algorithm of this kind separates some of the constituent grains from the agglomerate, even in the case that the porosity of the agglomerate is not extremely high (see Fig.\,\ref{zubko-et-al2009:f2}).

Alternatively, one could first create constituent grains in preferable sizes and shapes and then assemble a cluster of them in an arbitrary configuration.
In this way, numerical calculations of electromagnetic waves scattered not only by a cluster of spheres, but also by a cluster of tetrahedrons has been implemented to simulate light scattering by dust agglomerates in planetary systems \citep{xing-hanner1997,yanamandrafisher-hanner1999}.
It is certainly of great interest from a mathematical point of view to consider arbitrary configuration and shape of constituent grains in dust agglomerates.
However, one should keep in mind that the application of such an agglomerate to dust particles in planetary systems is not necessarily justified.

\subsection{Coagulation growth of pristine constituent grains}

\begin{figure*}
\center
\resizebox{1.0\textwidth}{!}{%
  \includegraphics{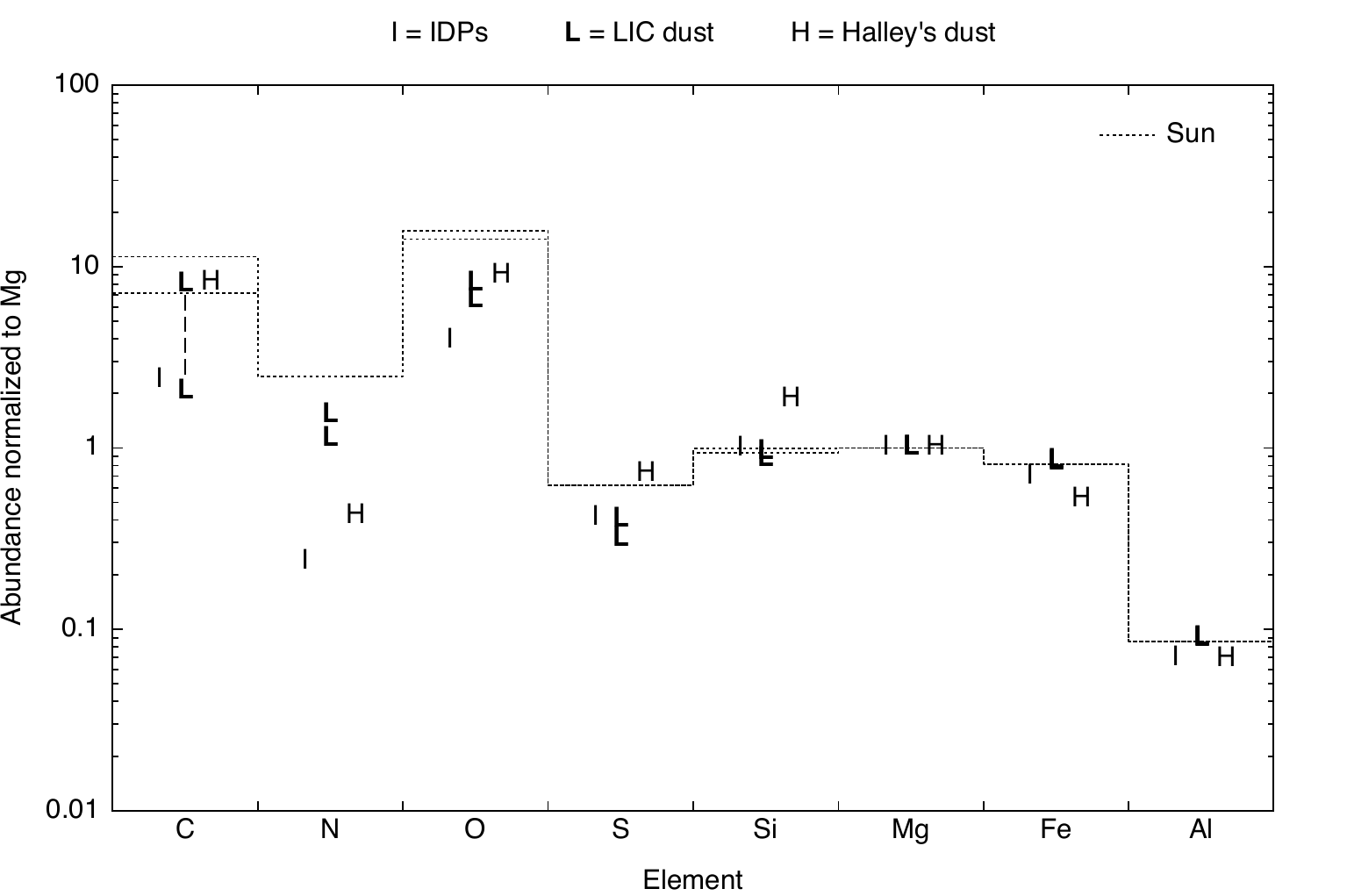}
}
\caption{The chemical compositions of cometary dust, interplanetary dust, interstellar dust as well as the solar photosphere, normalized to their Mg abundances. 
The data for cometary dust (H), interplanetary dust (I), interstellar dust (L) correspond to those of dust in comet 1P/Halley, the chondritic porous subset of interplanetary dust particles, and dust in the Local Interstellar Cloud, respectively.
The {\it two dotted lines} for the photosphere of the Sun reflect uncertainties in the solar photospheric composition, which also affect the composition of interstellar dust.
From \citet{kimura-et-al2003a}}
\label{kimura-et-al2003a:f3}       
\end{figure*}

Agglomerates of submicron constituent grains make up the anhydrous chondritic porous (CP) subset of interplanetary dust particles (IDPs) collected in the stratosphere as well as Antarctic ice and snow \citep{brownlee1985,noguchi-et-al2015}.
CP IDPs are found to consist of chondritic grains embedded in organic-rich carbonaceous material and most likely originate from short-period comets.
As a result, their chemical composition bears strong resemblance to the composition of cometary dust measured in situ for comet 1P/Halley as shown in Fig.\,\ref{kimura-et-al2003a:f3} \citep{jessberger1999,kimura-et-al2003a}.
A dust agglomerate consisting of submicron grains is a natural consequence of dust growth due to collisions of submicron pristine condensates in a protoplanetary disk.
If the collision takes place at a low relative velocity ($v \leq 1\,\mathrm{m\,s^{-1}}$), an agglomerate grows without restructuring of constituent grains and possess a fractal geometry \citep{wurm-blum1998}.
The structure of such an agglomerate is characterized by a fractal dimension $D$, which is defined as
\begin{equation}
m \propto a_\mathrm{g}^{D},
\end{equation}
where $m$ and $a_\mathrm{g}$ are the mass and gyration radius of the agglomerate, respectively.
The radius of gyration $a_\mathrm{g}$ is given by
\begin{equation}
a_\mathrm{g} = \left[{\frac{1}{2N^{2}}\sum_{i}^{N} \sum_{j}^{N} \left({\vec{r}_\mathrm{i}-\vec{r}_\mathrm{j}}\right)^{2}}\right]^{1/2},
\end{equation}
where $\vec{r}_\mathrm{i}$ and $\vec{r}_\mathrm{j}$ are the position vectors of the $i$-th and $j$-th constituent grains, respectively, and $N$ is the total number of the grains.
At a higher velocity, a collision results in restructuring constituent grains and highly compressed agglomerates, or even destructing agglomerates \citep{dominik-tielens1997}.
Even at the maximum compression of agglomerates due to high-velocity collisions, possession of a fractal structure has been manifested by highly sophisticated numerical simulations \citep{wada-et-al2008}.
Therefore, it is reasonable as a first step to assume fractal agglomerates consisting of submicron grains when simulating light scattering and thermal emission of primitive dust particles in planetary systems.

\begin{figure*}
\center
\resizebox{0.6\textwidth}{!}{%
  \includegraphics{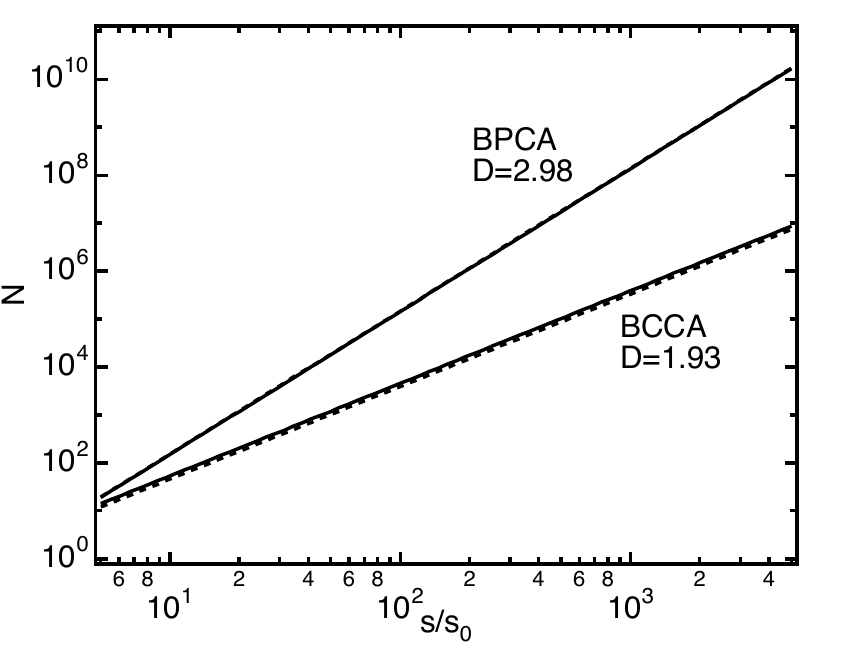}
}
\caption{The number $N$ of constituent grains in fractal agglomerates formed under coagulation growth of the BPCA and the BCCA processes as a function of its characteristic radius normalized to the radius of constituent grains. {\it Solid line}: $\log\,k_0 = -0.576 D + 0.915$ \citep{kimura-et-al1997}; {\it dashed line}: $\log\,k_0 = -0.5 D + 0.7$ \citep{mukai-et-al1992}. From \citet{kimura-et-al1997}}
\label{kimura-et-al1997:f1}       
\end{figure*}
If constituent grains in an agglomerate are identical, the number $N$ of the grains and the characteristic radius ${a}_\mathrm{c}$ of the agglomerate fulfill the relation:
\begin{equation}
N = k_0 \left({\frac{{a}_\mathrm{c}}{a_0}}\right)^{D},
\end{equation}
where $a_0$ is the radius of the constituent grains and $k_0$ is a proportionality constant of order unity, which depends on the coagulation process (see Fig.\,\ref{kimura-et-al1997:f1}).
The characteristic radius $a_\mathrm{c}$ of an agglomerate is defined by
\begin{equation}
a_\mathrm{c} = \sqrt{\frac{5}{3}}\, a_\mathrm{g},
\end{equation}
which is known to well describe an apparent radius of agglomerates.
In reality, constituent grains are not identical, but a nucleation theory of pristine condensates in a protoplanetary disk predicts a size distribution of pristine grains with a single peak in an extremely narrow size range \citep{yamamoto-hasegawa1977}.
In addition, the major constituents of CP IDPs called GEMS (glass with embedded metal and sulfides) grains are typically limited to $a_0 = 0.05$--$0.25\,\mathrm{\mu m}$ \citep{keller-messenger2011}.
Therefore, the radius of constituent grains in an agglomerate does not seem to vary considerably as far as primitive dust particles in planetary systems are concerned.
Consequently, it is popular practice to use agglomerates of identical constituent grains in an approximation of primitive dust particles in planetary systems.
One should, however, keep in mind that the size distribution of constituent grains influences light-scattering and thermal emission properties of agglomerates to a certain extent.

\begin{figure*}
\resizebox{1.0\textwidth}{!}{%
  \includegraphics{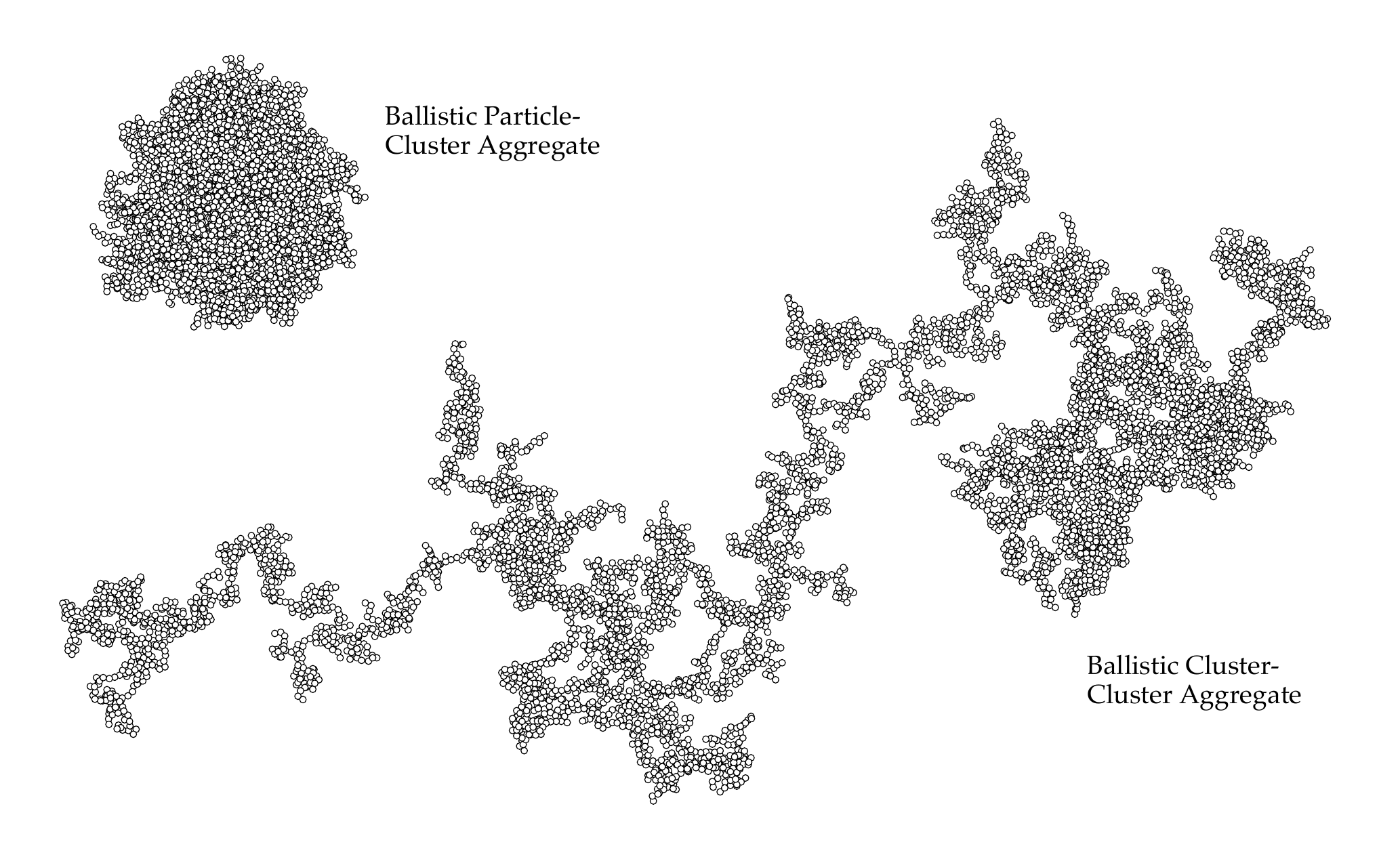}
}
\caption{Two realizations of dust agglomerates that formed by the BPCA and the BCCA processes. 
The coagulation growth was conducted to hit and stick on their respective contacts, while $2^{13}$ constituent grains were assumed to be identical spheres. 
From \citet{kimura2001}}
\label{kimura2001:f1}       
\end{figure*}
In popular open-source light-scattering codes available to date, the size parameter of an agglomerate is defined as $x_\mathrm{v} = 2 \pi a_\mathrm{v} / \lambda$ where $a_\mathrm{v}$ is the radius of volume-equivalent sphere and $\lambda$ is the wavelength of interest.
Note that both $a_\mathrm{v}$ and $x_\mathrm{v}$ are quantities that are independent of the coagulation process, because $a_\mathrm{v} = {N}^{1/3} a_0$.
As a result, an apparent radius of agglomerates cannot be described by $a_\mathrm{v}$ at all, but $a_\mathrm{v}$ is a useful quantity if the mass or volume of the agglomerates is of interest.

\begin{figure*}
\center
\resizebox{0.75\textwidth}{!}{%
  \includegraphics{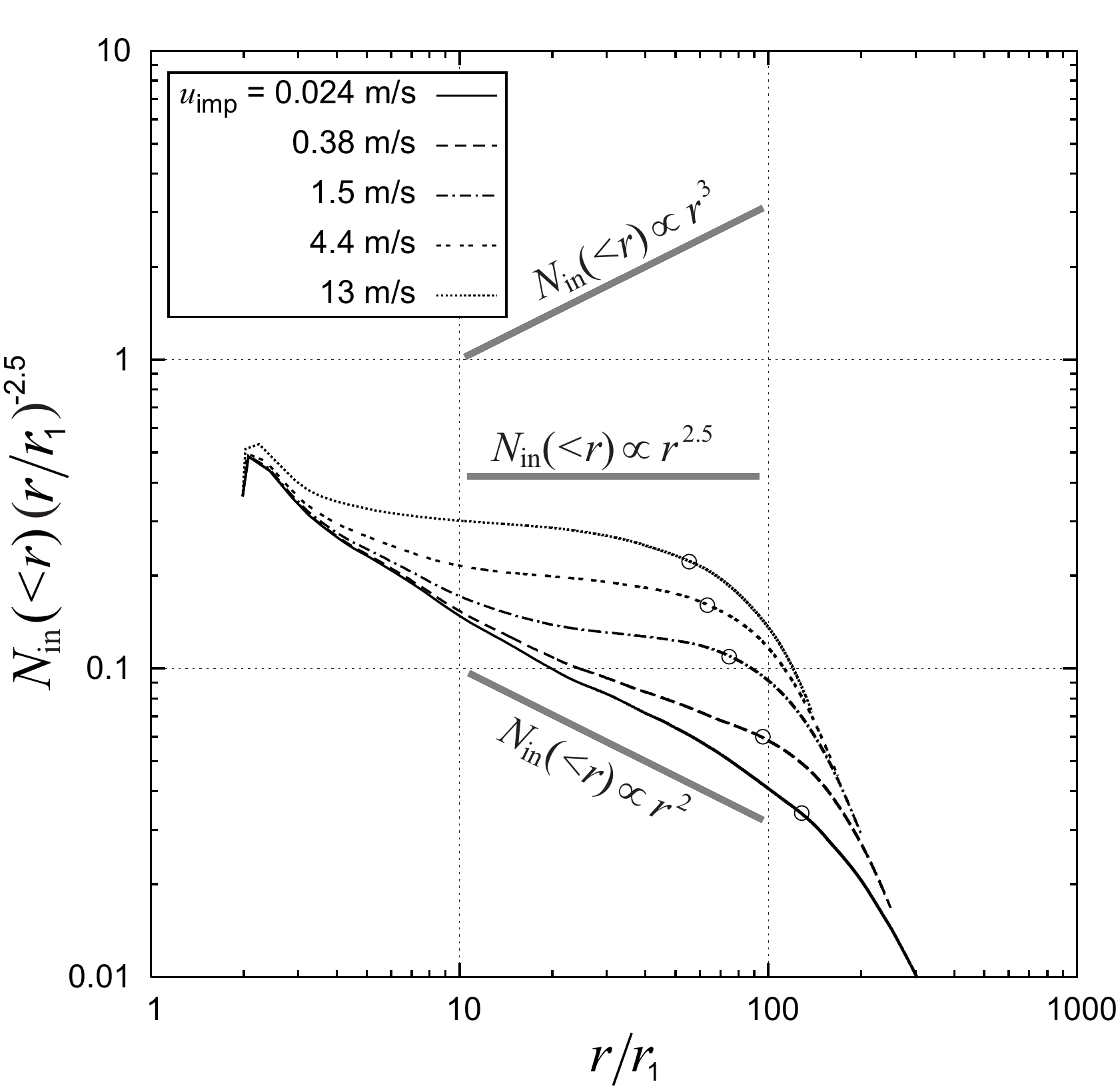}
}
\caption{The number of constituent grains within a distance $r$ from the center of an agglomerate as a function of the distance. 
Here the number is normalized by the distance to the power of $2.5$ and the distance is normalized by the radius of constituent grains so that the fractal dimension $D = 2.0$, $2.5$, and $3.0$ gives a slope of $-0.5$, $0$, and $0.5$, respectively. 
The agglomerates were produced by numerical simulations on mutual collisions of identical BCCA particles consisting of $2^{13}$ grains at various collision velocities based on contact mechanics.
From \citet{wada-et-al2008}}
\label{wada-et-al2008:f7}       
\end{figure*}

Figure\,\ref{kimura2001:f1} illustrates an example of fractal agglomerates consisting of $2^{13}$ spherical grains formed under hit-and-stick coagulation processes along ballistic trajectories.
The agglomerates were grown by either the ballistic cluster-cluster agglomeration (BCCA) or the ballistic particle-cluster agglomeration (BPCA) of constituent grains.
The fractal dimensions of BCCA and BPCA particles have been determined numerically to be approximately $D \approx 2$ and $\approx 3$, respectively \citep{meakin1984}.
Agglomerates grown by the diffusion limited agglomeration (DLA) process are also fractals with $D \approx 2.5$ if their constituent grains are assumed to hit and stick on contact \citep{meakin1984}.
Although the DLA process is not relevant to coagulation growth in a protoplanetary disk, high-velocity collisions of BCCA particles result in a fractal dimension $D \approx 2.5$ at their maximum compression as shown in Fig.\,\ref{wada-et-al2008:f7} \citep{wada-et-al2008}.
DLA particles could, therefore, be used to represent highly compressed BCCA particles if the fractal dimension plays a vital role in the determination of their light-scattering and thermal-emission properties.

It is worth noting that the configuration of constituent grains in an agglomerate is far from well known for primitive dust particles in planetary systems, although the morphology of CP IDPs is sometimes considered typical of the particles.
We argue that CP IDPs represent the most compact structure of primitive dust particles in planetary systems, since there is natural selection with respect to the morphology.
If CP IDPs are primitive dust particles released from comets or asteroids, then they must have been injected into a bound orbit around the Sun.
It is well known that dust particles in orbit around a star could spiral to the central star by a relativistic effect, called the Poynting-Robertson effect \citep{robertson1937,wyatt-whipple1950,kimura-et-al2002a}.
However, the fluffier the morphology of dust agglomerates is, the more efficiently the stellar radiation pressure blows out the agglomerates into the interstellar medium.
As a result, it is dynamically infeasible that dust agglomerates with a more fluffy structure stay in a bound orbit around the Sun.
Therefore, CP IDPs must be so compact that they could stay in a bound orbit after their release from their parent bodies and reached the Earth by the Poynting-Robertson effect.

In 1990s, the radius of constituent grains in primitive dust agglomerates was often assumed to be ${a_0} \sim 0.01~\mathrm{\mu m}$, simply to alleviate computational limitations \citep[e.g.,][]{mukai-et-al1992,kozasa-et-al1993,kimura-et-al1997}.
However, laboratory studies on the morphology of CP IDPs as well as theoretical studies on the optical properties of cometary dust and dust in debris disks indicate that the radius of constituent grains is on the order of submicrometers \citep{brownlee1985,kimura-et-al2003c,graham-et-al2007}.
In addition, amoeboid olivine aggregates in the Allende meteorite and aggregate-type particles in the samples of asteroid 25143 Itokawa are characterized by constituent grains whose radii are on the order of micrometers \citep{grossman-steele1976,yada-et-al2014}.
Therefore, we regard ${a_0} \sim 0.1$--$1\,\mathrm{\mu m}$ as a stringent constraint on a plausible model of dust agglomerates in planetary systems.

\section{Light-scattering techniques for dust agglomerates}
\label{sec:3}

\subsection{T-matrix method and generalized multiparticle Mie solution}

The superposition T-matrix method (TMM) is robust when numerical simulation of light scattering is performed for a cluster of spheres particularly in random orientations \citep{mackowski-mishchenko1996}.
Although the advantage of the TMM is its efficiency in analytical averaging of scattering characteristic over particle orientations, huge random-access memory (RAM) is a requisite for the publicly available code {\sf scsmtm1} developed by \citet{mackowski-mishchenko1996} when used for large agglomerates.
Therefore, \citet{okada2008} proposed to use another code {\sf scsmfo1b} developed by \citet{mackowski-mishchenko1996}, which gives a solution in a fixed orientation, and to numerically perform orientation averaging with a quasi-Monte Carlo (QMC) method.
Because the requisite RAM is smaller in {\sf scsmfo1b} than {\sf scsmtm1}, the QMC method provides an opportunity of dealing with large agglomerates of submicron constituent grains.
\citet{penttilae-lumme2011} recommend the optimal cubature on the sphere (OCoS) method to perform fast numerical orientation averaging along with the {\sf scsmfo1b} code.
The new publicly available {\sf MSTM} (Multi Sphere T Matrix) code in Fortran-90 is intended to replace the {\sf scsmtm1} and {\sf scsmfo1b} codes in Fortran-77, so that the {\sf MSTM} code optimally uses the memory and processor resources \citep{mackowski-mishchenko2011}.


Similar to the superposition T-matrix method, the generalized multiparticle Mie solution (GMM) is an exact method for computing light-scattering properties of dust agglomerates consisting of spherical grains \citep{xu1995,xu-gustafson2001}.
A T-matrix formulation of the GMM provides fast computing of light scattering by an agglomerate of spherical constituent grains in random orientation \citep{xu2003}.
In 2000s, a prominent feature of GMM was the ability of providing a solution to light-scattering problems for an agglomerate consisting of spherical core-mantle grains with the use of the publicly available code {\sf gmm02TrA} \citep{xu-khlebtsov2003}. 
Another code {\sf gmm02f} for a fixed orientation is expected to deal with larger agglomerates of core-mantle spheres than {\sf gmm02TrA} for random orientation, but we are aware that the {\sf gmm02f} code contains a bug, which has not been resolved to date.
In 2010s, all the features of the {\sf gmm02TrA} code can be provided by the {\sf MSTM} code of the TMM along with the ability of parallel computing.

\subsection{Discrete dipole approximation}

The discrete dipole approximation (DDA) is a flexible and powerful method for computing light-scattering properties of arbitrary shaped dust particles \citep{purcell-pennypacker1973,draine-flatau1994}.
In the DDA, dust particles are represented by an array of point electric dipoles and thus the determination of dipole polarizability plays a key role in determining the responses of the dipoles to electromagnetic waves \citep{draine-goodman1993,collinge-draine2004}.
\citet{hage-greenberg1990} demonstrated that the dipole polarizability determined by the digitized Green function/volume integral equation formulation (DGF/VIEF) gives more accurate results than the Clausius-Mossotti relation with the radiative reaction correction employed by \citet{draine1988}.
The lattice dispersion relation (LDR) has been implemented as a default prescription for the dipole polarizabilities in the publicly available code {\sf DDSCAT} developed by \citet{draine-flatau1994}.
\citet{okamoto1995} demonstrated that the $a_1$-term method is superior to the LDR method for agglomerates of spherical constituent grains, since the former takes into account the boundary condition of spheres \citep[see also][]{okamoto-xu1998}.
Therefore, one should keep in mind that the choice of polarizability prescription is not a minor issue for the accuracy of numerical results computed by the DDA.
Another open-source code for the DDA is {\sf ADDA} that has been intended to run on a multiprocessor memory-distributed system \citep{yurkin-hoekstra2011}.
The advantage of the DDA over the TMM is the ability to deal with not only arbitrary shapes of constituent grains in an agglomerate but also a contact spot formed by adhesion between elastic constituent grains.

The DDA has been formulated in a way that it provides a solution to a fixed orientation of a particle with respect to the direction of incident electromagnetic wave.
As a result, it is common practice to numerically perform orientation averaging of scattering properties by solving the equations for each orientation sequentially.
However, for highly fluffy agglomerates, the solution is very sensitive to their orientations and thus orientation averaging in the DDA is time consuming.
\citet{singham-et-al1986} provide an exact analytic expression for the orientational average of scattering matrix elements in the DDA, but the implementation of the analytic averaging in modern DDA formulations is not straightforward \citep{yurkin-hoekstra2007}.
\citet{mackowski2002} proposed to use the DDA for computing the T-matrix of arbitrary shaped particles and then apply analytic formulas to the T-matrix for orientation averaging.
This is most likely the best method for orientation averaging of scattering properties in the framework of the DDA, but this method has not been applied to dust agglomerates in planetary systems.

\subsection{Effective medium approximations}

An effective medium approximation (EMA) is computer friendly and thus has been a popular method to estimate light-scattering and thermal-emission properties of dust agglomerates in combination with Mie theory.
Among EMAs, the Maxwell Garnett mixing rule and the Bruggeman mixing rule are the two most popular methods in astronomy.
In the Maxwell Garnett mixing rule, dust agglomerates of small constituent grains are represented by a hypothetical inhomogeneous sphere with inclusions of constituent grains in the matrix of vacuum. 
In the Bruggeman mixing rule, agglomerates of small constituent grains are assumed to be a hypothetical inhomogeneous sphere consisting of constituent grains and ``vacuum'' grains.

\begin{figure*}
\resizebox{1.0\textwidth}{!}{%
  \includegraphics{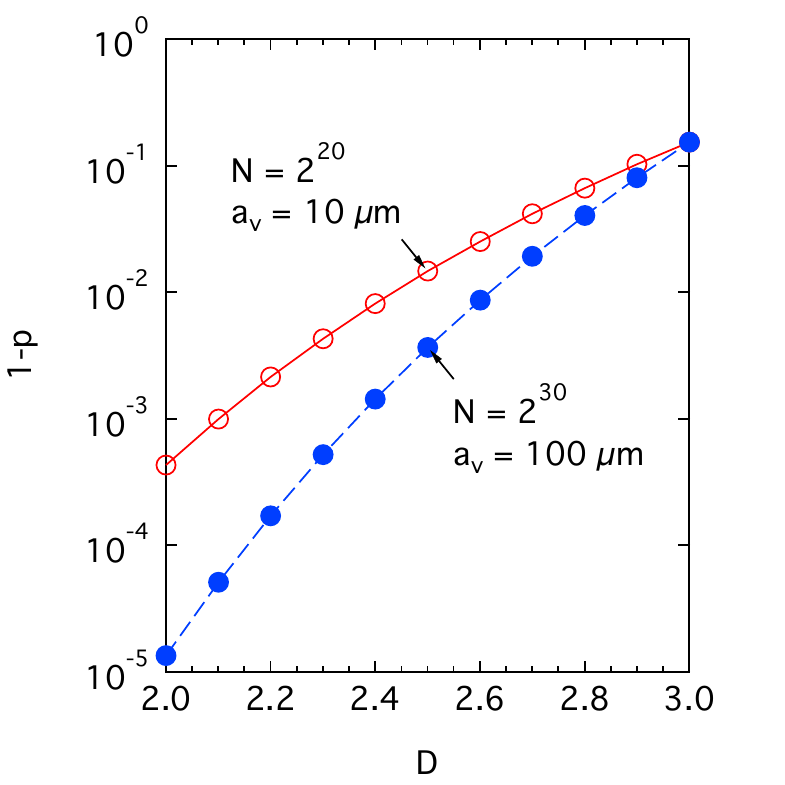}
  \includegraphics{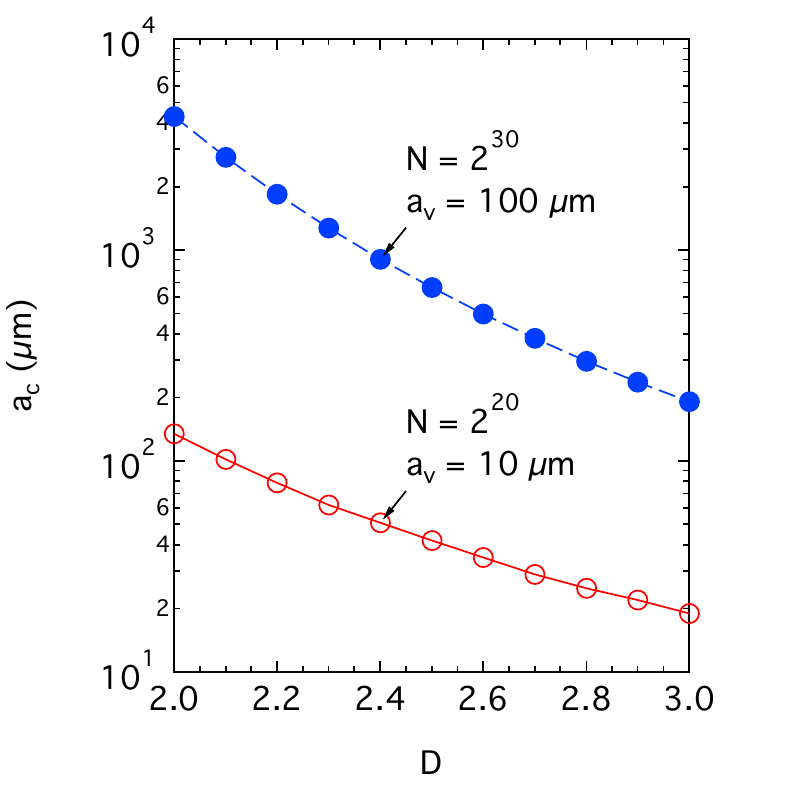}
}
\caption{The volume fraction $1-p$ and the characteristic radius ${a}_\mathrm{c}$ of fractal agglomerates consisting of $0.1\,\mathrm{\mu m}$-radius spherical grains as a function of fractal dimension $D$. 
{\it Open circles with a solid line} and {\it filled circles with a dashed line} are the corresponding values at $N = 2^{20}$ ($a_\mathrm{v} = 10\,\mathrm{\mu m}$) and $N = 2^{30}$ ($a_\mathrm{v} = 100\,\mathrm{\mu m}$), respectively. Reproduced from Table\,1 of \citet{kimura-et-al2009}}
\label{kimura-et-al2009:t1}       
\end{figure*}
It is important for EMAs to properly define the radius of a hypothetical porous sphere and its porosity, since the definitions of these quantities are not unique. 
\citet{mukai-et-al1992} proposed to use the characteristic radius $a_\mathrm{c}$ of an agglomerate and the porosity defined by
\begin{equation}
p = 1 - \frac{V}{V_\mathrm{c}},
\label{porosity}
\end{equation}
where $V$ and $V_\mathrm{c}$ are the total volume of constituent grains and the volume of a hypothetical sphere with a radius $a_\mathrm{c}$.
It turned out that these definitions enable EMAs to well reproduce the radiation pressure on agglomerates computed by the DDA, as far as agglomerates with $a_0 = 0.01\,\mathrm{\mu m}$ are concerned \citep{mukai-et-al1992,kimura-et-al2002a}. 
Note that Eq.\,(\ref{porosity}) can be written as
\begin{equation}
p = 1 - k_0 \left({\frac{{a}_\mathrm{c}}{a_0}}\right)^{D-3} ,
\end{equation}
in case of fractal agglomerates.
This clarifies that the porosity for agglomerates with $D=3$ is independent of ${a}_\mathrm{c}$ and the porosity for agglomerates with $D<3$ increases with ${a}_\mathrm{c}$ (see Fig.\,\ref{kimura-et-al2009:t1}).
Therefore, as illustrated in Fig.\,\ref{kimura2001:f1}, fractal agglomerates with $D \approx 3$ are relatively compact compared to those with $D \approx 2$.

\section{Light scattering by dust agglomerates}
\label{sec:4}

\subsection{Single scattering}
\label{sec:4.1}

Most of light-scattering phenomena associated with primitive dust particles in planetary systems are well described by scattering of stellar radiation on single particles.
The geometric albedo of dust particles is one of the quantities that can be derived from both numerical simulations and astronomical observations of light scattering by the particles.
It is defined by the ratio of stellar radiation scattered by the particles at zero phase angle (exact backscattering) to that scattered by a white Lambertian flat disk of the same geometric cross section.
The geometric albedos of micrometeoroids, dust particles in the zodiacal cloud and the coma of comet 1P/Halley have been determined to be $<0.1$, $0.06$, $0.04$, respectively \citep{hanner1980,ishiguro-et-al2013,lamy-et-al1989}.
These values reveal visibly very dark appearances of primitive dust particles at least in the Solar System, implying that the particles are dominated by carbonaceous materials.
\citet{hanner-et-al1981} extended the definition of geometric albedo so that the brightness of fluffy agglomerates at any scattering angle is normalized to the geometric albedo at backscattering.
In this way, the geometric albedo $A_\mathrm{p}$ of a particle at a wavelength of $\lambda$ is proportional to the $(1,1)$ element of the Mueller matrix, $S_{11}$, as $A_\mathrm{p}=S_{11} \lambda^2/(4 \pi G)$ where $G$ is the geometric cross section of the particle.
The degree of linear polarization is another quantity that gives a direct comparison between numerical simulations and observations.

With the help of the DDA, \citet{west1991} has pioneered numerical simulation of light scattering by DLA ($D \approx 2.5$) particles consisting of submicrometer-sized grains.
It is most likely that he employed the Clausius-Mossotti relation with the radiative reaction correction given by \citet{draine1988} to determine the polarizability of each dipole, since he used the early version of the {\sf DDSCAT} code.
The number of the constituent grains was set to either $N=170$ or $8$ and the size parameter $x_{0}$ of the grains, which is defined as $x_{0}=2\pi a_{0}/\lambda$, ranged from $x_{0} = 0.19$ to $0.57$ for $N=170$ and $x_{0} = 0.6$--$1.8$ for $N=8$.
The intensity of light scattered by agglomerates shows an enhancement toward forward scattering that is characteristic of large particles, but agglomerates do not display undulations in its angular dependence as opposed to spheres of similar geometrical cross sections.
Interestingly, the angular dependence of linear polarization is well characterized by the size of constituent grains in agglomerates, rather than the overall size of the agglomerates.

\citet{kozasa-et-al1992,kozasa-et-al1993} intensively studied light-scattering properties of BPCA ($D \approx 3$) and BCCA ($D \approx 2$) particles consisting of tiny constituent grains with $a_{0} = 0.01\,\mathrm{\mu m}$.
They implemented DDA computations using the {\sf DDSCAT} code (ver.\,4a) along with the DGF/VIEF method for the determination of dipole polarizability.
\citet{kozasa-et-al1992} presented the dependences of absorption and scattering cross sections and asymmetry parameter on wavelength, while \citet{kozasa-et-al1993} considered the dependences of intensity and linear polarization on scattering angle for silicate and magnetite agglomerates at various numbers of the constituent grains up to $N = 4096$.
They also investigated the validities of empirical formulas as well as the Maxwell Garnett mixing rule for the light-scattering properties of the agglomerates.
Although the size of their constituent grains at a wavelength $\lambda = 0.6\,\mathrm{\mu m}$ lies in the Rayleigh scattering regime ($x_{0} = 0.1$), their results for BPCA ($D \approx 3$) particles revealed that the maximum degree of linear polarization decreases with the number of constituent grains.
Consequently, we may expect that the degree of linear polarization strongly depends on the size of constituent grains, but to a lesser degree the apparent size of agglomerates.

\citet{lumme-et-a1997} provided a survey of light-scattering properties with agglomerates of submicron spherical constituent grains relevant to primitive dust particles in planetary systems. 
They accomplished the survey using their own DDA code with icy and silicate agglomerates consisting of $N=200$ spherical grains.
The size parameter $x_{0}$ of constituent grains were assumed to be $x_{0} = 1.2$ or $1.9$, which corresponds to submicron constituent grains in the visible wavelength range. 
The arrangement of the spheres in an agglomerate follows either the DLA process or a removal of spheres from a maximum packed cluster of spheres.
The $a_1$-term method was applied to determine the dipole polarizability, although the polarizability was truncated to the first three terms in the expansion of the $a_1$-term with respect to the size parameter of the dipoles.
Their results show no clear difference in the intensity between ice and silicate agglomerates, but otherwise they provide a supplement to the results by \citet{west1991}.
They suggested that the DDA and the TMM should be combined with ray-tracing techniques if one considers a broad range of agglomerate sizes.

\citet{xing-hanner1997} considered agglomerates of tetrahedral constituent grains as well as those of spherical constituent grains in order to study the effect of grain's shapes on the angular dependences of intensity and polarization at $\lambda = 0.6\,\mathrm{\mu m}$.
Their numerical simulations were performed using the {\sf DDSCAT} code (ver.\,4b) with the LDR method for the determination of dipole polarizabilities.
Agglomerates of glassy carbon were assumed to consist of $N = 4$ or $10$ grains with a radius of $a_{0} = 0.25$ or $0.5\,\mathrm{\mu m}$, which are either touching, overlapping, or separated, while silicate agglomerates were assumed to consist of $10$ touching tetrahedral grains with $a_{0} = 0.25\,\mathrm{\mu m}$.
In comparison to the degree of linear polarization, the angular dependence of intensity showed less variations with the size and shape of the agglomerates and the shape of the grains.
Although they concluded that a mixture of carbon and silicate agglomerates represents cometary dust, the presence of ripples in the angular dependences of intensity and polarization has not been detected by observations to date.
 
\citet{levasseurregourd-et-al1997} computed the angular dependences of intensity and linear polarization for BPCA ($D \approx 3$) and BCCA ($D \approx 2$) particles of spherical constituent grains at $\lambda = 0.62\,\mathrm{\mu m}$.
They employed the {\sf DDSCAT} code (ver.\,4b) with presumably the LDR method to compute the light-scattering properties of the agglomerates.
The number of constituent grains was fixed to $N = 512$, but the size of the agglomerates lay in the range from $x_\mathrm{v} = 0.90$ to $6.02$, in other words, $x_\mathrm{0} = 0.11$--$0.75$.
Because the angular dependences of intensity and polarization for BPCA particles exhibit ripples at $x_\mathrm{v} > 2$ regardless of the refractive index, they concluded that primitive dust particles in the Solar System are agglomerates with $D \sim 2$.
On the one hand, we are unable to agree with their conclusion, since their results have not been reproduced by any of the later studies with BPCA particles at $x_\mathrm{v} > 2$ \citep[cf.][]{kimura2001,kimura-et-al2003c,kimura-et-al2006,bertini-et-al2007,kolokolova-mackowski2012}.
On the other hand, we notice that compact agglomerates at $x_\mathrm{v} > 2$ may exhibit ripples in their angular dependences of intensity and polarization if their constituent grains are located on a periodic lattice or distributed in a more symmetric manner \citep[cf.][]{petrova-et-al2000,petrova-et-al2001a,kolokolova-mackowski2012}.
It is most likely that their computations were performed with fast-Fourier-transform methods that require dipoles to be located on a periodic lattice \citep[see][]{draine-flatau1994}.
Because constituent grains were replaced by single dipoles in their computations, we cannot help wondering whether the ripples resulted from the displacement of constituent grains on a periodic lattice.

To better understand optical observations of zodiacal light at small phase angles, \citet{nakamura-okamoto1999} studied the angular dependences of intensity and polarization for BPCA ($D \approx 3$) particles consisting of spherical silicate grains.
They used the {\sf DDSCAT} code (ver.\,4a) with the $a_1$-term method by assuming $a_{0} = 0.03\,\mathrm{\mu m}$ at $\lambda = 0.5\,\mathrm{\mu m}$ so that $x_{0} < 1$.
Their results revealed that the gegenschein and the negative branch of linear polarization are insensitive to the number of constituent grains, which lay in the range from $N = 8000$ (${a}_\mathrm{v} = 0.6\,\mathrm{\mu m}$) to $N = 27000$ (${a}_\mathrm{v} = 0.9\,\mathrm{\mu m}$).
They attributed quantitative disagreements between their numerical results and zodiacal light observations to the size and shape of the constituent grains.
However, for a quantitative discussion on a model of zodiacal light, one should integrate light scattered by agglomerates along the line of sight.
Furthermore, it is clear that large agglomerates (${a}_\mathrm{v} \gg 1\,\mathrm{\mu m}$) are required to model the strong forward scattering of zodiacal light.

\begin{figure*}
\center
\resizebox{1.0\textwidth}{!}{%
  \includegraphics{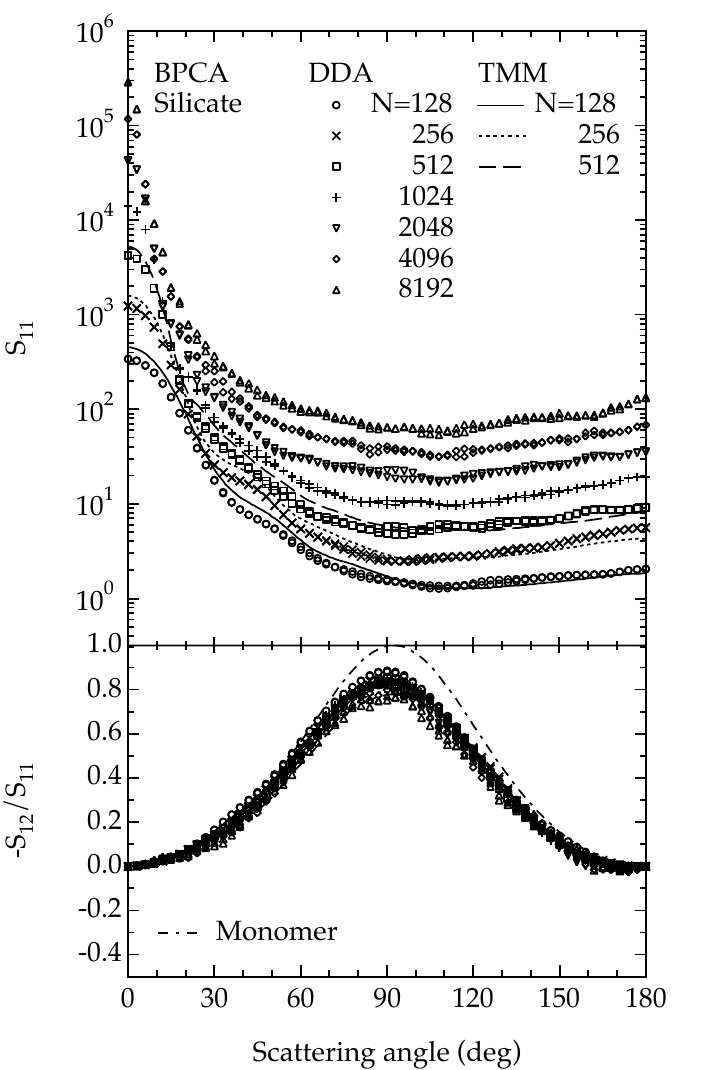}
  \includegraphics{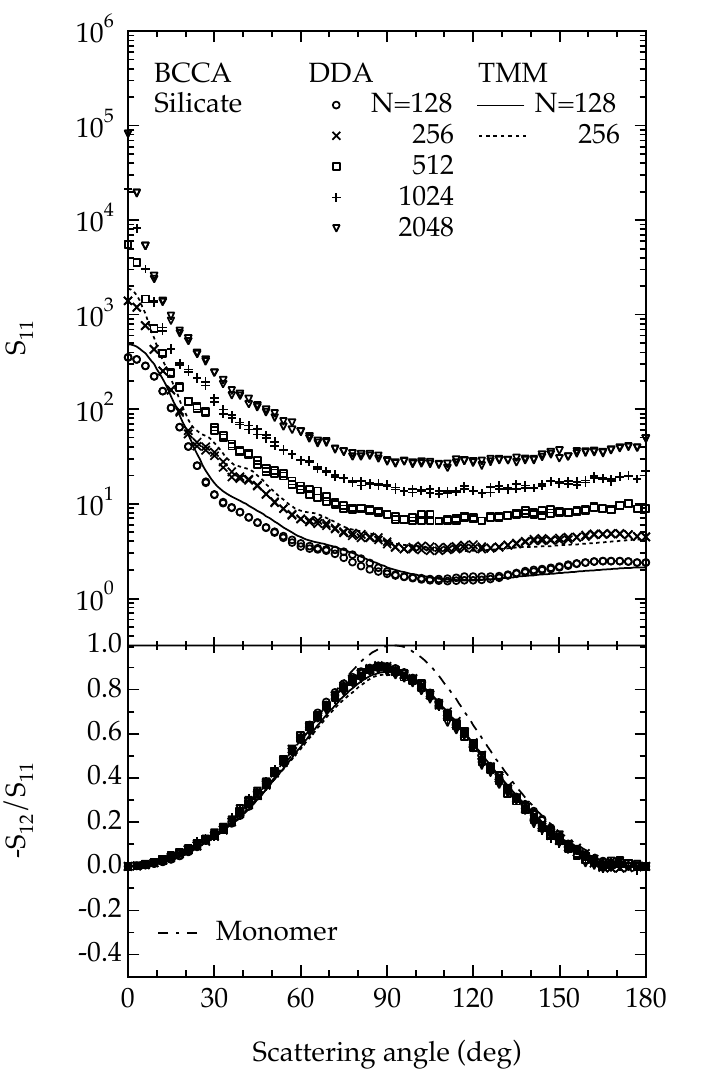}
}
\resizebox{1.0\textwidth}{!}{%
  \includegraphics{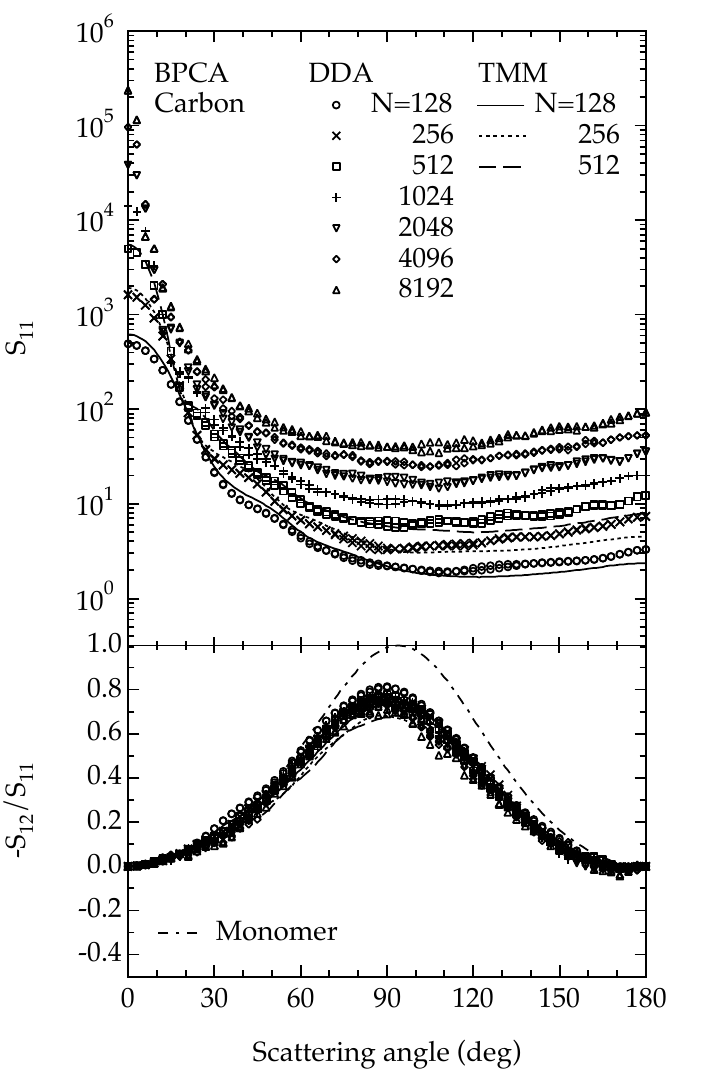}
  \includegraphics{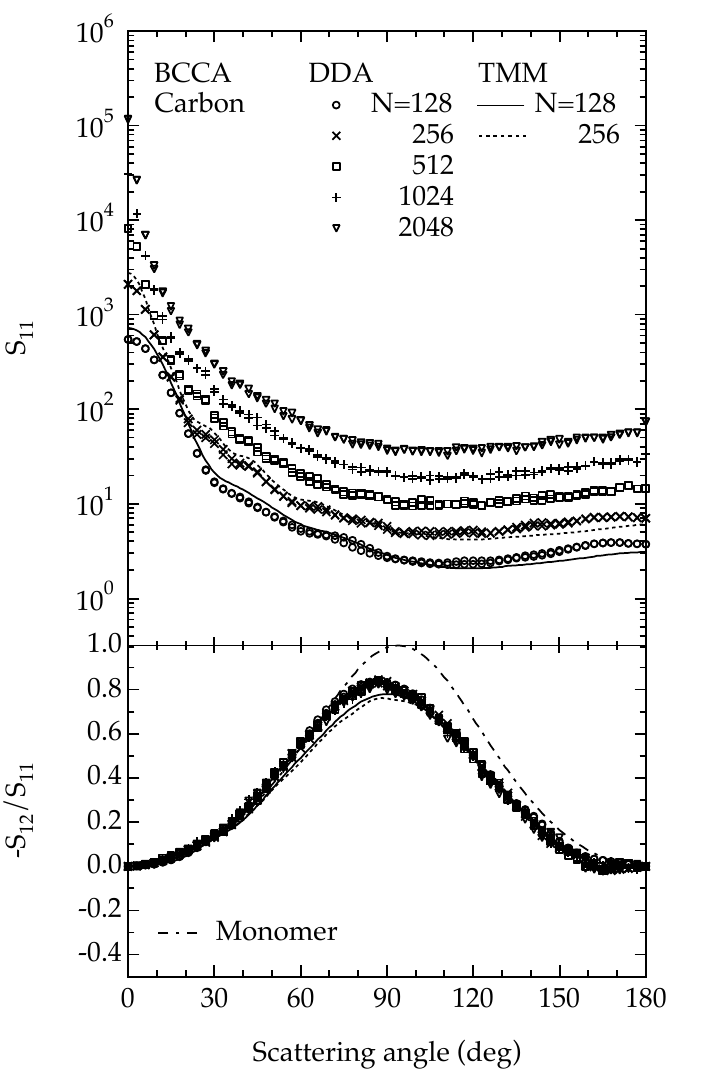}
}
\caption{The dependences of intensity $S_{11}$ and linear polarization $-S_{12}/S_{11}$ on scattering angle at a wavelength $\lambda = 0.6\,\mathrm{\mu m}$ for BPCA particles ({\it left}) and BCCA particles ({\it right}) consisting of silicate grains ({\it upper panels}) and carbon grains ({\it lower panels}) with $a_{0} = 0.07\,\mathrm{\mu m}$. Also plotted as {\it dash-dotted lines} are the $-S_{12}/S_{11}$ values for the constituent grains. From \citet{kimura2001}}
\label{kimura2001:f3}       
\end{figure*}

\citet{kimura2001} investigated intensively how the angular dependences of intensity and linear polarization for fractal agglomerates of spherical constituent grains depend on the grain size at a wavelength $\lambda = 0.6\,\mathrm{\mu m}$.
The radius of constituent grains lay in the range of $a_{0} = 0.01$--$0.15\,\mathrm{\mu m}$ ($x_{0} = 0.10$--$1.57$) and the configuration of the grains was determined by either the BPCA ($D \approx 3$) or the BCCA ($D \approx 2$) process.
The {\sf DDSCAT} code (ver.\,4a) with the $a_1$-term method and the {\sf scsmtm1} code were applied to agglomerates of up to $8192$ grains with $a_{0} = 0.01\,\mathrm{\mu m}$, $512$ grains with $a_{0} = 0.07\,\mathrm{\mu m}$, and $256$ grains with $a_{0} = 0.15\,\mathrm{\mu m}$.
It turned out that the morphology of agglomerates is of importance for the smallest grains with $a_{0} = 0.01\,\mathrm{\mu m}$, but otherwise the angular dependences of intensity and linear polarization are independent of coagulation process.
A comparison of the results between silicate and carbon agglomerates indicates that the composition of the constituent grains strongly affects the optical properties of agglomerates with $a_{0} = 0.15\,\mathrm{\mu m}$ ($x_{0} = 1.57$) as opposed to the comparison between ice and silicate agglomerates studied by \citet{lumme-et-a1997}.
Interestingly, the degree of linear polarization becomes negative at backscattering not only for silicate agglomerates but also for carbon agglomerates if $x_{0} \sim 1$ and $x_\mathrm{v} \gg 1$ (see Fig.\,\ref{kimura2001:f3}).
This result shed new light on the presence of the negative polarization branch at small phase angles for primitive dust particles whose geometric albedo is very small.

\citet{petrova-et-al2000,petrova-et-al2001a} used the {\sf scsmtm1} code to study the optical properties of silicate agglomerates consisting of spherical grains that were arranged in a tetrahedral or cubic lattice.
The size parameter of the constituent grains lies in the range of $x_{0} = 0.7$--$2.5$, which corresponds to $a_{0} = 0.07$--$0.25\,\mathrm{\mu m}$ at a wavelength of $\lambda = 0.63\,\mathrm{\mu m}$ and $a_{0} = 0.06$--$0.21\,\mathrm{\mu m}$ at $\lambda = 0.54\,\mathrm{\mu m}$.
A special feature of their work is that a size distribution of the agglomerates was taken into consideration, although their agglomerates were composed of only $1$--$43$ grains.
They claim that the intensity and the polarization of their agglomerates and cometary dust are similar with respect to their angular dependences, if the grain radius lies in the range of $x_{0} = 1.3$--$1.65$.
As the authors noticed, however, their results on the intensity and the polarization do not provide conclusive evidence for their wavelength dependences, owing to a lack of large agglomerates in their models.

\citet{petrova-et-al2004} extended their study on angular dependences of intensity and polarization to larger silicate agglomerates of $12$--$150$ constituent grains.
They used BPCA and DLA particles consisting of spherical grains with $a_{0} \approx 0.1\,\mathrm{\mu m}$ and considered a power-law size distribution of the particles with the power of $-3$. 
The agglomerates were constructed on the assumption that the constituent grains hit and stick on contact only if the grains have two or more contacts.
Although the agglomerates were still small, they were characterized by a fractal dimension $D \approx 2$--$3$ and a porosity $p \approx 0.95$--$0.63$. 
Their results show that the degree of the maximum polarization is higher and the scattering angle of the maximum polarization is larger than observed for comets.
They attributed the discrepancies between their model results and observations to insufficiency of the number of constituent grains, but part of the discrepancies most likely resulted from low refractive indices assumed in their paper.

\begin{figure*}
\resizebox{1.0\textwidth}{!}{%
  \includegraphics{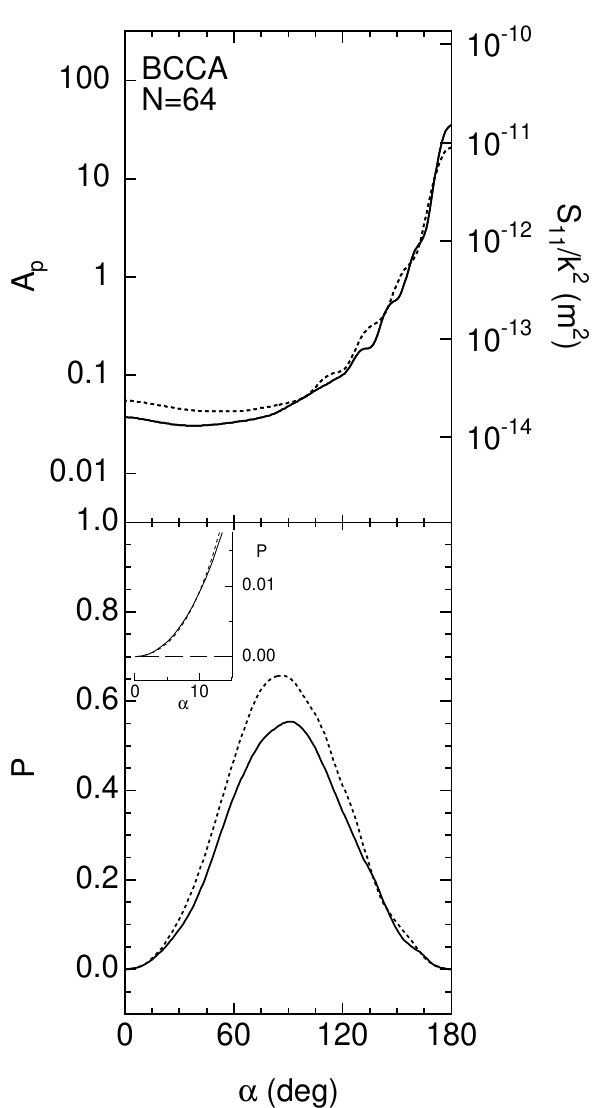}
  \includegraphics{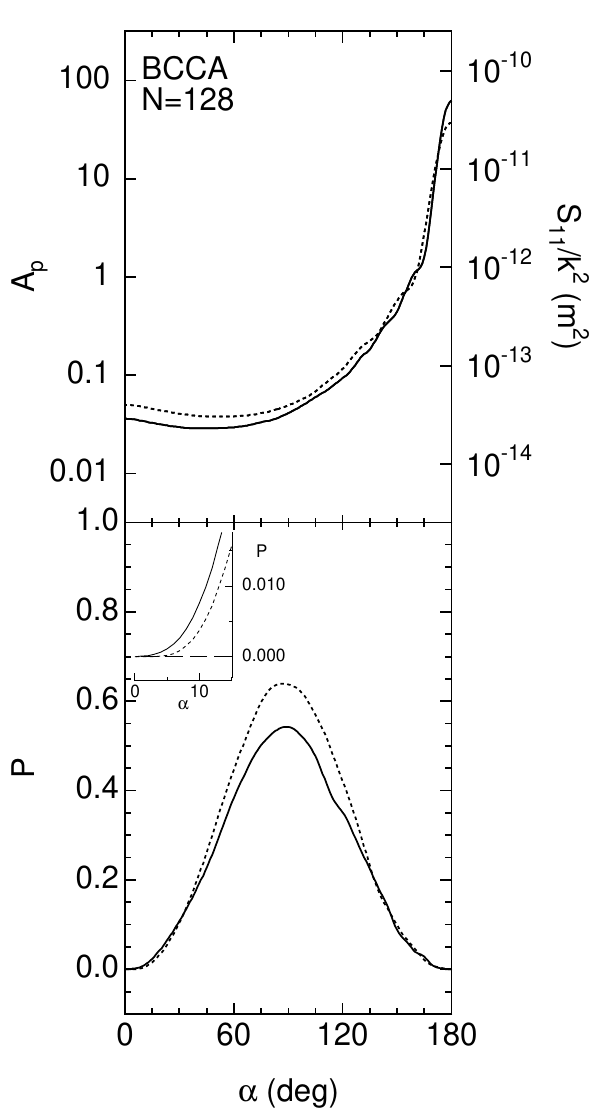}
  \includegraphics{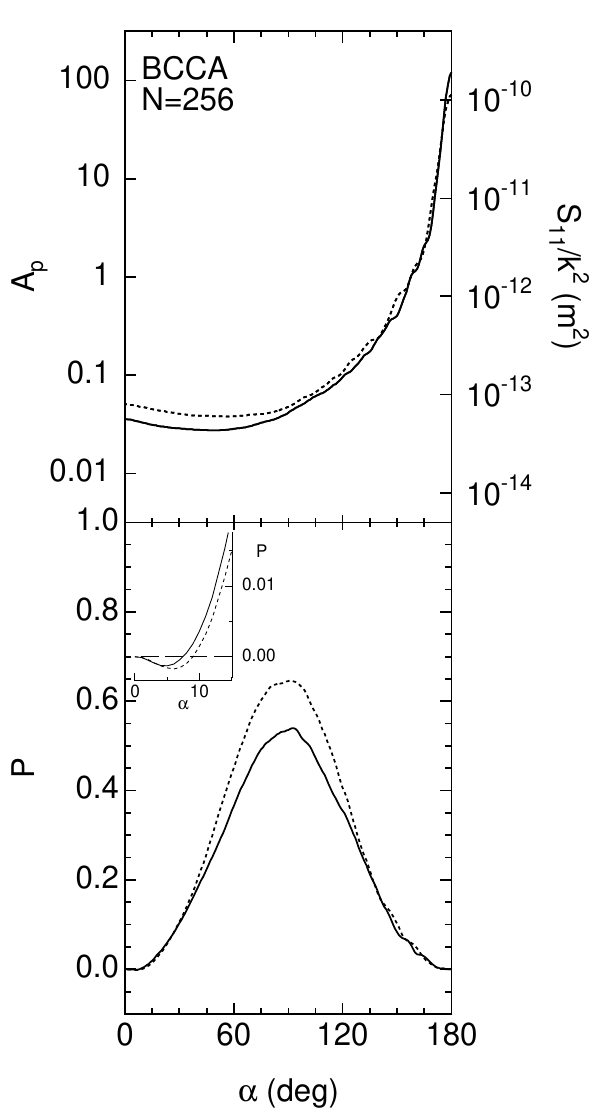}
}
\caption{The geometric albedo $A_\mathrm{p}$ and linear polarization $P$ of BCCA ($D \approx 2$) particles consisting of monodisperse spherical grains as a function of phase angle $\alpha$. 
The constituent grains have a radius $a_{0} = 0.1\,\mathrm{\mu m}$ and a composition inferred from the elemental abundances measured for comet 1P/Halley.
{\it Solid lines} and {\it dotted ones} are the corresponding values at a wavelength $\lambda = 0.45\,\mathrm{\mu m}$ and $\lambda = 0.60\,\mathrm{\mu m}$, respectively. From \citet{kimura-et-al2003c}}
\label{kimura-et-al2003c:f1}       
\end{figure*}

It is worth noting that the dependences of geometric albedo and linear polarization on scattering angle and wavelength for cometary dust show common characteristics among a variety of comets \citep{gustafson-kolokolova1999}.
\citet{kimura-et-al2003c} succeeded for the first time in qualitatively reproducing all the common characteristics of cometary dust simultaneously using BPCA ($D \approx 3$) and BCCA ($D \approx 2$) particles of optically dark submicron constituent grains (see Fig.\,\ref{kimura-et-al2003c:f1}).
The authors first determined the average refractive indices of the constituent grains by applying the Maxwell Garnett mixing rule to a mixture of silicate, iron, organic refractory, and amorphous carbon.
The volume fractions of these substances were derived from the elemental abundances of dust in comet 1P/Halley that were measured in situ by PUMA-1 onboard VeGa 1 \citep[see][for the elemental abundances]{jessberger-et-al1988}.
The average refractive indices were then used for their numerical computations with the {\sf scsmtm1} code for agglomerates of identical spherical constituent grains with $a_{0} = 0.1\,\mathrm{\mu m}$.
Their results did not show a clear dependence on the structure of agglomerates between the BPCA and BCCA processes nor the number of constituent grains.
Therefore, their success in reproducing the common characteristics can be attributed to the use of optically dark material, which is expected from the composition of cometary dust \citep[see also][]{mann-et-al2004,kolokolova-et-al2005}.
In conclusion, the presence of common light-scattering characteristics in cometary dust reflects the fact that comets formed out of the same protoplanetary disk materials with the solar composition.

\citet{kimura-et-al2006} have thoroughly explored which size, number, composition, and configuration of constituent grains reproduce the common light-scattering characteristics of cometary dust.
The authors applied the {\sf scsmtm1} code to numerical simulation of light scattering by BPCA ($D \approx 3$) and BCCA ($D \approx 2$) particles of up to 256 spherical constituent grains.
Their numerical results were used to place constraints on the size and composition of constituent grains, confirming the successful model of \citet{kimura-et-al2003c} based on the properties of CP IDPs and dust in comet 1P/Halley.
The contribution of amorphous carbon to the results is so significant that the common light-scattering characteristics of cometary dust could be reproduced even by agglomerates without organic materials, in particular, agglomerates of amorphous carbon.
It turned out that the size and composition of constituent grains play a crucial role in the determination of optical properties.
Although the size and composition of constituent grains are well constrained in the framework of the model, quantitative fits to the observed angular dependence of polarization seem to require a larger number and non-spherical shape of constituent grains \citep[cf.][]{kimura-mann2004}.

A similar numerical approach to constraining the morphology and composition of cometary dust was taken by \citet{bertini-et-al2007}.
They used the {\sf DDSCAT} code (ver.\,5a10) presumably with the LDR method to study light scattering by BPCA particles of up to $100$ spherical particles with $a_{0} = 0.13$--$0.16\,\mathrm{\mu m}$ at $\lambda = 0.535$, $0.6274$, and $1.5\,\mathrm{\mu m}$.
They assumed the agglomerates to be composed of silicates, organic materials, or a mixture of silicates and organic materials with equal mass.
Unfortunately, they failed to find a solution for the morphology and composition of agglomerates that simultaneously reproduces all the common characteristics of cometary dust.
Nevertheless, they suggest that two distinct types of linear polarization among different comets suggested by \citet{levasseurregourd-et-al1996} arise from the difference in the radius of constituent grains. 
However, as discussed in \citet{kolokolova-et-al2007}, the presence of two polarimetric classes might be artifact due to a contribution of gas emission to the continuum measured with broad-band filters.

\begin{figure*}
\center
\resizebox{0.75\textwidth}{!}{%
  \includegraphics{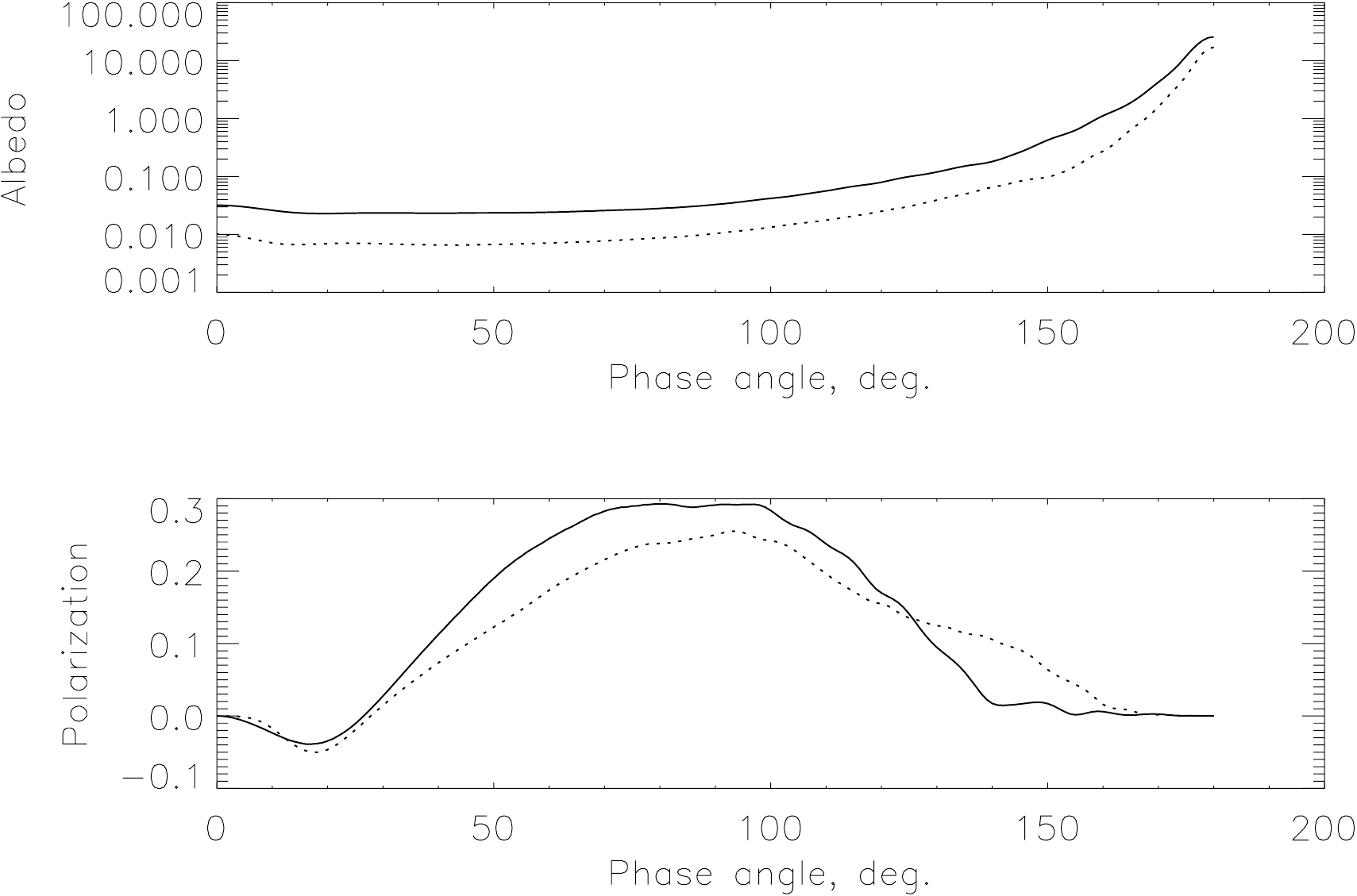}
}
\caption{The dependences of geometric albedo $A_\mathrm{p}$ and linear polarization $P$ on phase angle $\alpha$ for a mixture of BCCA ($D \approx 2$) particles consisting of $N=256$ spherical grains with $a_{0} = 0.1\,\mathrm{\mu m}$ and compact spheroidal particles with a power-law size distribution. 
{\it Dotted lines} and {\it solid ones} are the corresponding values at wavelengths $\lambda = 0.45\,\mathrm{\mu m}$ and $\lambda = 0.60\,\mathrm{\mu m}$, respectively. 
The agglomerates are either an admixture of silicates, metals, and carbonaceous materials, or pure organic refractory material, while the spheroids are pure silicates.
From \citet{kolokolova-kimura2010}}
\label{kolokolova-kimura2010:f1}       
\end{figure*}

Using the {\sf scsmtm1} code, \citet{kolokolova-kimura2010} modeled cometary dust with two types of BCCA particles consisting of 256 identical spherical grains with $a_{0} = 0.1\,\mathrm{\mu m}$, along with a multi-shaped, polydisperse mixture of spheroids.
The authors considered not only the elemental abundances of dust in comet 1P/Halley, but also the mineralogical classification of the dust \citep[see][for the classification]{fomenkova1999}.
In the model, the agglomerates are composed of either organic refractory material alone or a mixture of silicate, metal, organic refractory, and amorphous carbon, and the spheroids are composed of silicate alone.
The three types of particles were intended to model organic-rich, silicate-poor particles, organic-poor, silicate-rich particles, and compositionally intermediate particles. 
The results have shown that the model is successful in reproducing all the common characteristics of cometary dust not only qualitatively, but also quantitatively (see Fig.\,\ref{kolokolova-kimura2010:f1}).
This might indicate that a mixture of compositionally different particles is a key to the common characteristics of cometary dust, instead of large agglomerates suggested by \citet{kimura-et-al2006}.

\citet{moreno-et-al2007} applied the {\sf DDSCAT} code (ver.\,6.0) with the LDR method to simulate light scattering by cometary dust with DLA particles ($D \approx 2.5$) of up to 256 constituent grains. 
They assumed that either constituent grains are cubes with the side length of $0.15\,\mathrm{\mu m}$ or they are spheres with $a_{0} = 0.075\,\mathrm{\mu m}$. 
Using the same refractive indices as \citet{kimura-et-al2003c}, their results show that the maximum degree of linear polarization for agglomerates of cubes is lower than that for agglomerates of spheres, while the maximum takes place at a smaller scattering angle for cubic constituent grains than spherical ones.
However, it should be noted that the diameter of spherical constituent grains equals the side length of cubic constituent grains, implying that the volume of the spheres is smaller than that of the cubes. 
Therefore, it is not clear whether the difference in the results between cubic constituent grains and spherical ones arises from the shape or the size of the grains.

\citet{lumme-penttilae2011} employed not only agglomerates of spherical constituent grains but also those of Gaussian random constituent grains with the same volume as the spherical grains.
They used the {\sf scsmfo1b} code with the OCoS method to the agglomerates of spherical grains and the {\sf ADDA} code presumably with the LDR method to the agglomerates of Gaussian random grains.
The morphology of agglomerates was determined either by the BPCA process with controls on the ballistic trajectories or by the BCCA process with an assemblage of 16 BPCA particles at a single point.
The radius, number, and composition of constituent grains lay in the range of $0.05 \le a_{0} \le 0.25\,\mathrm{\mu m}$ ($0.5 \le x_{0} \le 2.5$), $128 \le N \le 512$, and ice to silicate, respectively.
Both the constituent grains and the agglomerates were assumed to either be monodisperse or have a size distribution. 
A comparison of the results between agglomerates of spherical constituent grains and those of Gaussian random constituent grains shows similar angular dependences of intensity and polarization, although there are noticeable differences in their magnitudes.
They claim the presence of a strong correlation between the real part and the imaginary part of refractive index that results in almost the same angular dependences of intensity and polarization.
If this correlation is universal, then any attempts to fit observational data without the geometric albedo will not provide a unique solution.

\citet{zubko-et-al2011,zubko-et-al2012,zubko-et-al2013,zubko-et-al2014} consider cometary dust as agglomerates of irregularly shaped polydisperse constituent grains, along with a power-law size distribution of the agglomerates \citep[see also][]{zubko2012,zubko2013}. 
They apply the DDA with the LDR method to solve light-scattering problems using their own code {\sf ZDD}, instead of the publicly available {\sf DDSCAT} and {\sf ADDA codes} \citep{penttilae-et-al2007,zubko-et-al2010}.
The main problem of their approach is how the size distribution of the agglomerates was achieved, since they used the same agglomerates to scale larger and smaller sizes of agglomerates.
Namely, the larger is the agglomerate, the larger are its ``constituent grains'', but in all other respects all the agglomerates are identical to one another.
In reality, larger agglomerates have beyond a shadow of doubt a larger number of constituent grains, not the larger size of the grains. 

\citet{lasue-levasseurregourd2006} presented their numerical simulations on light scattering by BPCA ($D \approx 3$) and BCCA ($D \approx 2$) particles of up to $128$ spheres or prolate spheroids with the axis ratio of two. 
The constituent grains have a radius of $a_{0} = 0.1\,\mathrm{\mu m}$ and consist of pure silicate, pure organics, or a mixture of silicate in the core and organics in the mantle.
Their computational results with the {\sf DDSCAT} code (ver.\,5a10) presumably with the LDR method showed a great similarity in the optical properties of agglomerates between the organic-coated silicate constituent grains and the pure organic ones. 
The paper also provides a model of fractal agglomerates with a size distribution that fits the observational data on the angular dependence of linear polarization for comet C/1995 O1 (Hale-Bopp). 
Despite the best fit to the observed polarization in the range of phase angles from 0 to $50^\circ$, the geometric albedo of the agglomerates seems to be higher than observed.

\citet{lasue-et-al2009} extended their light-scattering modeling of cometary dust using agglomerates of up to $N=256$ constituent grains of $a_0 = 0.1\,\mathrm{\mu m}$ and fitting to the polarimetric data for both comets C/1995 O1 (Hale-Bopp) and 1P/Halley. 
They used the {\sf DDSCAT} code (ver.\,6.1) presumably with the LDR method for their numerical simulations with agglomerates of spherical or spheroidal constituent grains.
Although they obtained a good fit to the angular dependence of linear polarization for comet Hale-Bopp using a mixture of agglomerates and compact spheroids, they failed to reach the observed values of negative polarization for comet 1P/Halley.
It is unfortunate that this study provides no clue about the geometric albedo nor the color of their agglomerates and that the size of the constituent grains is scaled to represent the size distribution of the agglomerates.

\citet{golimowski-et-al2006} presented multiband coronagraphic images of the debris disk around the A-type star $\beta$ Pictoris.
They modeled the observed red colors of the disk using light-scattering properties of porous agglomerates computed by \citet{wolff-et-al1998} and \citet{voshchinnikov-et-al2005}.
On the one hand, \citet{wolff-et-al1998} used the {\sf DDSCAT} code (ver.\,4c) with the LDR method for agglomerates consisting of pure silicate grains encased in a spherical volume with $60\%$ of vacuum.
On the other hand, \citet{voshchinnikov-et-al2005} used the {\sf DDSCAT} code (ver.\,6.0) with the LDR method for agglomerates consisting of silicate grains and graphite grains half and half encased in a spherical volume with $33\%$ or $90\%$ of vacuum.
\citet{golimowski-et-al2006} reconciled model results with the observed color of the disk except for highly porous agglomerates consisting of silicate grains and graphite grains.
It should be, however, noted that the observed red color does not refute a predominance of highly porous agglomerates in the disk, because fractal agglomerates of submicron constituent grains with a rocky core and a carbonaceous mantle exhibit red colors \citep[cf.][]{kimura-et-al2003c,kimura-et-al2006}. 

\citet{graham-et-al2007} derived the phase function and polarization of dust particles in the debris disk of the young (12\,Myr) nearby M-type star AU Microscopii as a function of scattering angle from optical measurements by the Hubble Space Telescope.
The scattering-angle dependences of phase function and polarization were compared to light-scattering properties of porous spheres and those of agglomerates computed by \citet{petrova-et-al2000} and \citet{kimura-et-al2006}.
It turned out that the phase function and the polarization are better reproduced by the optical properties of BPCA particles consisting of $128$ silicate grains with $x_0 = 0.9$.
The color of silicate BPCA particles is blue in the visible wavelength range, which agrees with the observed color of the AU Mic debris disk \citep[cf.][]{krist-et-al2005}.
These results are entirely consistent with a picture that dust particles in the AU Mic disk originate from planetesimals, similar to asteroids in the Solar System.

The debris disk of the nearby young (8\,Myr) A-type star HR\,4796A appears to be a circumstellar dust ring around 70\,au from the central star \citep{schneider-et-al1999}.
The ring with a red color in the visible and a gray color in the near-infrared wavelength range
led \citet{debes-et-al2008} to argue for the presence of complex organic materials like Titan's tholins.
In contrast, \citet{koehler-et-al2008} modeled the visible to near-infrared spectrum of the ring with porous agglomerates of amorphous silicate, amorphous carbon, and water ice.
They considered that spherical constituent grains of $a_0 = 0.1\,\mathrm{\mu m}$ coagulated into agglomerates with a porosity $p = 0.73$.
They used the so-called Henyey-Greenstein phase function, which is an analytic function specified by the single scattering albedo and the asymmetry parameter \citep{henyey-greenstein1941}.
The single scattering albedo and the asymmetry parameter of the porous agglomerates were computed by the Bruggeman mixing rule.
It should be noted that the Bruggeman mixing rule does not properly describe the light-scattering properties of agglomerates, because it ignores the interaction between constituent grains which plays a vital role in light scattering \citep[cf.][]{kimura-mann2004,kolokolova-kimura2010b}.

\citet{shen-et-al2009} studied light-scattering properties of BPCA particles with different degrees of restructuring and applied their results to the AU Mic debris disk and cometary comae.
The agglomerates were composed either of silicate alone or a mixture of silicate and graphite in equal volumes, although both the compositions are inconsistent with our current understandings of cometary dust.
The number and radius of spherical constituent grains were considered to lie in the range of $N=32$--$1024$ and $a_0 = 0.0143$--$0.16\,\mathrm{\mu m}$.
The {\sf DDSCAT} code (ver.\,7.0) was used along with the ``modefied'' LDR method, in which \citet{gutkowiczkrusin-draine2004} corrected a subtle error in the LDR determined by \citet{draine-goodman1993}.
Contrary to previous studies, the computational results at $\lambda = 0.1$--$3.981\,\mathrm{\mu m}$ have shown that the size of constituent grains does not play a vital role in the light-scattering properties of agglomerates, if the mass and porosity of agglomerates are fixed.
However, the results in \citet{kimura2001} reveal that light-scattering properties of agglomerates varies with the size of constituent grains, even if the mass and porosity of agglomerates are fixed.
Therefore, the agglomerates used in \citet{shen-et-al2009} might be so dense that their light-scattering properties represent those of porous spheres than those of fluffy agglomerates.

\begin{figure*}
\center
\resizebox{1.0\textwidth}{!}{%
  \includegraphics{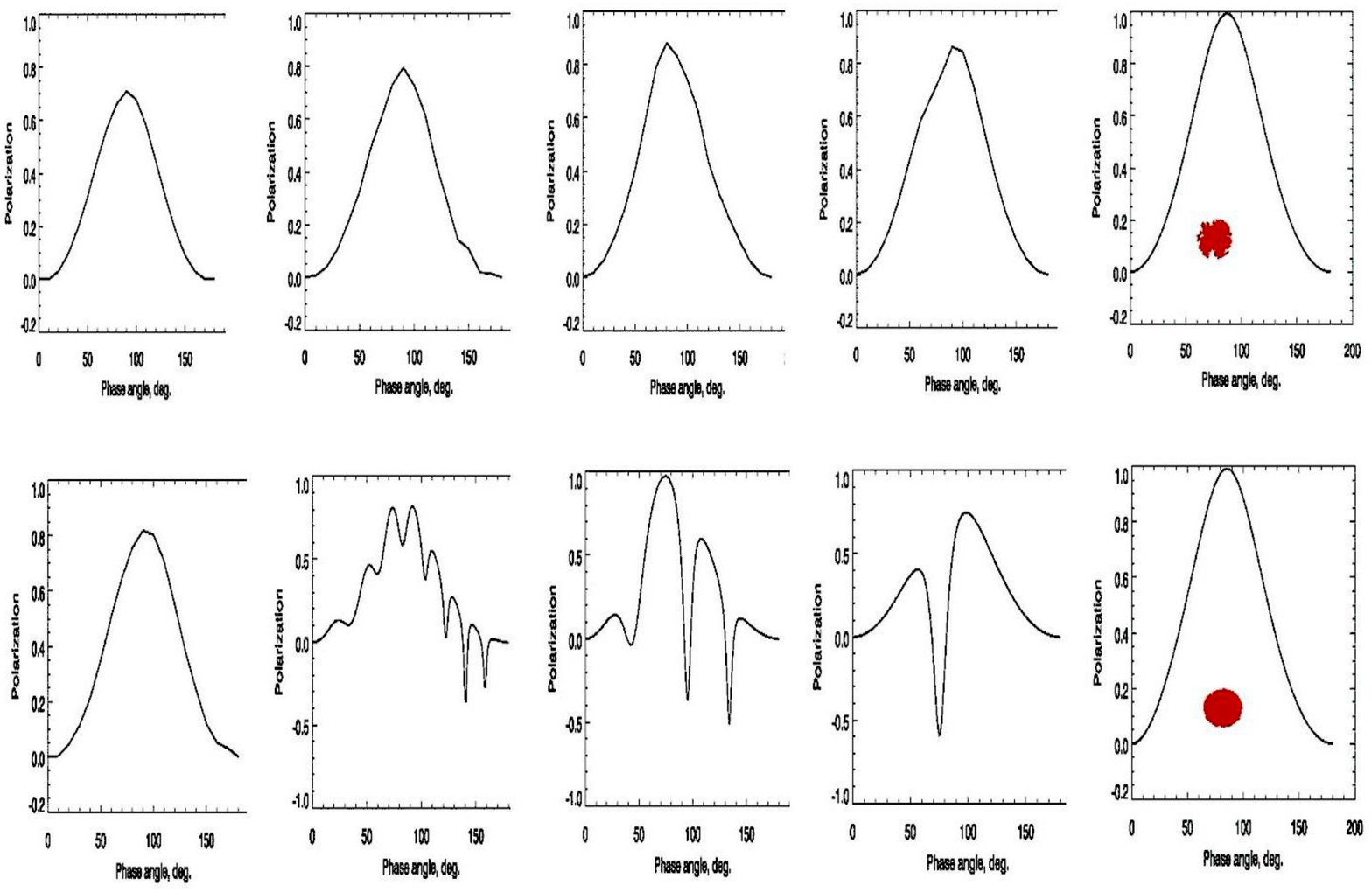}
}
\caption{The degree of linear polarization $P$ as a function of phase angle $\alpha$ for BPCA ($D \approx 3$) particles consisting of $N=1024$ spherical grains with $a_{0} = 0.1\,\mathrm{\mu m}$ ({\it upper panel}) and equi-dimensional agglomerates consisting of $N=1024$ spherical grains with $a_{0} = 0.1\,\mathrm{\mu m}$ randomly distributed in a spherical volume with equal porosity ({\it lower panel}) at wavelengths $\lambda = 0.6\,\mathrm{\mu m}$, $1.1\,\mathrm{\mu m}$, $2.2\,\mathrm{\mu m}$, $4.4\,\mathrm{\mu m}$, and $6.0\,\mathrm{\mu m}$ (from {\it left} to {\it right}). From \citet{kolokolova-mackowski2012}}
\label{kolokolova-mackowski2012:f6}       
\end{figure*}

\citet{kolokolova-mackowski2012} used the {\sf MSTM} code (ver.\,3.0) to intensively study linear polarization of light scattered by agglomerates of $N=1024$ spherical constituent grains with $a_0 = 0.1\,\mathrm{\mu m}$.
They have shown how the porosity and the overall size of agglomerates can be constrained by spectroscopic observations of linear polarization from the visible to the near-infrared wavelength range.
Regardless of the wavelength, the refractive index was fixed at the value that was suggested for cometary dust at $\lambda = 0.45\,\mathrm{\mu m}$ by \citet{kimura-et-al2003c}.
As demonstrated in Fig.\,\ref{kolokolova-mackowski2012:f6}, misleading numerical results on the degree of linear polarization might arise from an artificial configuration of constituent grains.
Consequently, it is essential for a correct understanding of light scattering by primitive dust particles in planetary system to properly model dust agglomerates based on their formation mechanisms.

\citet{videen-muinonen2015} applied a radiative transfer technique with coherent backscattering to compute light scattering by sparse agglomerates of identical spherical constituent grains whose positions were chosen randomly and uniformly within a spherical volume.
They considered two porosities of 0.94 and 0.97, and assumed the constituent grains to have a size parameter $x_0 = 1.76$ and be non-absorbing.
To our knowledge, this is the first study that was successful in computing light scattering by large agglomerates of $N \sim 3 \times {10}^8$ despite the fact that the assumption of non-absorbing material is inappropriate for primitive dust particles in planetary systems.
However, their computations show highly unrealistic phase functions where the forward-scattering peak does not appear for large agglomerates of $N > {10}^5$.
It is unfortunate that the effects of diffraction, which dominate the forward-scattering region, have not been incorporated in their model.

\subsection{Multiple scattering}
\label{sec:4.2}

Multiple scattering may be of crucial importance in the vicinity of a source region (e.g., a comet nucleus) where dust particles are released from their parent bodies as well as on the surface (e.g., a regolith layer) of their parent bodies.
Impacts of micrometeoroids onto the surfaces of asteroids are common phenomena that produce not only regolith particles but also impact craters and ejecta curtains.
Dusty ejecta clouds in Saturn's rings produced by impacts of meteoroids onto the rings have been imaged by Cassini's Imaging Science Subsystem \citep{tiscareno-et-al2013}.
NASA's Deep Impact mission excavated the surface of comet 9P/Tempel 1 by an artificial impactor to produce an impact crater on the comet as well as an ejecta curtain \citep{ahearn-et-al2005}.
A similar impact experiment and a subsequent imaging observation of an ejecta curtain are planed on asteroid 1999JU3 by the Japanese Hayabusa-2 mission \citep{arakawa-et-al2013}.
Recently, radiative transfer computations to model such an ejecta curtain have been performed with assumptions about the phase function, single scattering albedo, and asymmetry parameter of the ejecta particles \citep[cf.][]{nagdimunov-et-al2014,shalima-et-al2015}.

\begin{figure}
\center
\resizebox{1.0\textwidth}{!}{%
  \includegraphics{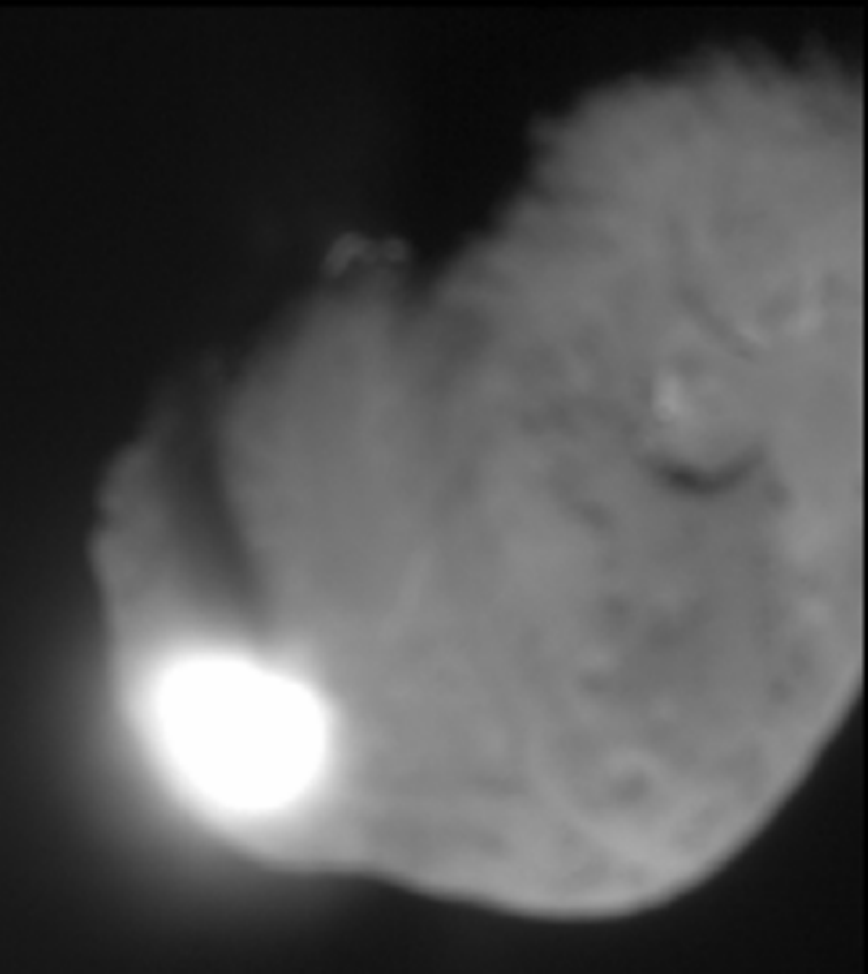}
  \includegraphics{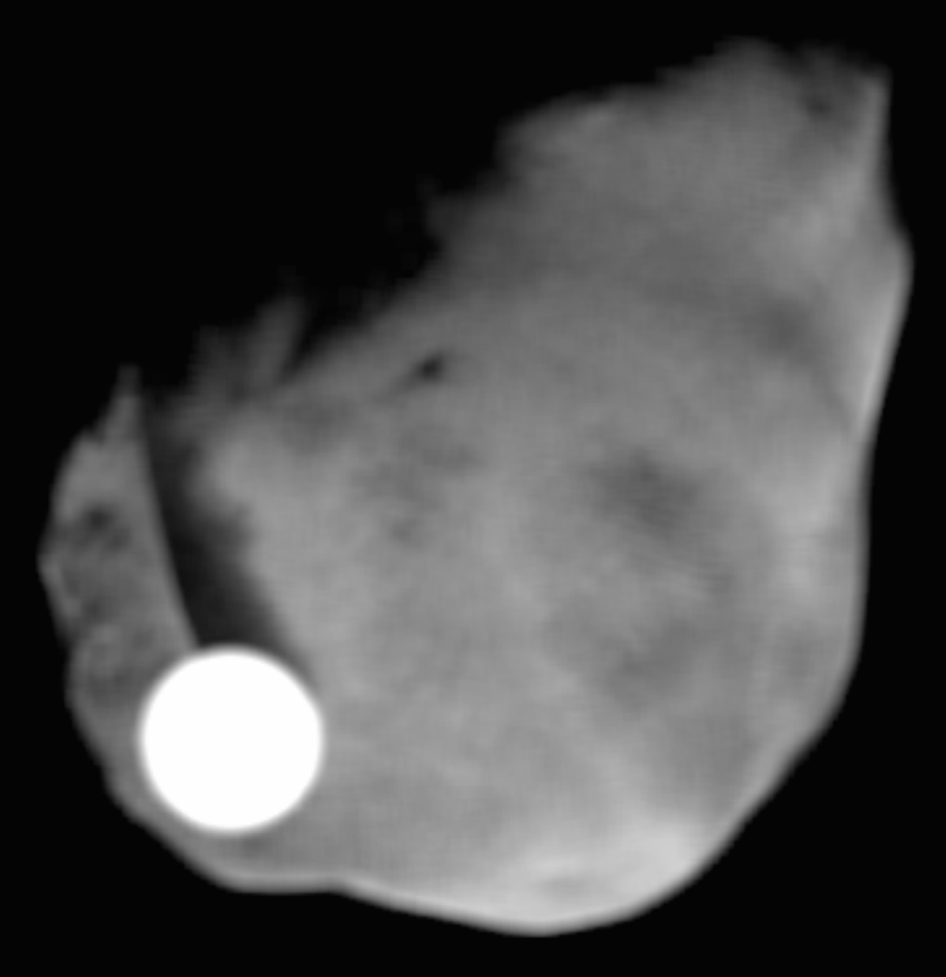}
}
\caption{The real image of the ejecta plume from the surface of comet 9P/Tempel 1 and its shadow on the surface taken by the Deep Impact High Resolution Instrument ({\it left}) and the simulated image of the ejecta plume and its shadow computed by radiative transfer computations ({\it right}). 
From \citet{nagdimunov-et-al2014}}
\label{nagdimunov-et-al2014:f1+3}       
\end{figure}

\citet{nagdimunov-et-al2014} have succeeded in modeling both an optically thick ejecta plume and its shadow on the surface of comet 9P/Tempel 1 produced and imaged by the Deep Impact mission (see Fig.\,\ref{nagdimunov-et-al2014:f1+3}).
They considered the ejecta particles to be porous spheres composed of ice, silicate, amorphous carbon, organic refractory, and ``vacuum'' materials.
The porous spheres were used to simulate light-scattering properties of fluffy agglomerates with a wide size range in the framework of EMAs.
They used the Maxwell Garnet mixing rule to compute the single scattering albedo, asymmetry parameter, and extinction cross section of the agglomerates at a wavelength $\lambda = 0.65\,\mathrm{\mu m}$.
However, they adopted the Henyey-Greenstein phase function to determine the scattering angles of photons probabilistically along the path of the photons in their radiative transfer computations.
Because the single scattering albedo and the asymmetry parameter are associated with the phase function of porous agglomerates based on the Maxwell Garnett mixing rule, the use of the Henyey-Greenstein phase function is self-contradictory.
It is, therefore, unfortunate that the optical properties of dust agglomerates have not yet been incorporated into available radiative transfer modelings for ejecta curtains in a self-consistent way.

To simulate light scattering by a regolith layer on the surface of asteroids, \citet{petrova-et-al2001b} solved the radiative transfer equation along with the optical properties of agglomerates that were computed by \citet{petrova-et-al2001a}.
The agglomerates were assumed to have a power-law size distribution and consist of $N=8$--$43$ spherical grains with $x_0 = 1.50$ or $1.65$.
The optical depth $\tau$ of a plane-parallel regolith layer, in which agglomerates were embedded, lay in the range of $\tau = 0.2$--$50$.
Their results have shown that the absolute value of the degree of linear polarization tends to decrease with $\tau$ in all scattering angles, although the results at $\tau > 10$ merge into a single curve.
It turned out that the negative and positive branches of linear polarization become shallower and higher as the imaginary part of refractive index decreases.
Their results may serve as an explanation for polarimetric observations of asteroids, in which S-type asteroids show a shallower negative branch and a higher positive branch, compared to C-type asteroids.

\section{Thermal emission from dust agglomerates}
\label{sec:5}

\subsection{Spectral energy distribution}
\label{sec:5.1}

One of the primary observational data available to constrain the physics of primitive dust particles in planetary systems is its spectral energy distribution (SED) constructed using broad-band photometric measurements from mid-infrared to millimeter wavelengths. 
The SED is dominated by silicate features in the mid-infrared wavelength range from $10\,\mathrm{\mu m}$ up to $70\,\mathrm{\mu m}$ and provides basic constraints on the dust temperature and optical properties. 
There is generally no unique solution to SED fitting problems owing in particular to the degeneracies between the optical properties of dust particles and the disk geometry. 
For example, dust particles in the vicinity of the central star are hotter than the particles of the same size and composition located far from the star, while small opaque particles are hotter than larger ones at the same distance from the star and large transparent particles are hotter than smaller ones at the same distance.

\begin{figure}
\center
\resizebox{0.75\textwidth}{!}{%
  \includegraphics{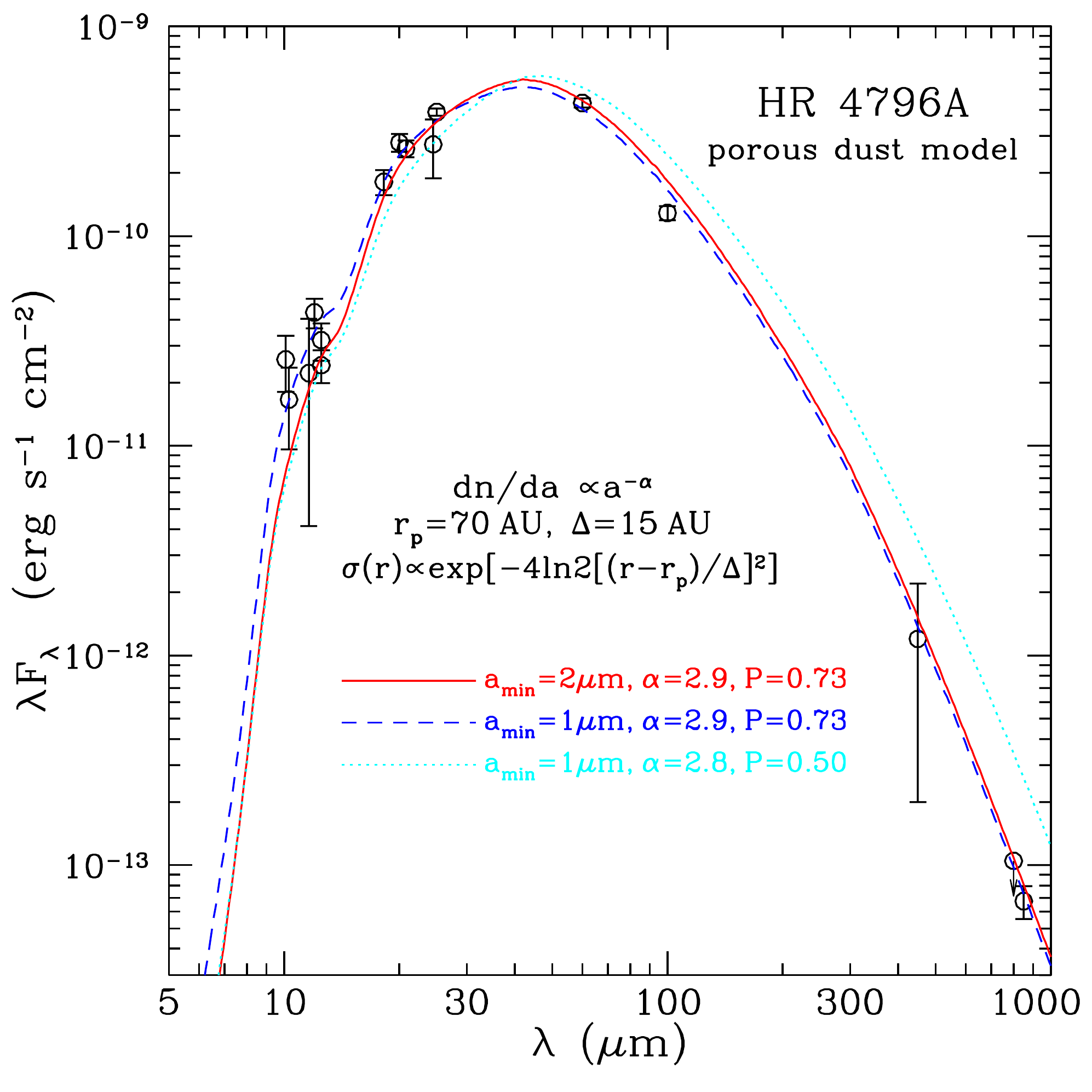}
}
\caption{Comparison of the observed infrared emission of the HR\,4796A debris disk to the model spectra calculated from the so-called porous dust model of \citet{li-greenberg1998}, in which the constituent grains of an agglomerate are composed of amorphous silicate, carbonaceous material, and ice. 
From \citet{koehler-et-al2008}}
\label{koehler-et-al2008:f3}       
\end{figure}

\citet{li-greenberg1998} reproduced the $10\,\mathrm{\mu m}$ silicate emission and the SED of the dust disk around $\beta$ Pic using a model of very porous agglomerates, similar to cometary dust.
They applied the Maxwell Garnett mixing rule to the computations of the SED from agglomerates of crystalline silicate constituent grains as well as agglomerates of constituent grains with an amorphous silicate core and an organic refractory mantle. 
Furthermore, H$_2$O ice was considered to encase the silicate core-organic mantle grains in the outer cold region of the disk ($r \ge 100\,\mathrm{au}$).
They were able to place a tight constraint on the porosity of the agglomerates being approximately $p=0.95$ or as high as $p=0.975$.
Such highly porous agglomerates have also been successful in fitting the SEDs of dust rings around the A-type star HR\,4796A, which is depicted in Fig.\,\ref{koehler-et-al2008:f3} and the K-type star $\epsilon$ Eridani \citep{koehler-et-al2008,li-lunine2003a,li-et-al2003}.
Therefore, it seems plausible that fluffy agglomerates may well represent primitive dust particles not only in the Solar System but also in debris disks.

In a similar approach, \citet{augereau-et-al1999} modeled the HR\,4796A debris disk with two distinct populations of dust agglomerates consisting of small grains with a silicate core and an organic mantle.
More precisely, agglomerates with a porosity of $p \sim 0.6$ in a cold annulus around $70\,\mathrm{au}$ from the star contain amorphous silicate grains and have sizes of ${a}_\mathrm{c} > 10\,\mathrm{\mu m}$, while agglomerates with a higher porosity of $p \sim 0.97$ in a warm annulus at about $9\,\mathrm{au}$ contain crystalline silicate grains.
They used the Maxwell Garnett mixing rule to compute scattering and absorption cross sections and the asymmetry parameter of the agglomerates, but the Henyey-Greenstein phase function to simulate the scattered light images.
Although the phase function determined by the Maxwell Garnett mixing rule was dismissed, the simultaneous use of the Henyey-Greenstein phase function with the other quantities determined by the Maxwell Garnett mixing rule makes the model not self-consistent.

\begin{figure}
\center
\resizebox{0.75\textwidth}{!}{%
  \includegraphics{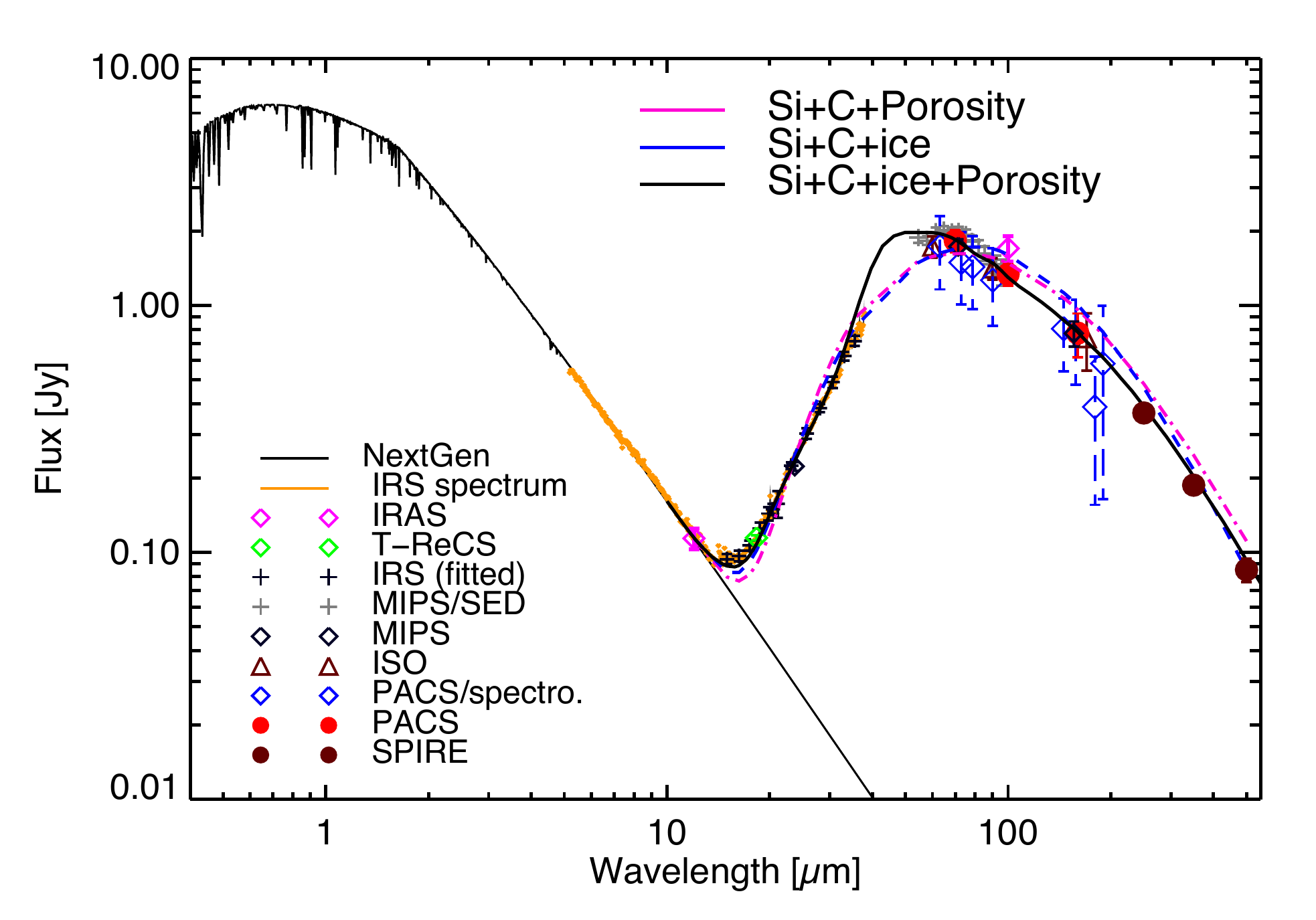}
}
\caption{The spectral energy distribution (SED) of the debris disk around HD\,181327 constructed by photometric data ({\it red crosses} with {\it error bars}) as well as model SEDs.
The best SED fit is achieved by a model of dust agglomerates that are composed of amorphous silicate, carbonaceous material, and ice, while a model without ice or porosity fails to reproduce the observed SED.
From \citet{lebreton-et-al2014}}
\label{lebreton-et-al2014:f2}       
\end{figure}

The young (20\,Myr) F-type star HD\,181327 was observed by the Herschel Space Observatory in the far-infrared to submillimeter domain providing a detailed coverage of the SED from its cold dust belt \citep{lebreton-et-al2012,lebreton-et-al2014}. 
A simulation of the SED using the Bruggeman mixing rule demonstrated that agglomerates in the dust belt contain $67\pm7\%$ of ice and are characterized by the porosity $p=0.63\pm0.21$, if the ice is mixed with 2/3 of silicate and 1/3 of amorphous carbon (see Fig.\,\ref{lebreton-et-al2014:f2}). 
However, \citet{lebreton-et-al2012} recognized that the resulting asymmetry parameter of the agglomerates from the Bruggeman mixing rule is significantly different from the asymmetry parameter determined by a model of scattered light images with the Henyey-Greenstein phase function.
While the Henyey-Greenstein phase function does not describe well the phase function of fluffy agglomerates, an estimate of the asymmetry parameter for fluffy agglomerates of submicron constituent grains goes beyond the applicability of the Bruggeman mixing rule.

The same approach was applied to the 30\,Myr-old A-type star HD\,32297 by \citet{donaldson-et-al2013} who have shown that this system contains fluffy agglomerates with $p=0.9$.
The agglomerates were modeled as a mixture of ices, silicate, and carbon with the volume fraction of 1/2, 1/6, and 1/3, respectively, which resembles dust in comet 1P/Halley. 
However, \citet{rodigas-et-al2014} have shown that this cometary dust model fails to reproduce the surface brightness of the disk in the near-infrared wavelength range.
The cometary dust model predicts a blue color of the disk in the near-infrared, but the observed brightness of the disk is gray in this wavelength range.
It turned out that the near-infrared spectrum of the HD\,32297 debris disk is better reproduced by compact ($p=0$) pure water-ice particles, as far as the near-infrared wavelength range is concerned.
This is a typical example of how SED fitting models encounter difficulties in providing a warranty for the uniqueness of solutions to the SED fitting problems.

To model the SED of the debris disk around AU Mic from near-infrared to millimeter wavelengths, \citet{fitzgerald-et-al2007} used a Monte Carlo radiative transfer code, which relies on the optical properties of spheres calculated by the Mie theory.
Unfortunately, we cannot figure out the reason that such an optically thin debris disk was modeled with the radiative transfer code developed by \citet{pinte-et-al2006} for optically thick protoplanetary disks.
According to a model of interstellar dust by \citet{mathis-whiffen1989}, they assumed porous agglomerates whose constituent grains are composed of silicate, carbon, ices, or ``vacuum'' and used the refractive indices of the mixture derived from the Bruggeman mixing rule.
They also modeled the dependences of intensity and polarization on projected distance from the central star and rejected a model of compact particles.
Furthermore, they emphasized the importance of polarimetric data to constrain the composition and distribution of dust particles, but we do not expect that the application of an EMA to agglomerates provides correct understandings for the degree of linear polarization.

\subsection{Characteristic features of minerals}
\label{sec:5.2}

\begin{figure}
\center
\resizebox{0.75\textwidth}{!}{%
  \includegraphics{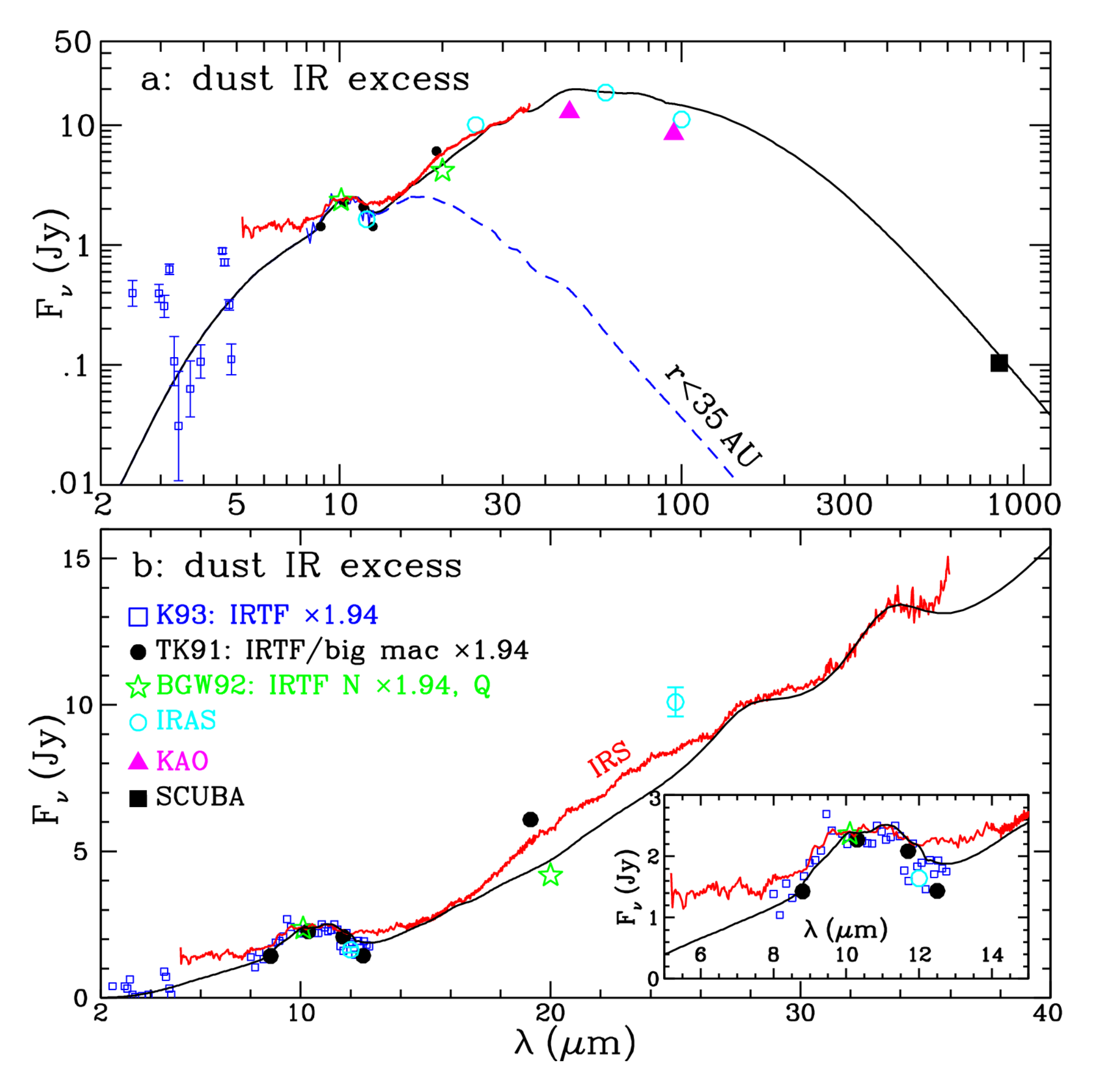}
}
\caption{The spectral energy distribution of the $\beta$ Pic debris disk and its Spitzer/IRS spectrum ({\it red line}).
In addition to the $10\,\mathrm{\mu m}$ amorphous silicate feature and the $11.2\,\mathrm{\mu m}$ crystalline silicate feature, the $28$ and $33.5\,\mathrm{\mu m}$ crystalline silicate features are also prominent in the Spitzer/IRS spectrum.
Also shown is the dust infrared emission calculated from the porous dust model of \citet{li-greenberg1998}. 
From \citet{chen-et-al2007}}
\label{chen-et-al2007:f2}       
\end{figure}

When high resolution spectra are obtained in the infrared wavelength range, conspicuous emission features characteristic of a certain mineral may appear in the spectra.
The central wavelength of emission features in the infrared spectra of dust particles is diagnostic of mineral species that are contained in the particles.
The most common mineral in primitive dust particles in planetary systems is magnesium-rich olivine, in particular, forsterite, which shows a prominent peak around $\lambda = 11.2\,\mathrm{\mu m}$ \citep[e.g.,][]{campins-ryan1989,knacke-et-al1993}.
Prominent spectral features of forsterite in primitive dust particles appear not only in the $10\,\mathrm{\mu m}$ wavelength range, but also in the $20$--$30\,\mathrm{\mu m}$ range (see Fig.\,\ref{chen-et-al2007:f2}).
A direct link between infrared spectral features and minerals has been confirmed by mineralogical analyses and infrared spectra of CP IDPs \citep{brunetto-et-al2011}.

\citet{hage-greenberg1990} used the Maxwell Garnett mixing rule to compute mid-infrared spectra of fluffy agglomerates consisting of cubical shaped grains, which are randomly distributed in the agglomerates.
They considered agglomerates of 1000--8000 cubical identical homogeneous grains in terms of their size, composition, and orientation. 
In comparison with the DGF/VIEF method of the DDA, they examined the validity of the Maxwell Garnett mixing rule, in which the porosity was defined by the fractional volume of vacuum within the smallest convex volume encasing the agglomerates.
They concluded that the porosity of agglomerates in the coma of comet 1P/Halley exceeds 0.97 if agglomerates of $a_\mathrm{v} = 10\,\mathrm{\mu m}$ contribute to the mid-infrared silicate feature.
However, they modeled the $10\,\mathrm{\mu m}$ feature with amorphous silicate, while the silicate emission band observed for comet 1P/Halley indicates that the silicates are crystalline \citep{bregman-et-al1987}.
Therefore, their study cannot be used to provide conclusive constraints on the porosity of dust agglomerates in the coma of comet 1P/Halley.

\citet{okamoto-et-al1994} studied the dependences of infrared spectra for fractal agglomerates on the porosity of the agglomerates and the number of their constituent grains.
They applied the Maxwell Garnett mixing rule to calculations of absorption cross sections for agglomerates consisting of spherical grains with $a_{0} = 0.01\,\mathrm{\mu m}$.
On the one hand, the infrared spectral features of olivine become weak for large agglomerates with $a_\mathrm{c} > 10\,\mathrm{\mu m}$ if the agglomerates are relatively compact ($D \sim 3$).
On the other hand, the olivine features remain noticeable even for large agglomerates with $a_\mathrm{c} > 10\,\mathrm{\mu m}$, if the agglomerates are fluffy ($D \sim 2$).
This indicates that the size distribution of agglomerates does not influence their infrared spectra, if the agglomerates underwent the growth by the BCCA process because of $D \approx 2$.
Therefore, if comets do not exhibit any silicate emission features in their infrared spectra, we may expect that their comae are dominated by large agglomerates of $a_\mathrm{c} > 10\,\mathrm{\mu m}$ with a relatively compact structure.

\citet{nakamura1998} attempted to model the characteristic features of olivine in the mid-infrared spectra of the $\beta$ Pic debris disk using BPCA ($D \sim 3$) particles of $N = 27000$ spherical constituent grains.
He assumed the agglomerates to consist of olivine grains and graphite grains half and half with either $a_{0} = 0.2\,\mathrm{\mu m}$ or $a_{0} = 0.5\,\mathrm{\mu m}$.
The {\sf DDSCAT} code (ver.\,4a) with the $a_1$-term method was used to compute the wavelength dependence of absorption cross section for the agglomerates.
The olivine features appear around $\lambda = 10\,\mathrm{\mu m}$ and $11\,\mathrm{\mu m}$ for agglomerates with $a_{0} = 0.2\,\mathrm{\mu m}$ ($a_\mathrm{v} = 6\,\mathrm{\mu m}$), while the features are weak for those with $a_{0} = 0.5\,\mathrm{\mu m}$ ($a_\mathrm{v} = 15\,\mathrm{\mu m}$).
However, \citet{golimowski-et-al2006} claim that agglomerates consisting of silicate grains and graphite grains present neutral colors, contrary to red colors observed for the $\beta$ Pic debris disk.
Therefore, there is a room for improvement in modeling of dust particles in the debris disk around $\beta$ Pic by viewing the observational evidence from various angles.

\begin{figure*}
\resizebox{1.0\textwidth}{!}{%
  \includegraphics{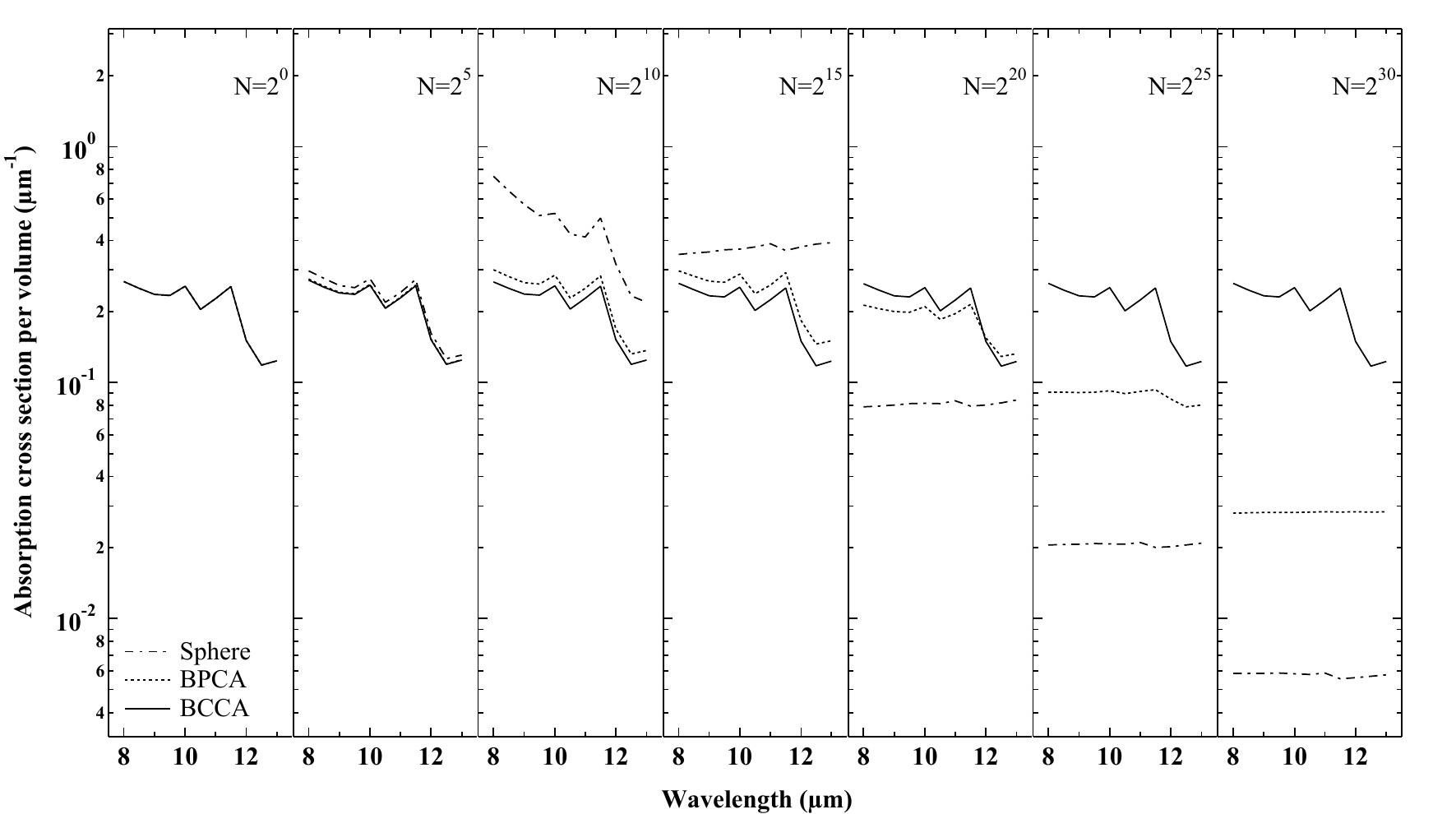}
}
\caption{The dependences of olivine features in the infrared spectra of fractal agglomerates on the number $N$ of constituent grains in an agglomerate and on its structure.
The {\it solid curves}, {\it dotted ones}, and {\it dash-dotted ones} are the infrared spectral of BCCA ($D \approx 2$) particles, BPCA ($D \approx 3$) particles, and compact spherical particles, respectively.
From \citet{kolokolova-et-al2007}}
\label{kolokolova-et-al2007:f3}       
\end{figure*}

\citet{kolokolova-et-al2007} considered the evolution of dust mantles on the surface of comets based on the infrared spectra for BPCA ($D \approx 3$) and BCCA ($D \approx 2$) particles consisting of spherical grains with $a_{0} = 0.1\,\mathrm{\mu m}$ computed by the Maxwell Garnett mixing rule.
The validity of the Maxwell Garnett mixing rule has been confirmed with the {\sf scsmtm1} code at $N=1024$ and the {\sf DDSCAT} code (ver.\,4a) with the $a_1$-term method at $N=32768$.
Their results have also shown that olivine features in the infrared spectra of fractal agglomerates disappear at $N > 2^{20}$ ($a_\mathrm{c} > 19\,\mathrm{\mu m}$) for BPCA particles of submicron constituent grains (see Fig.\,\ref{kolokolova-et-al2007:f3}).
The mid-infrared spectral features of olivine are less pronounced for short-period comets with small semi-major axes than those with large semi-major axes \citep{kolokolova-et-al2007}.
This implies that short-period comets with small semi-major axes have more compact agglomerates on their surfaces than those with large semi-major axes.
This is consistent with the picture of dust mantle formation on the surface of comets, which is characterized by a deficit of highly porous agglomerates on the surface of cometary nuclei due to longer or stronger solar irradiation.

\citet{yamamoto-et-al2008} used fractal agglomerates to interpret the strength of a silicate emission feature as well as the color temperature observed for the ejecta of comet 9P/Tempel 1 during the Deep Impact mission.
By taking into account the formation and evolution of dust mantles on comets, they assumed that the dust mantle of the comet consist of agglomerates with $D=2.5$ and the interior of the comet consists of agglomerates with $D=1.9$.
They used the Maxwell Garnett mixing rule to compute the infrared spectra for the agglomerates whose spherical constituent grains of $a_{0} = 0.1\,\mathrm{\mu m}$ have a structure of a forsterite and amorphous silicate core and an organic refractory mantle.
Their success in modeling the infrared spectroscopic observation of the ejecta from comet 9P/Tempel 1 confirmed that the surface layer of periodic comets has been processed and does not maintain their primordial structures and compositions \citep[cf.][]{kolokolova-et-al2007}.

It is well known that the position of any emission feature characteristic of a certain mineral in general depends on the shape of the mineral grains \citep[e.g.,][\S\,12.2.7]{bohren-huffman1983}.
\citet{yanamandrafisher-hanner1999} employed the {\sf DDSCAT} code (ver.\,4a) with the LDR method to study the effect of grain shapes on the positions of silicate features in the infrared spectra.
They considered small agglomerates of $2$--$5$ submicron silicate constituent grains whose shapes are either spherical and tetrahedral.
Their results show strong shape effects on the positions of forsterite features as expected, but the shape effects are weak for infrared spectral features of amorphous silicate.
Since absorption bands take place at wavelengths where the real part of complex dielectric function is negative, shape effects are not important if the negative value is confined in a very narrow wavelength range.
Infrared spectral observations of comets and debris disks have shown that the positions of forsterite features appear at similar wavelengths, irrespective of their different circumstances \citep[e.g.,][]{knacke-et-al1993}.
Therefore, the complex dielectric function of primitive dust particles in planetary systems most likely does not have a deep negative value compared to pure forsterite, indicating that the particles in planetary systems are not composed of bare mineral grains.

\begin{figure*}
\center
\resizebox{0.5\textwidth}{!}{%
  \includegraphics{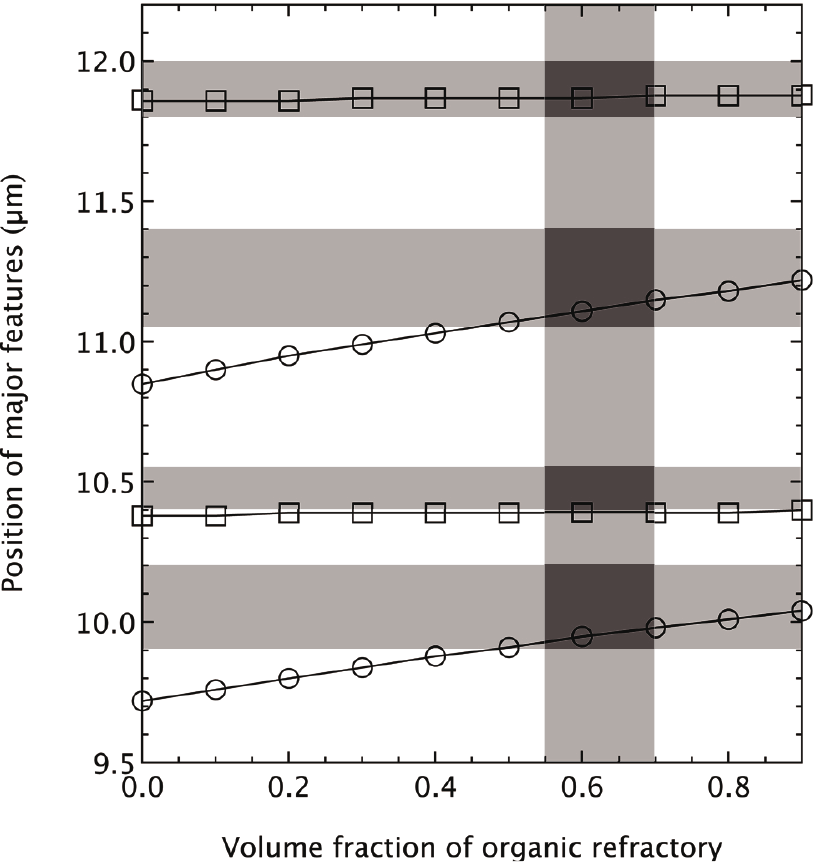}
}
\caption{The central wavelengths of forsterite features in simulated infrared spectra of BCCA ($D \approx 2$) particles versus the volume fraction of an organic refractory mantle that encases a forsterite core in spherical constituent grains. 
The {\it horizontally long shaded area} indicates the central wavelengths of forsterite features in observed infrared spectra of primitive dust particles in planetary systems. The {\it vertically long shaded bar} is the plausible range of volumetric organic fraction inferred from the cosmic abundance constraints. From \citet{kimura2013}}
\label{kimura2013:f1}       
\end{figure*}

The presence of organic refractory material encasing mineral grains plays a vital role in the positions of the peaks that appear in the infrared spectra \citep{kimura2013}.
This is known as the matrix effect, which has been experimentally proven to shift the positions of olivine features to longer wavelengths \citep{day1975,dorschner-et-al1978}.
\citet{kimura2013} computed the infrared spectra of BCCA ($D \approx 2$) particles consisting of silicate-core, organic-mantle spherical grains with $a_{0} = 0.1\,\mathrm{\mu m}$ using the Mawell-Garnett mixing rule.
It turned out that the positions of olivine features are consistent with the infrared spectra of dust in cometary comae, debris disks, and protoplanetary disks, if the organic volume fraction fulfills the cosmic abundance constraints (see Fig.\,\ref{kimura2013:f1})
This indicates that the composition of agglomerates is not completely a free parameter in modeling primitive dust particles in planetary systems.

\begin{figure*}
\resizebox{1.0\textwidth}{!}{%
  \includegraphics{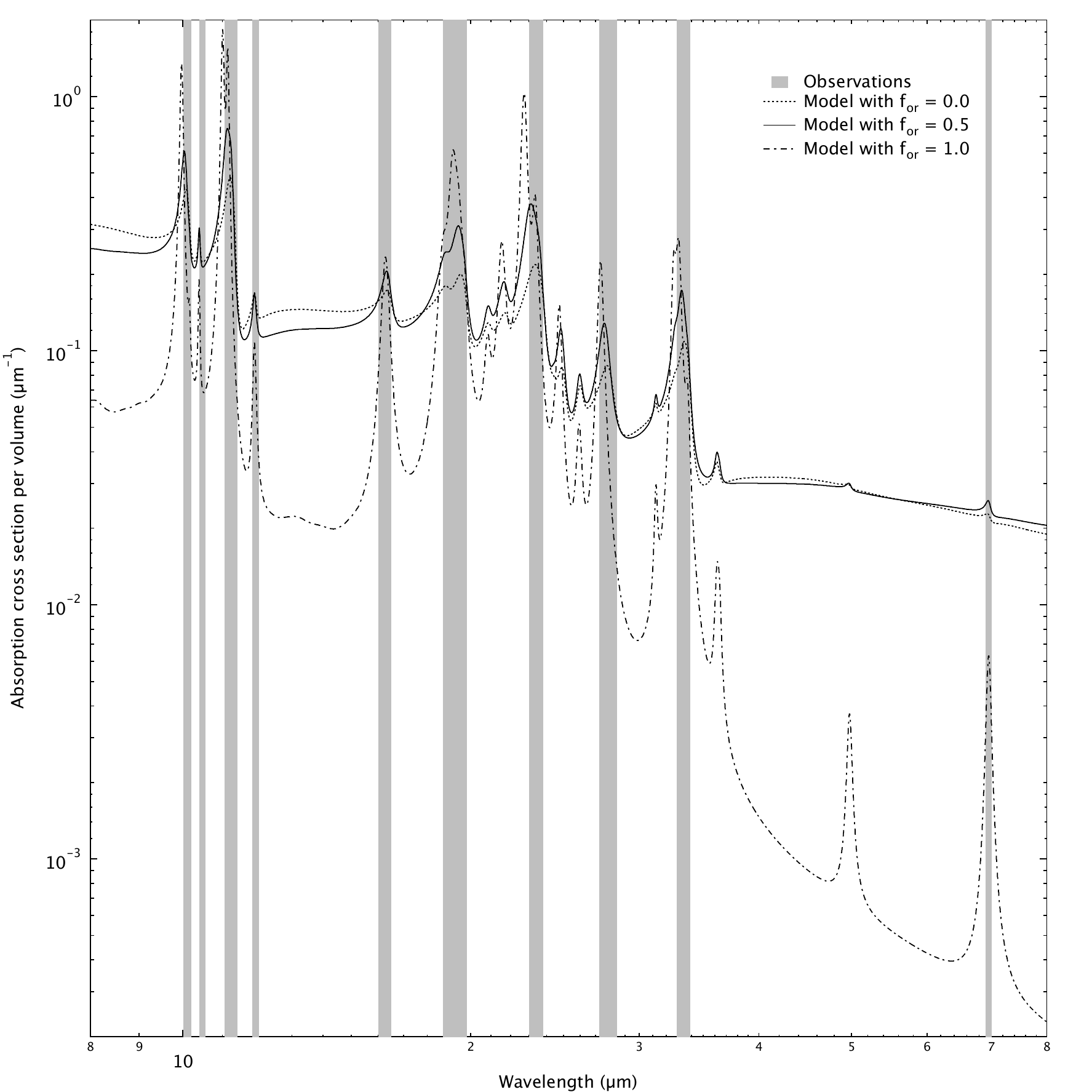}
}
\caption{Infrared spectra of dust agglomerates consisting of $2^{30}$ submicron grains with a forsterite core and an organic refractory mantle. 
The carbonization of organic refractory mantle is characterized by $1-f_\mathrm{or}$ where $f_\mathrm{or}$ denotes the volume fraction of organic refractory material within the carbonaceous mantle.
The {\it shaded bars} indicate the central wavelengths that have been observed to exhibit noticeable emission features from dust particles in comets, debris disks, and protoplanetary disks.
From \citet{kimura2014}}
\label{kimura2014:f2}       
\end{figure*}

It is worth noting that organic refractory material could be carbonized due to the preferential loss of hydrogen, nitrogen, and oxygen by ultraviolet irradiation and ion bombardments \citep{jenniskens1993,jenniskens-et-al1993}.
Since the carbonization of organic refractory material changes its chemical composition, the refractive indices of organic-rich carbonaceous material depends on the degree of carbonization.
Therefore, the degree of carbonization affects the positions of mineral features in the infrared spectra, if the mineral is encased in organic-rich carbonaceous material.
Using the Mawell-Garnett mixing rule, \citet{kimura2014} has studied the effect of carbonization on the infrared spectral features of olivine for BCCA ($D \approx 2$) particles of $a_{0} = 0.1\,\mathrm{\mu m}$ spherical constituent grains with a forsterite core and an organic-rich carbonaceous mantle (see Fig.\,\ref{kimura2014:f2}).
The forsterite feature at $\lambda = 11.1,\mathrm{\mu m}$ observed in the debris disk of $\beta$ Pic indicates that dust particles in the $\beta$ Pic disk did not suffer from severe carbonization, compared to cometary dust in the Solar System.
Therefore, infrared spectra of dust particles with high spectral resolution are highly useful to diagnose the carbonization degree of organic materials in the particles.

\begin{figure}
\center
\resizebox{0.5\textwidth}{!}{%
  \includegraphics{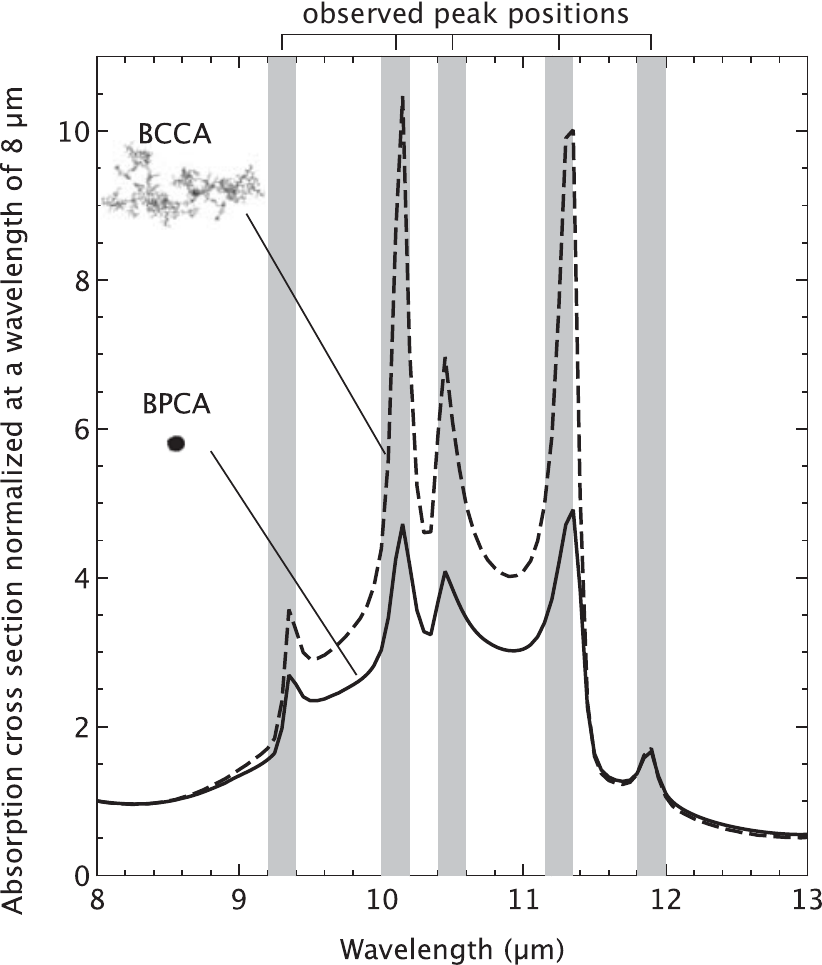}
}
\caption{Infrared spectra of dust agglomerates consisting of $2^{20}$ submicron grains with an amorphous silicate core and an organic refractory mantle after processing by exothermic chemical reactions. 
The processing of such a core-mantle grain forms a thin forsterite layer on the surface of amorphous silicate cores. 
The {\it shaded bars} indicate the central wavelengths that have been observed to exhibit noticeable emission features from dust particles in cometary comae.
From \citet{kimura-et-al2008}}
\label{kimura-et-al2008:f2}       
\end{figure}

Infrared spectroscopic observations of comets reported on the discovery of a $9.3\,\mathrm{\mu m}$ feature in the coma of comets C/1995 O1 (Hale-Bopp), C/2001 Q4 (NEAT), and 9P/Tempel 1 \citep[e.g.,][]{wooden-et-al1999,wooden-et-al2004,harker-et-al2007}. 
Although the $9.3\,\mathrm{\mu m}$ feature is not always present in the spectra, the feature has been attributed to magnesium-rich pyroxene.
\citet{kimura-et-al2008} demonstrated that the $9.3\,\mathrm{\mu m}$ feature could also be produced by forsterite, if amorphous silicate constituent grains are covered by a forsterite layer encased in organic refractory mantle (see Fig.\,\ref{kimura-et-al2008:f2}).
While the presence of pyroxene in primitive dust particles could be supported by mineralogical investigation of CP IDPs, one should keep in mind that the mineralogical identification with an infrared feature alone is not always unique.

\section{Integral optical quantities of dust agglomerates}
\label{sec:6}

\subsection{Bolometric albedo}
\label{sec:6.1}

The bolometric albedo of dust particles is an integrated quantity over wavelengths, defined as the fraction of stellar luminosity scattered by the particles \citep{gehrz-ney1992}.
Therefore, the SED could be used to estimate the bolometric albedo of dust particles, when observational data on scattered light are available.
The bolometric albedos of dust particles in comet 1P/Halley and the debris disks of $\beta$ Pictoris, HD 207129 and HD 92945 have been determined to be $0.32$, $0.35$, $0.051$, $0.10$, respectively \citep{gehrz-ney1992,backman-et-al1992,krist-et-al2010,golimowski-et-al2011}.

It is sometimes claimed that numerical simulation of the SED with an EMA predicts a much higher albedo than the bolometric albedo from the SED \citep{krist-et-al2010,lebreton-et-al2012}.
This discrepancy has been attributed to the inapplicability of the EMA to properly describe the light-scattering properties of agglomerates in planetary systems.
We have no objection to the note of caution that EMAs are not applicable to agglomerates of submicrometer-sized constituent grains in the visible wavelength range.
However, we notice that the simulated albedo is not the bolometric albedo, but the single-scattering albedo, which is defined by the ratio of the scattering cross section to the extinction cross section.
Therefore, to the best of our knowledge, there has not yet been a direct comparison of the bolometric albedo of primitive dust particles in planetary systems between numerical simulations and astronomical observations.

\subsection{Radiation pressure}
\label{sec:6.2}

Stellar radiation inevitably exerts a force on a dust particle in planetary systems and thus an estimate of radiation pressure has been of great importance to understand the dynamics of dust particles in planetary systems.
An estimate of radiation pressure requires integration of radiation pressure cross sections weighted by stellar radiation spectrum over wavelengths from ultraviolet to far-infrared. 
Stellar radiation pressure also retards the motion of dust particles and, in consequence, the particles spiral to the central star, referred to as the Poynting-Robertson effect \citep{robertson1937}.
In particular, stellar radiation pressure controls the dynamics of dust particles and their spatial distribution, unless the stellar mass-loss rate is more than a few times larger than that of the stellar wind or the total mass of a circumstellar dust disk is heavier than one millionth of the Earth \citep{minato-et-al2006}. 

It is a common practice to discuss the importance of radiation pressure in terms of $\beta$ defined by the ratio of stellar radiation pressure to gravitational attraction.
A pioneering work on numerically evaluating the $\beta$ values for fluffy agglomerates was performed by \citet{mukai-et-al1992}. 
They applied the Maxwell Garnett mixing rule to estimate effective refractive indices for fractal agglomerates and then computed radiation pressure cross sections in the framework of Mie theory. 
\citet{mukai-et-al1992} assumed $a_0 = 0.01\,\mathrm{\mu m}$, because the Maxwell Garnett mixing rule is applicable under the condition that the radius of inclusions is much smaller than the wavelength of light. 
It turned out that the $\beta$ values computed by the EMA are in good agreement with those computed by the {\sf DDSCAT} code (ver.\,4a) with the DGF/VIEF method for $N \le 1024$, except for highly porous ($D \approx 2$) agglomerates composed of silicate constituent grains \citep{mukai-et-al1992}. 
\citet{kimura-et-al1997} applied the Bruggeman mixing rule to compute the $\beta$ values for BPCA ($D \approx 3$) and BCCA ($D \approx 2$) particles of $0.01\,\mathrm{\mu m}$-radius constituent grains. 
Their results are also shown to agree with the results computed by the {\sf DDSCAT} code (ver.\,4a) with the $a_1$-term method, except for silicate BCCA particles \citep{kimura-et-al2002a}. 
Therefore, EMAs provide reasonably accurate results for $\beta$ values of agglomerates as far as a relatively compact agglomerate composed of absorbing material is concerned.

In general, the dependence of $\beta$ values on the size of agglomerates becomes weaker as the fractal dimension of the agglomerates decreases \citep{mukai-et-al1992,kimura-et-al1997,kimura-et-al2002a}.
Therefore, $\beta$ values of agglomerates approach those of their constituent grains with increase in the porosity of the agglomerates, even if the agglomerates are not fractals.
This statement agrees with the results of \citet{saija-et-al2003} who applied the T-matrix method to compute the $\beta$ values for {non-fractal agglomerates.
They considered} 200 constituent grains of silicate or amorphous carbon with $a_{0} = 0.005\,\mathrm{\mu m}$ and the cluster of the grains with $p = 0.842$--$0.995$ in circumstellar environments with stellar temperatures of 2700, 5800, and 10000\,K.
As a result, the $\beta$ values of agglomerates are smaller than those of volume-equivalent compact spheres in the submicron-size range and larger in the sizes above tens of microns.

\citet{wilck-mann1996} used the Maxwell Garnett mixing rule for their computations of the $\beta$ values for porous spheres that mimic fluffy agglomerates in the Solar System.
A similar approach to the computation of $\beta$ values using the Maxwell Garnett mixing rule has been adopted for porous spheres around A-type stars \citep{artymowicz-clampin1997,grigorieva-et-al2007}.
\citet{kirchschlager-wolf2013} used the {\sf DDSCAT} code (ver.\,7.1) with the LDR method to compute the $\beta$ values for porous silicate spheres with a fixed porosity in the environments of the Sun and other main-sequence stars.
\citet{kimura-mann1999a} have shown that the $\beta$ values for small ($N \le 2048$) BPCA ($D \approx 3$) and BCCA ($D \approx 2$) particles of amorphous carbon and silicate computed by the {\sf DDSCAT} code (ver.\,4a) with the $a_1$-term method could be well represented by those of porous spheres with a fixed porosity computed by the Bruggeman mixing rule as far as small constituent grains of $a_{0} = 0.01\,\mathrm{\mu m}$ are concerned.
It is worth noting that the $\beta$ values of porous spheres with a fixed porosity are inversely proportional to the radius of the spheres in the sizes above tens of microns, independent of the porosity.
Such a size dependence of porous spheres differs from that of fractal agglomerates, in particular, highly porous ones ($D \approx 2$) whose $\beta$ values tend to be nearly independent of their sizes.

The major drawback of the above mentioned studies is the assumption that the radius of constituent grains is smaller than the wavelength of ultraviolet light, owing to the limitation for the EMAs, the DDA $a_1$-term method, or computational resources such as RAM and CPU time.
However, the constituent grains of agglomerates in planetary systems are one order of magnitude larger than $a_0 = 0.01\,\mathrm{\mu m}$, as pointed out in Sect.~\ref{sec:2}.
A deficit of small dust particles in planetary systems takes place around sizes near the maximum $\beta$ values due to mutual collisions of the particles and subsequent blowouts by stellar radiation pressure \citep{ishimoto-mann1999,krivov-et-al2000}.
In conclusion, correct estimates of the $\beta$ values for agglomerates of submicron constituent grains are essential to better understand not only the dynamical behavior but also the spatial distribution of dust particles in a planetary system.

\citet{kimura-et-al2002b} computed the $\beta$ values for agglomerates of submicron silicate constituent grains using the {\sf scsmtm1} code in order to simulate dust particles released from sungrazing comets. 
Although the number of constituent grains was limited to $N \le 32$ by computing resources for the {\sf scsmtm1} code, their results show that agglomerates of $a_0 \approx 0.1\,\mathrm{\mu m}$ are consistent with the dynamical constraint of $\beta \le 0.6$ for dust particles in the tails of mini ($a \sim 1$--$10\,\mathrm{m}$) sungrazers \citep[cf.][]{sekanina2000}. 
However, this does not mean that dust particles in sungrazing comets are too peculiar to lack organic refractory materials.
The assumption of bare silicate grains is easily justified, since the organic refractory component sublimates almost instantly in the vicinity of the Sun.
In the tail of the larger ($a \sim 100\,\mathrm{m}$) sungrazing comet C/2011 W3 (Lovejoy), the values of $2.5 > \beta \ge 0.6$ were identified before perihelion, although only $\beta \approx 0.6$ appeared after perihelion \citep{sekanina-chodas2012}.
Therefore, the high $\beta$ values can be associated with organic-rich carbonaceous materials that are less refractory than silicates with $\beta \approx 0.6$.
We cannot help wondering whether sublimation of organic materials is also responsible for unusual degrees of linear polarization observed for sungrazing comets \citep[cf.][]{weinberg-beesonn1976,thompson2015}.

\citet{mukai-okada2007} studied the size dependence of the $\beta$ values for BPCA ($D \approx 3$) particles of silicate constituent grains with $N=2048$ and $16384$, by varying the radius of constituent grains.
They used the Maxwell Garnett mixing rule to compute the radiation pressure cross sections of the agglomerates, although the radius of constituent grains ranged from nanometer to millimeter.
Because the size dependence of the $\beta$ values in their results originates from the variation in the radius of constituent grains, the $\beta$ values were independent of $N$.
By extending the study of \citet{mukai-okada2007} to agglomerates of amorphous carbon constituent grains, \citet{levasseurregourd-et-al2007} concluded that the $\beta$ values depend on the composition of the grains even for millimeter-sized agglomerates.
Although their results would have important implications for the dynamics of millimeter-sized dust particles in cometary trails, we should point out that the size dependence of the $\beta$ values with a fixed number of constituent grains differs from that with a fixed radius of constituent grains. 

\citet{koehler-et-al2007} extended $\beta$ values of agglomerates with $a_{0} = 0.1\,\mathrm{\mu m}$ up to $N =512$ for BPCA ($D \approx 3$) and BCCA ($D \approx 2$) particles of silicate constituent grains or carbon constituent grains.
They used both the {\sf DDSCAT} code (ver.\,6.1) with the LDR method and the {\sf scsmtm1} code for the computations of radiation pressure cross sections.
The radiation pressure cross sections were computed in the range of wavelengths from ultraviolet to far-infrared for small agglomerates of $N \le 32$, but only at a wavelength $\lambda = 0.6\,\mathrm{\mu m}$ for larger agglomerates of $32 < N \le 512$.
The approximation to the $\beta$ values at a single wavelength of $0.6\,\mathrm{\mu m}$ was shown to reproduce those computed with a full wavelength range within the accuracy of 30\%.
Furthermore, \citet{koehler-et-al2007} extrapolated the results to much larger ($N \le {10}^{17}$) agglomerates on the assumption that the radiation pressure cross sections for large agglomerates are proportional to $N^{2/D}$ in the geometrical optics regime.
The overall size dependence of the $\beta$ values for agglomerates of submicron grains has significant implications for a better understanding of their dynamical behaviors in planetary systems.

\begin{figure*}
\resizebox{1.0\textwidth}{!}{%
  \includegraphics{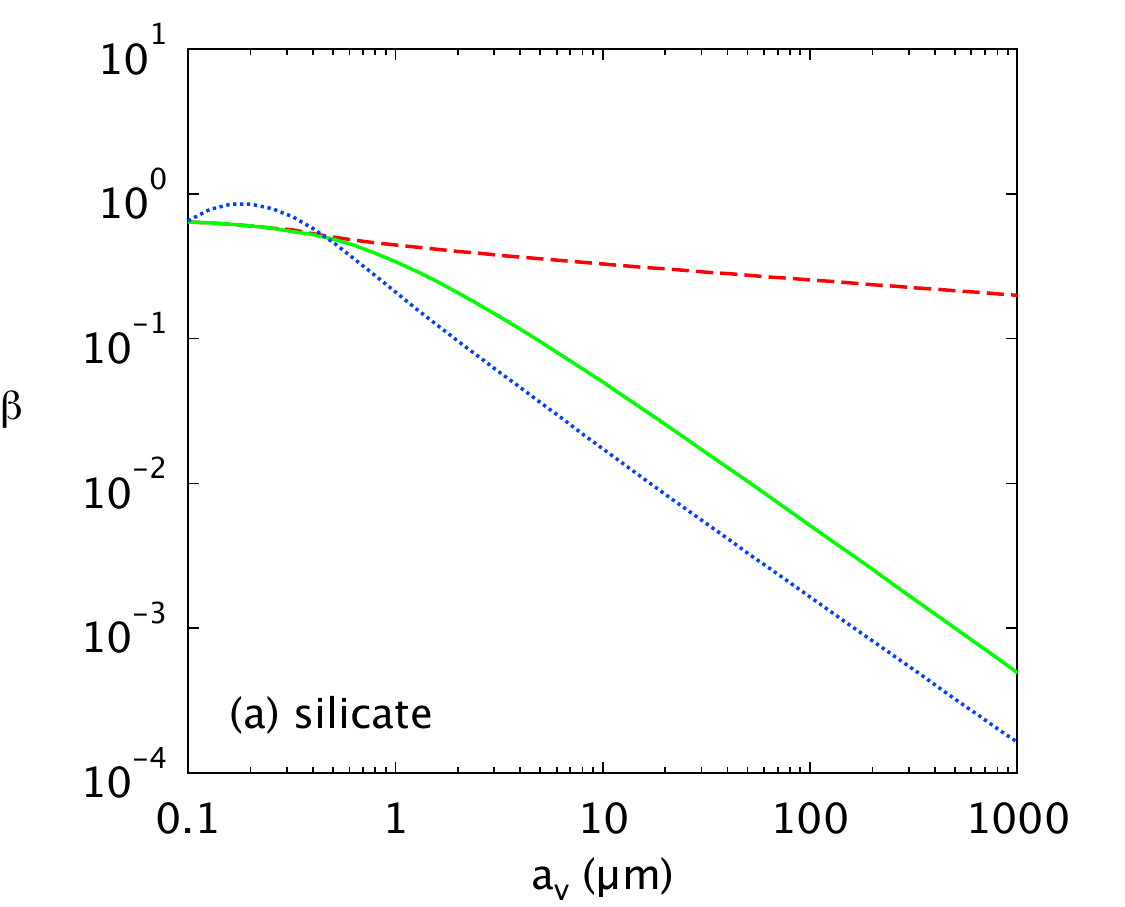}
  \includegraphics{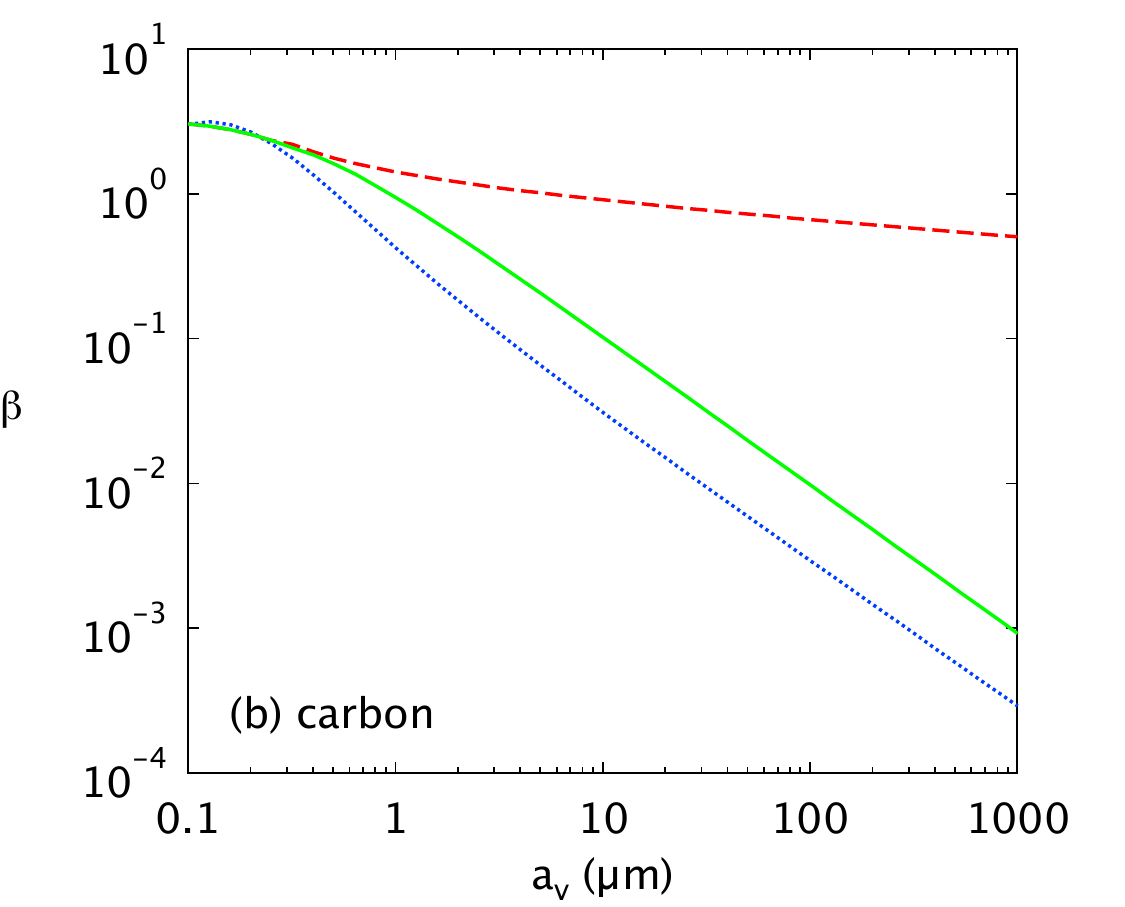}
}
\caption{The ratios of radiation pressure to gravitational attraction, $\beta$, of the Sun acting on (a) silicate particles and (b) carbon particles as a function of $a_\mathrm{v}$, the radius of volume-equivalent spheres. 
{\it Solid line}: fractal agglomerates consisting of $0.1\,\mathrm{\mu m}$-radius particles grown by ballistic particle-cluster agglomeration ($D \approx 3$); {\it dashed line}: fractal agglomerates consisting of $0.1\,\mathrm{\mu m}$-radius particles grown by ballistic cluster-cluster agglomeration ($D \approx 2$); {\it dotted line}: compact spheres.
Reproduced from Fig.\,5 of \citet{minato-et-al2006}}
\label{minato-et-al2006:fx}       
\end{figure*}

It has been known that the geometric cross sections of fractal agglomerates show a deviation from the ideal proportionality to $N^{2/D}$ \citep{meakin-donn1988,ossenkopf1993}.
Therefore, \citet{minato-et-al2006} took into account the geometric cross sections of the agglomerates to improve the results of \citet{koehler-et-al2007} for $N > 512$. 
In Fig.\,\ref{minato-et-al2006:fx}, we reproduce their results\footnote{The size dependence of the $\beta$ values for agglomerates of submicron constituent grains has not explicitly been shown in \citet{minato-et-al2006}, but could be deduced from their figures.} on the overall size dependence of $\beta$ values for agglomerates of submicron silicate constituent grains (left) and those of submicron carbon constituent grains (right).
Their results were also applied to estimate the ratio of solar radiation pressure to the local gravity of sub-km sized asteroids acting on BPCA and BCCA particles and their $\beta$ values in the debris disk of the A-type star Fomalhaut \citep{kimura-et-al2014}.
Stellar radiation pressure could replenish BPCA particles of $a_\mathrm{v}=10$--$370\,\mathrm{\mu m}$ into the exo-zodiacal cloud of Fomalhaut, if their parent bodies orbit the star at 2\,au.
Accordingly, the size dependence of $\beta$ values for agglomerates of submicrometer-sized constituent grains play a vital role in better understanding of dust dynamics in planetary systems.

\citet{tazaki-nomura2015} applied the {\sf scsmfo1b} code with the QMC method to compute the $\beta$ values for agglomerates of spherical constituent grains in orbit around a star with a blackbody radiation of $5778\,\mathrm{K}$.
Note that the solar radiation spectrum is close to a spectrum of a $5778\,\mathrm{K}$ blackbody and thus their results can be applied to dust particles in the Solar System.
The agglomerates were assumed to consist of amorphous silicate spheres with a radius of $a_0 = 0.01$ or $0.1\,\mathrm{\mu m}$ and have grown under the BCCA ($D \approx 2$) process in the range of $N \le 1024$.
Their results coincide with those in \citet{kimura-mann1999a} and \citet{minato-et-al2006}, which were obtained by the {\sf DDSCAT} code (ver.\,4a) with the $a_1$-term method and the {\sf DDSCAT} code (ver.\,6.1) with the LDR methods, respectively.
This coincidence is a natural consequence, because the computational techniques are not critical to the integrated optical quantities such as $\beta$.

\begin{figure*}
\resizebox{1.0\textwidth}{!}{%
  \includegraphics{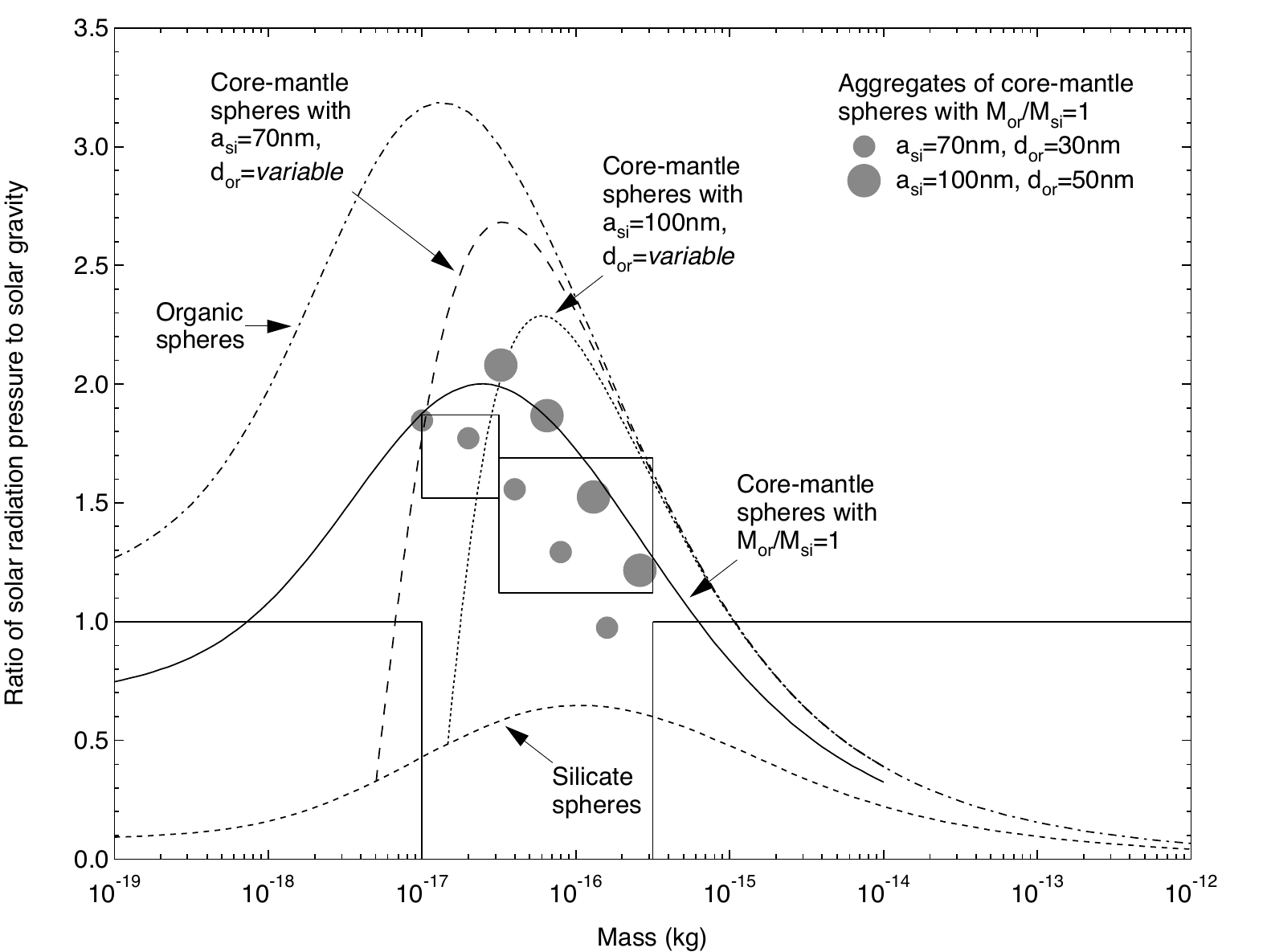}
}
\caption{The ratios of radiation pressure to gravitational attraction, $\beta$, of the Sun acting on BCCA ($D \approx 2$) particles consisting of submicron core-mantle particles. 
The core and the mantle of the constituent spherical grains are composed of amorphous silicate and organic refractory material, respectively, in equal mass.
Also plotted are the $\beta$ ratios for bare amorphous silicate spheres, bare organic refractory spheres, and silicate core-organic mantle spheres.
The {\it rectangles} indicate the expected ranges of $\beta$ values derived from in-situ impact data of interstellar dust by Ulysses.
From \citet{kimura-et-al2003b}}
\label{kimura-et-al2003b:f2}       
\end{figure*}

\citet{kimura-et-al2003b} applied the {\sf gmm02TrA} code to compute the $\beta$ values for BCCA ($D \approx 2$) particles consisting of submicron silicate-core, organic mantle grains in the Solar System.
They have shown that the $\beta$ values for the agglomerates of submicron silicate-core, organic mantle grains fulfill the dynamical constraint on the $\beta$ values for interstellar dust streaming into the Solar System (see Fig.\,\ref{kimura-et-al2003b:f2}).
However, it turned out that the agreement happened by coincidence, since interstellar dust in the Solar System does not contain organic material \citep{westphal-et-al2014,altobelli-et-al2016}.
Nevertheless, the results of \citet{kimura-et-al2003b} are applicable to cometary dust and dust in debris disks, although their results are available only up to $N = 16$.
Indeed, the $\beta$ values for the agglomerates of submicron silicate-core, organic mantle grains are consistent with $\beta < 2.5$ derived from coronagraphic images of comet C/2011 W3 (Lovejoy) by \citet{sekanina-chodas2012}.

\citet{silsbee-draine2016} computed the $\beta$ values for silicate agglomerates with $N=32$ and $256$ using a blackbody radiation of $5780\,\mathrm{K}$.
Although they aimed to study the dynamics of interstellar dust streaming into the Solar System, their computed $\beta$ values can be applied to dust particles in planetary systems.
They used the {\sf MSTM} code (ver.\,3.0) and the {\sf DDSCAT} code (ver.\,7.3) with the ``modefied'' LDR method for BPCA particles with and without restructuring.
The radius of the agglomerates was assumed to lie in the range of $a_\mathrm{v} = 0.01$--$1\,\mathrm{\mu m}$ for $N=32$ and $a_\mathrm{v} = 0.1$--$0.8\,\mathrm{\mu m}$ for $N=256$, implying $a_0 = 0.0032$--$0.32\,\mathrm{\mu m}$ and $a_0 = 0.016$--$0.126\,\mathrm{\mu m}$, respectively.
Their results show that $\beta$ values for silicate agglomerates do not exceed unity, in concord with previous studies.
They have also shown that the substitution of iron constituent grains for some of silicate constituent grains in an agglomerate enhances $\beta$ values. 
According to their results, the condition of $\beta > 1$ requires more than 35\% of the volume be iron, but this condition conflicts with the cosmic abundances of iron, silicon, and magnesium.

\begin{figure*}
\resizebox{1.0\textwidth}{!}{%
  \includegraphics{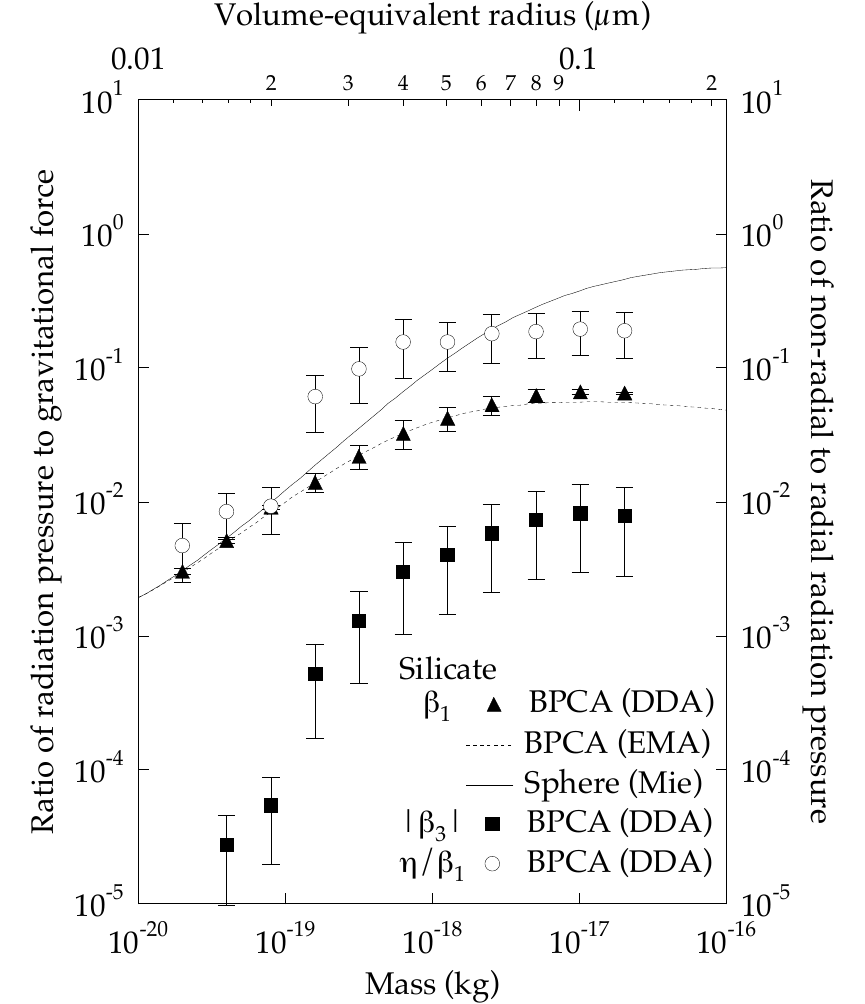}
  \includegraphics{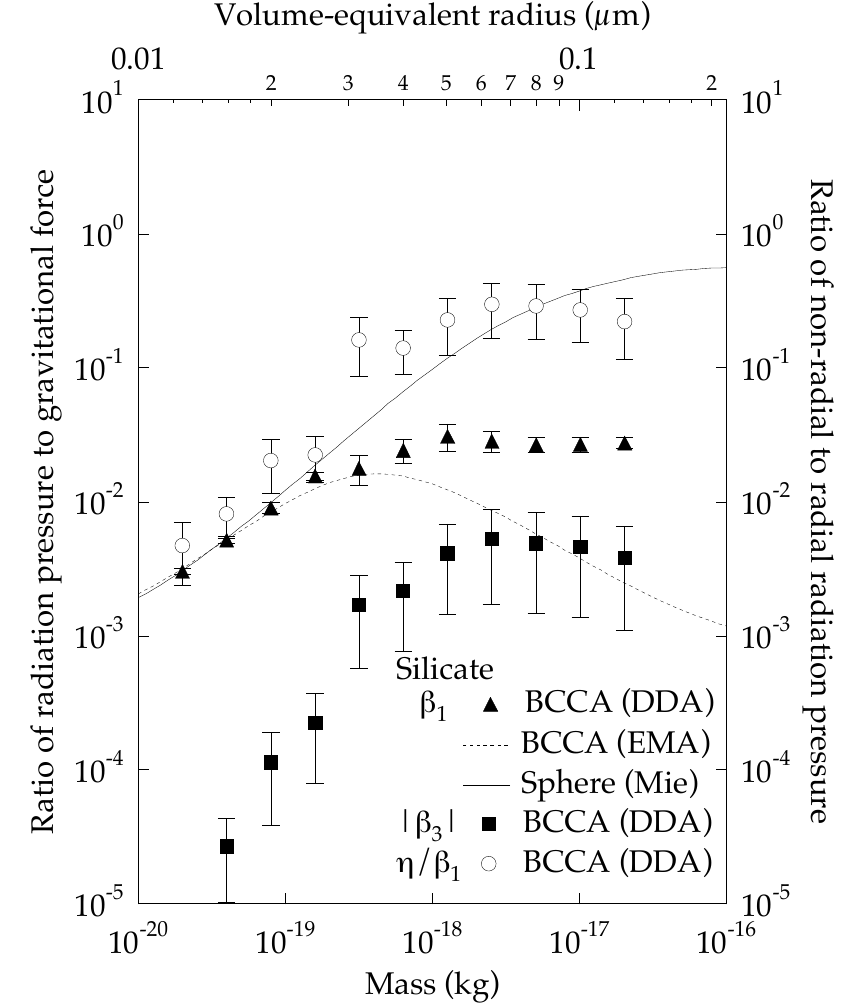}
}
\resizebox{1.0\textwidth}{!}{%
  \includegraphics{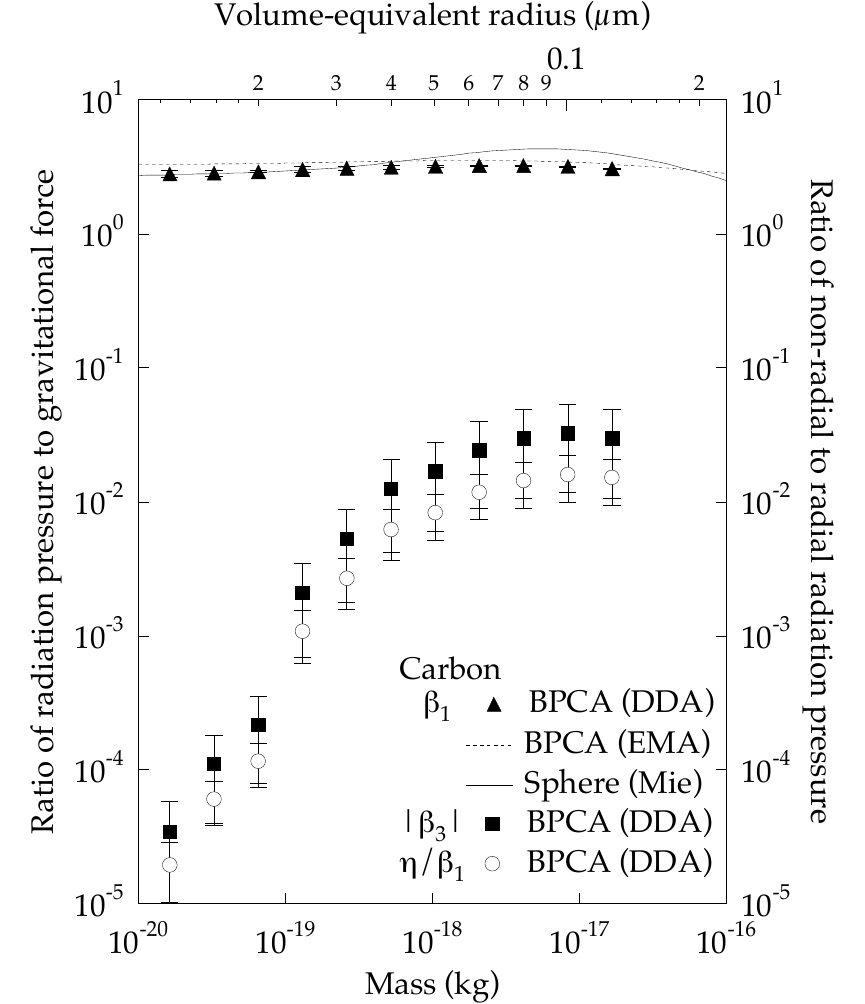}
  \includegraphics{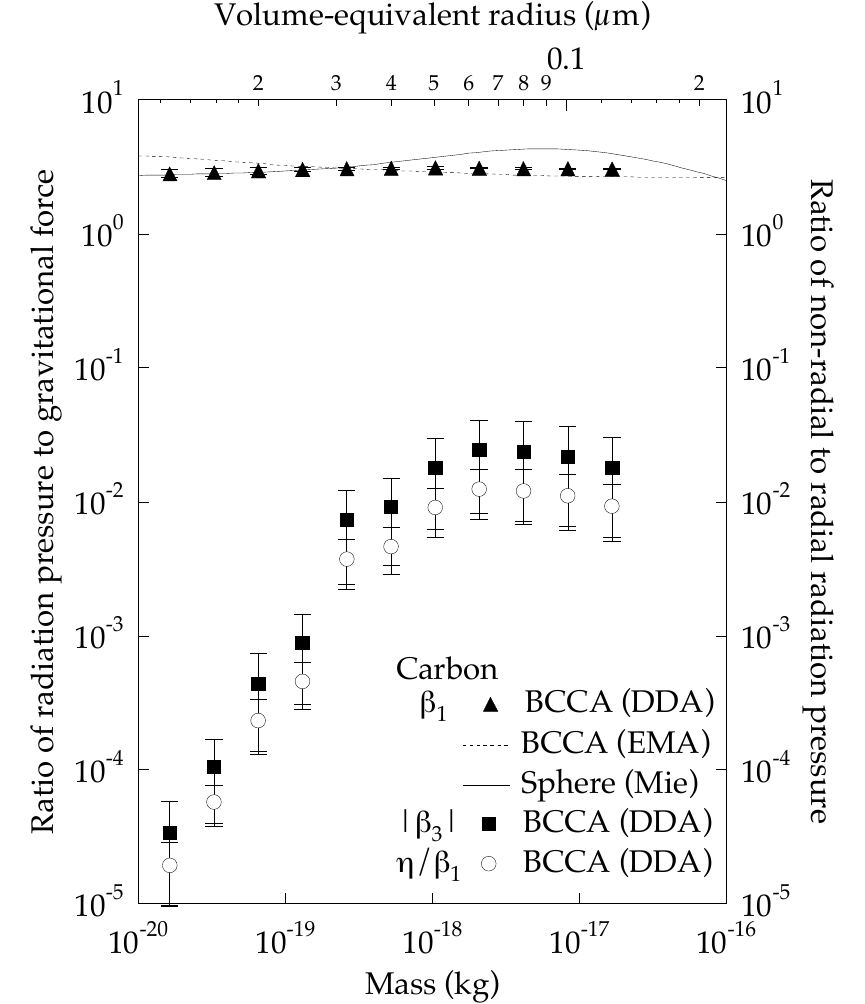}
}
\caption{The ratios of radiation pressure to gravitational attraction of the Sun acting on fractal agglomerates consisting of $0.01\,\mathrm{\mu m}$-radius grains. 
The {\it filled triangles} and {\it squares} are the ratios in the direction antiparallel to the solar gravity and the direction perpendicular to the orbital plane of the agglomerates. 
The {\it open circles} indicate the transverse component divided by the radial component of solar radiation pressure. 
The {\it dashed curves} are numerical results on the radial component of solar radiation pressure relative to the solar gravity estimated by the Bruggeman mixing rule. 
The {\it solid curves} are the radial components of solar radiation pressure relative to the solar gravity acting on compact spherical particles.
From \citet{kimura-et-al2002a}}
\label{kimura-et-al2002a:f2-f3}       
\end{figure*}

Stellar radiation exerts a force on a sphere in the direction antiparallel to the stellar gravity, but dust agglomerates are also subjected to a component of force perpendicular to the propagation of stellar radiation \citep[e.g.,][\S\,2.3]{vandehulst1957}. 
\citet{kimura-mann1998} applied the {\sf DDSCAT} code (ver.\,4a) with the $a_1$-term method to compute the radiation pressure cross sections for BPCA ($D \approx 3$) and BCCA ($D \approx 2$) particles of $0.01\,\mathrm{\mu m}$-radius constituent grains in the directions parallel and perpendicular to the propagation of stellar radiation.
\citet{kimura-et-al2002a} determined the parallel and perpendicular components of the $\beta$ values for BPCA ($D \approx 3$) and BCCA ($D \approx 2$) particles of $0.01\,\mathrm{\mu m}$-radius constituent grains in the Solar System using the {\sf DDSCAT} code (ver.\,4a) with the $a_1$-term method (see Fig.\,\ref{kimura-et-al2002a:f2-f3}). 
The {\sf DDSCAT} code (ver.\,4a) with the $a_1$-term method was also used to compute the parallel and perpendicular components of radiation pressure on BPCA ($D \approx 3$) and BCCA ($D \approx 2$) particles of $0.01\,\mathrm{\mu m}$-radius constituent grains in orbit around $\beta$ Pic \citep{kimura-mann1999b}.
The perpendicular components of radiation pressure are negligible for carbon agglomerates in comparison to the parallel component, but could be significant for silicate agglomerates.
However, we are not very confident that this conclusion holds for agglomerates consisting of submicron grains prior to a future study with submicron constituent grains.

\citet{silsbee-draine2016} applied the {\sf DDSCAT} code (ver.\,7.3) with the ``modefied'' LDR method to compute the transverse component of radiation pressure on silicate BPCA particles with restructuring in the Solar System.
They assumed that the agglomerates with $N=32$ have a radius $a_\mathrm{v} = 0.15$ and $0.6\,\mathrm{\mu m}$, implying $a_{0} = 0.047$ and $0.19\,\mathrm{\mu m}$, respectively.
Their results would suggest that the transverse component gives a more effective consequence for the dynamics of smaller agglomerates.
However, we cannot assert at this stage that this is a general trend, because it is not clear whether the difference originates from smaller $a_\mathrm{v}$ or smaller $a_{0}$.

\subsection{Equilibrium temperature}
\label{sec:6.3}

Dust particles in planetary systems attain equilibrium temperatures depending on their composition, structure, and distance from the central star.
Equilibrium temperature is a key parameter to determine the SEDs as well as the sublimation zone of dust particles in the vicinity of a star.
In equilibrium, the temperature of dust particles in planetary systems is determined by the balance of energies between absorption of stellar radiation and thermal emission and sublimation of the particles.
An estimate of equilibrium temperature requires integration of absorption cross sections weighted by stellar radiation spectrum and the Planck function over wavelengths from ultraviolet to far-infrared. 
It is, therefore, not an easy task to estimate the equilibrium temperature of agglomerates consisting of submicrometer-sized grains in particular.

\citet{kozasa-et-al1992} used an empirical formula to calculate the equilibrium temperature of BPCA ($D \approx 3$) and BCCA ($D \approx 2$) particles consisting of magnetite grains with a radius $a_0 = 0.01\,\mathrm{\mu m}$ at 1~au from the Sun.
They proposed a formula to approximate the absorption cross sections for agglomerates using the optical characteristics of a spherical cloud that encloses the constituent grains.
This formula reproduces the results from the {\sf DDSCAT} code (ver.\,4a) with the DGF/VIEF method within about 10\%, if the agglomerates are assumed to have the same radius as a sphere of equal geometric cross section.
It should be, however, noted that the empirical formula is limited to agglomerates of tiny constituent grains with $a_0 \le 0.01\,\mathrm{\mu m}$.

\citet{mann-et-al1994} estimated the equilibrium temperature of BPCA ($D \approx 3$) particles consisting of silicate grains and graphite grains with a radius $a_0 = 0.01\,\mathrm{\mu m}$ near the Sun using the Maxwell Garnett mixing rule.
A higher temperature was found for composite agglomerates of silicate grains and graphite grains in comparison to pure silicate agglomerates or pure graphite agglomerates.
The high temperature of an agglomerate consisting of silicate grains and carbon grains was confirmed by \citet{kimura-et-al1997} who applied the Bruggeman mixing rule, although the formula for energy of sublimation in \citet{kimura-et-al1997} is different from that in \citet{mann-et-al1994}.
Whereas \citet{mann-et-al1994} assumed that the contribution of each constituent grain to sublimation is given by its volume fraction to the power of 2/3, \citet{kimura-et-al1997} describe that the contribution of each constituent grain to sublimation is proportional to its volume fraction.
We should point out that the incorrectness of the formula in \citet{mann-et-al1994} is obvious, as it fails to work if two materials have an even volume fraction.

\citet{kimura-et-al1997} found that the equilibrium temperatures of fractal agglomerates with $a_0 = 0.01\,\mathrm{\mu m}$ become less dependent on their sizes as the fractal dimension decreases or the porosity increases.
As a result, the temperatures of fractal agglomerates asymptotically approach that of constituent grains in the limit of $D \to 2$.
This trend is associated with the fact that both the absorption cross section and surface area of fractal agglomerates are proportional to the number of constituent grains if $D = 2$.
Therefore, we expect for fractal agglomerates that the size dependence of their temperatures is weak and close to that for the constituent grains, despite the radius of the grains, as long as agglomerates of $D \approx 2$ are concerned.

Again, the assumption of $a_0 = 0.01\,\mathrm{\mu m}$ for the radius of constituent grains is the major drawback of the above mentioned studies as a simulation for primitive dust particles in planetary systems.
To our knowledge, \citet{greenberg-hage1990} performed the earliest study on the equilibrium temperature of fluffy agglomerates consisting of submicron grains in an application to dust particles in comet 1P/Halley at $0.9\,\mathrm{au}$ from the Sun.
They assumed that the constituent grains of $a_0 = 0.1\,\mathrm{\mu m}$ have a structure of a silicate core and an organic mantle and used the Maxwell Garnett mixing rule for wavelengths of $\lambda \ge 1\,\mathrm{\mu m}$ and, for shorter wavelengths, the optical characteristics of a spherical cloud that encloses the constituent grains.
They claimed the assumption of the cloud to be a good approximation as long as the size parameter of the cloud radius exceeds 10, but did not show any evidence for their claim.
The dependences of temperature on porosity and agglomerate size are qualitatively the same as, but quantitatively different from the case for $a_0 = 0.01\,\mathrm{\mu m}$.
Therefore, the use of submicrometer-sized constituent grains in an agglomerate for computation of equilibrium temperature is of crucial importance.

\citet{greenberg-li1998} calculated the equilibrium temperature of fluffy agglomerates consisting of spherical silicate core, organic refractory mantle grains with $a_0 = 0.1\,\mathrm{\mu m}$ at $1\,\mathrm{au}$ from the Sun.
They presented how the temperature depends on the mass and porosity of the agglomerates and on the mass ratio of the silicate core and the organic refractory mantle.
Their calculations were performed with the Maxwell Garnett mixing rule, but we notice that the Maxwell Garnett mixing rule alone is not applicable to calculations of temperatures for $a_0 = 0.1\,\mathrm{\mu m}$, because the calculations require the absorption cross sections at wavelengths of $\lambda < 1\,\mathrm{\mu m}$ where $x_0 \ge 1$.

\begin{figure*}
\center
\resizebox{0.8\textwidth}{!}{%
  \includegraphics{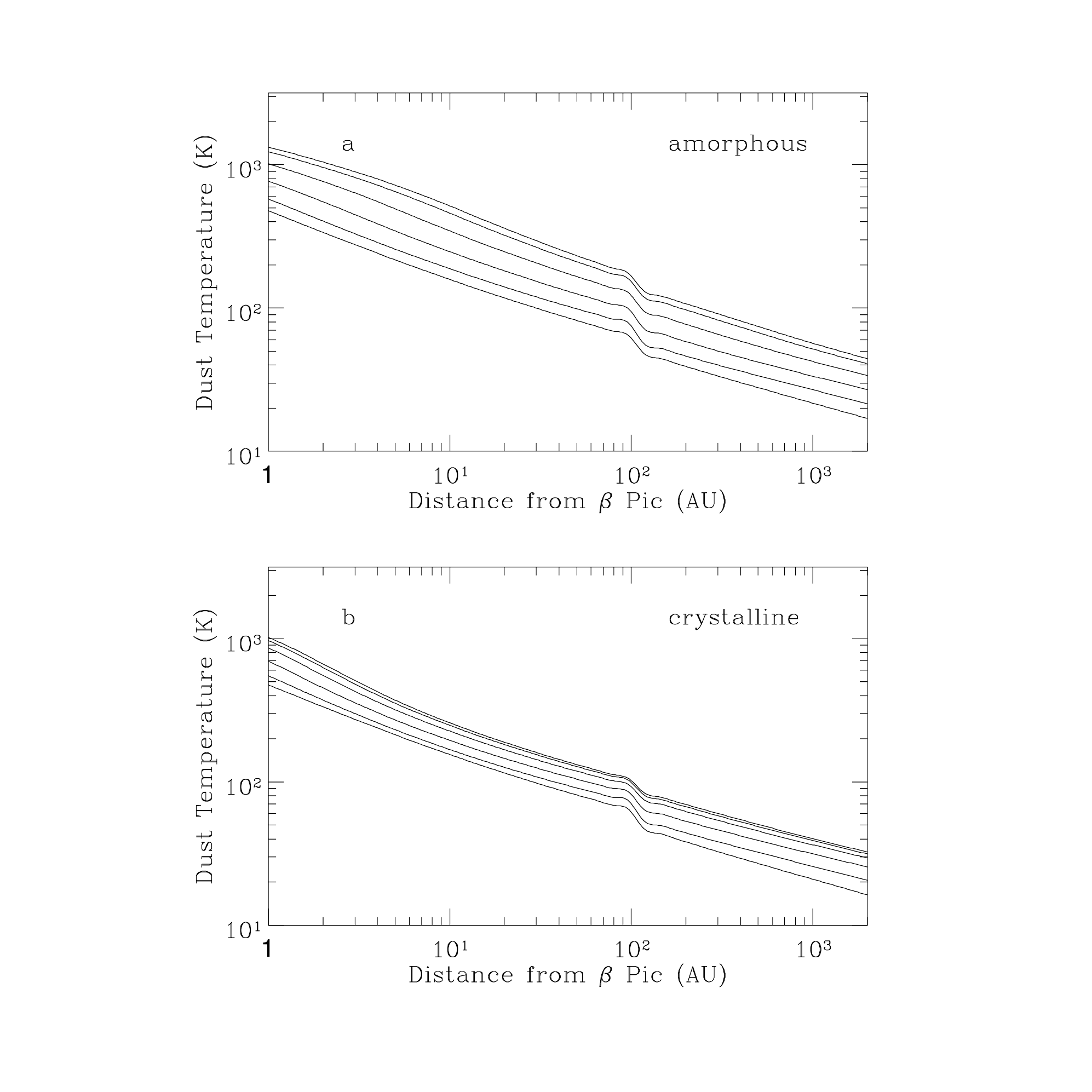}
}
\caption{The equilibrium temperature of porous agglomerates with masses of $m = 10^{-17}$, $10^{-15}$, $10^{-13}$, $10^{-10}$, $10^{-7}\,\mathrm{kg}$ (from {\it top} to {\it bottom}) as a function of distance from the central star in the debris disk of $\beta$ Pic. 
Spherical constituent grains of an agglomerate have a radius $a_0 = 0.1\,\mathrm{\mu m}$ and comprise either (a) amorphous silicate encased in an organic refractory material ({\it upper panel}) or (b) crystalline silicate ({\it lower panel}).
In the outer region of the disk at distances $r > 100\,\mathrm{au}$, the agglomerates are assumed to have an external layer of H$_2$O ice.
From \citet{li-greenberg1998}}
\label{li-greenberg1998:f1}       
\end{figure*}

\citet{li-greenberg1998} estimated the radial distribution of equilibrium temperatures for agglomerates of submicron constituent grains in $\beta$ Pic (see Fig.\,\ref{li-greenberg1998:f1}).
They assumed that the constituent grains with $a_0 = 0.1\,\mathrm{\mu m}$ are composed of either an amorphous silicate core and an organic mantle or bare crystalline silicate.
The presence of ice mantles are also considered if the agglomerates are located at a distance exceeding $100\,\mathrm{au}$ from the star, implying ice sublimation at $100\,\mathrm{au}$.
According to the method described in \citet{greenberg-hage1990}, they used the Maxwell Garnett mixing rule for $\lambda \ge 1\,\mathrm{\mu m}$ and considered a cloud of the constituent grains for $\lambda < 1\,\mathrm{\mu m}$.
Their results indicate that sublimation of ices in the inner disk enhances the temperatures of the agglomerates as a result of the increase in absorptivity.

\begin{figure*}
\resizebox{1.0\textwidth}{!}{%
  \includegraphics{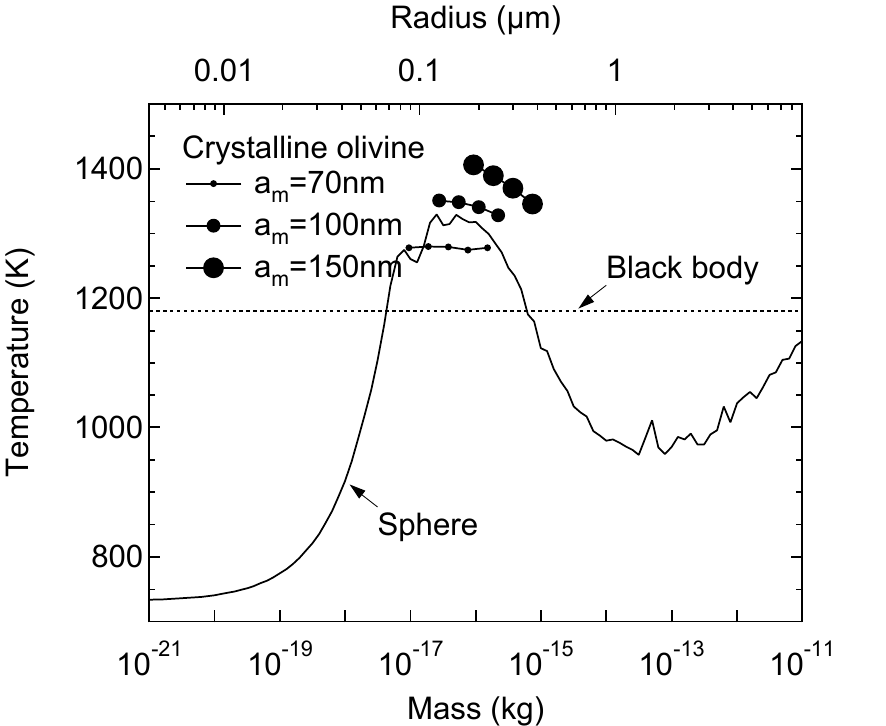}
  \includegraphics{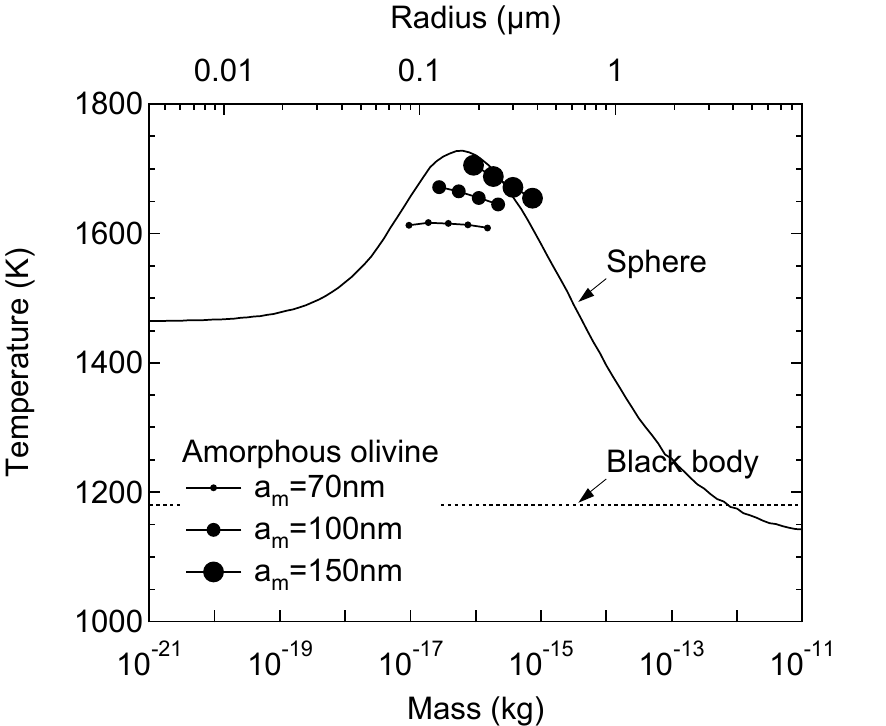}
}
\resizebox{1.0\textwidth}{!}{%
  \includegraphics{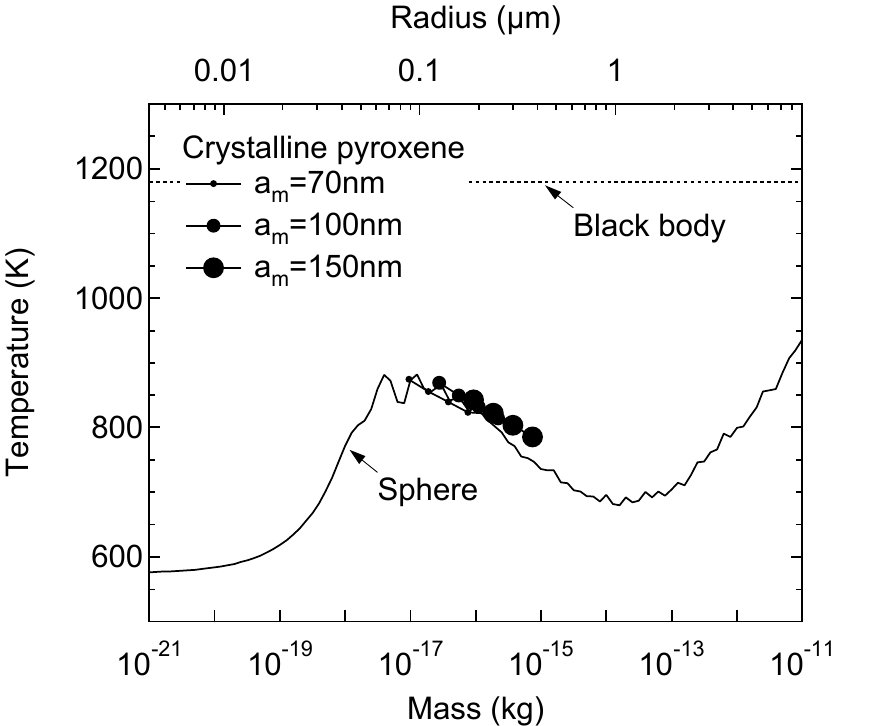}
  \includegraphics{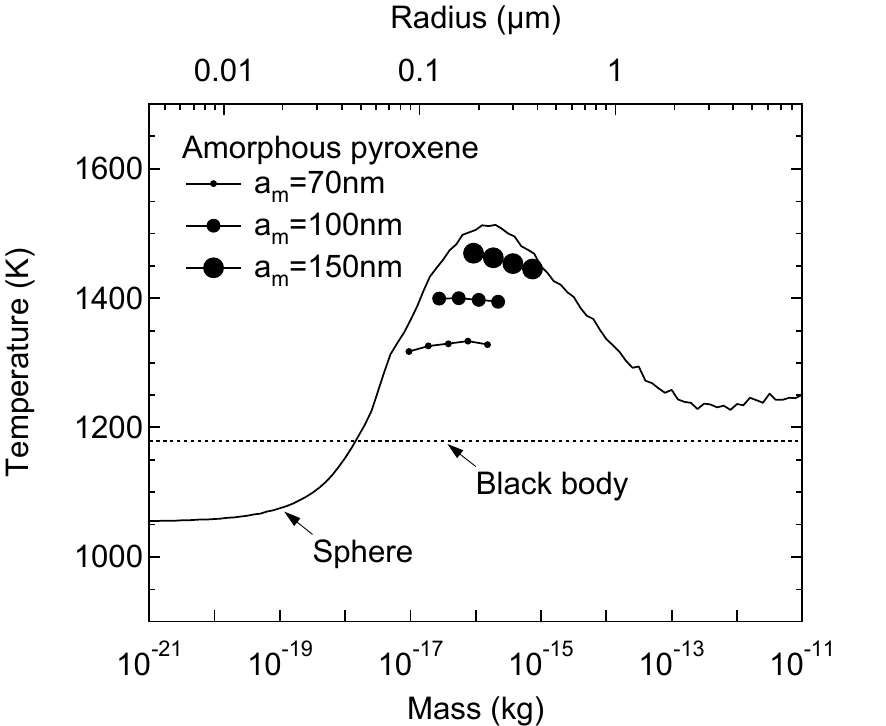}
}
\caption{The equilibrium temperature of fractal agglomerates consisting of spherical olivine and pyroxene grains at a heliocentric distance of $12\,\mathrm{R}_\odot$. 
The {\it largest}, the {\it medium}, and the {\it smallest filled circles} are the numerical values of the temperature with $a_0 = 0.07$, $0.10$, and $0.15\,\mathrm{\mu m}$, respectively.
The {\it solid curves} and {\it dotted ones} are the equilibrium temperature of compact spherical particles and the blackbody temperature.
From \citet{kimura-et-al2002b}}
\label{kimura-et-al2002b:f3}       
\end{figure*}

\citet{kimura-et-al2002b} computed the equilibrium temperatures of BCCA ($D \approx 2$) particles consisting of submicron silicate grains in the vicinity of the Sun using the {\sf scsmtm1} code. 
Although their computations with $a_0 = 0.1\,\mathrm{\mu m}$ were limited to $N \le 32$, their results confirmed that the temperatures of agglomerates show a weaker size dependence, compared to those of compact spheres (see Fig.\,\ref{kimura-et-al2002b:f3}).
They have shown that the temperatures of pyroxene agglomerates are 0.5--0.6 times lower than those of olivine agglomerates in the comae and tails of sungrazing comets.
Accordingly, their results predict that crystalline and amorphous pyroxene grains sublimate inside $5$ solar radii ($R_\odot$) from the Sun, while crystalline and amorphous olivine grains sublimate inside $10$--$11\,R_\odot$ and $12$--$13\,R_\odot$, respectively.
It turned out that sublimation of pyroxene agglomerates and olivine ones near the Sun well accounts for not only light curves of sungrazing comets observed by SOHO/LASCO but also neutral hydrogen Lyman $\alpha$ emission from sungrazers observed by SOHO/UVCS \citep[cf.][]{biesecker-et-al2002,bemporad-et-al2006,bemporad-et-al2007}.

\citet{xing-hanner1996,xing-hanner1997} estimated the dependence of equilibrium temperatures on heliocentric distance for agglomerates consisting of 10 submicron spherical carbon grains or tetrahedral ones with various separations of the grains.
They used the {\sf DDSCAT} code (ver.\,4b) with the LDR method for computing absorption cross sections of the agglomerates in the wavelength range from visible to far-infrared.
The absorption cross sections for ultraviolet and visible wavelengths were extrapolated from the cross section at $\lambda = 1\,\mathrm{\mu m}$ on the assumption that the absorption cross section at $\lambda = 0$ equals to the geometric cross section of the agglomerates.
Their results indicate that the temperature of agglomerates approaches that of constituent grains once the grains get separated, while the results for touching grains are similar to those for overlapping grains.
Agglomerates consisting of tetrahedral grains show a weaker dependence of temperature on the separation than agglomerates consisting of spherical grains.
In addition, the dependence of the temperature on heliocentric distance is weaker for agglomerates than for blackbody, irrespective of the separation between the grains.
The authors concluded that the agglomerates attain the blackbody temperature at small heliocentric distances, but this does not seem to be the case \citep[cf.][]{kimura-et-al1997,kimura-et-al2002b}.
Their misleading conclusion could be attributed to simply their crude method for estimates of absorption cross sections at short wavelengths.

\citet{mukai-okada2007} estimated equilibrium temperatures of BPCA ($D \approx 3$) particles consisting of silicate grains with $N=2048$ and $16384$.
They used the {\sf DDSCAT} code (ver.\,6.1) with the LDR method for computing the absorption cross sections of the agglomerates consisting of small constituent grains with $a_{0} < 0.1\,\mathrm{\mu m}$ and the geometric optics for large constituent grains with $a_{0} \ge 10\,\mathrm{\mu m}$.
They have shown that the results between the DDA and the geometric optics could be smoothly interpolated in the range of $0.1 \le a_{0} < 10\,\mathrm{\mu m}$, if they apply the Maxwell Garnett mixing rule to computations of absorption cross sections.
Therefore, they applied the Maxwell Garnett mixing rule to calculate the temperatures of the agglomerates, irrespective of the radius of constituent grains.
\citet{levasseurregourd-et-al2007} took the same procedures to compute equilibrium temperatures of BPCA ($D \approx 3$) particles consisting of carbon grains with $N=2048$ and $16384$.
Because they consider agglomerates up to $a_\mathrm{v} = 10^4\,\mathrm{\mu m}$, the radius of constituent grains reached $a_0 \approx 40$--$80\,\mathrm{\mu m}$.
Because their agglomerates do not have constituent grains of $a_0 \sim 0.1\,\mathrm{\mu m}$ except for $a_\mathrm{v} \sim 1$--$3\,\mathrm{\mu m}$, it does not seem plausible that their results on the size dependence of temperatures are applicable to primitive dust particles in a planetary system.

\citet{lasue-et-al2007} used the {\sf DDSCAT} code (ver.\,5a10) presumably with the LDR method for computing the temperatures of BPCA ($D \approx 3$) and BCCA ($D \approx 2$) particles as a function of heliocentric distance to simulate the radial dependence of temperature inferred from zodiacal light observations.
They considered agglomerates of 64 spheroidal constituent grains with $a_0 = 0.19\,\mathrm{\mu m}$ composed of silicate or organic material, as well as compact spheres and spheroids of silicates, organic material, or amorphous carbon.
Their results indicate that absorbing materials such as organics and amorphous carbon play an important role in the radial gradient of the temperature in the zodiacal light.
They concluded that the radial dependence of the temperature does not strongly depend on the shape and structure of the particle, but the size and material of the particle.
Their conclusion on the size dependence of equilibrium temperature was, however, not confirmed for agglomerates, as they studied the size dependence of equilibrium temperature only with compact spheres and spheroids.

\section{Concluding remarks}
\label{sec:7}

In-situ measurements in the Solar System have revealed that the total cross sections of primitive dust particles are dominated by large agglomerates of submicron constituent grains.
The mass distribution of interplanetary dust measured in situ at $1\,\mathrm{au}$ from the Sun shows that the cross sectional distribution has a peak at $m \approx 3 \times {10}^{-10}\,\mathrm{kg}$, which is equivalent to $a_\mathrm{v} \approx 30\,\mathrm{\mu m}$ and $N \approx 3\times {10}^7$ \citep{gruen-et-al1985}.
The predominance of large agglomerates with $m \ge {10}^{-13}\,\mathrm{kg}$ ($a_\mathrm{v} \ge 2\,\mathrm{\mu m}$) is supported by the mass distribution of dust particles in the inner coma of comet 1P/Halley measured in situ by the Dust Impact Detection System (DIDSY) onboard Giotto \citep{mcdonnell-et-al1987,kolokolova-et-al2007}.
This is in harmony with the characteristic radius of $a_\mathrm{c} \approx 20\,\mathrm{\mu m}$ ($N \approx 2 \times {10}^{6}$) for agglomerates of submicron constituent grains with $D = 2.5$ derived from $\beta \approx 0.4$ for dust particles in the ejecta plume of comet 9P/Tempel 1 \citep{kobayashi-et-al2013}.
Because meteoritic impact is the major process of releasing dust particles from asteroids, we expect that typical dust particles ejected from asteroids are also characterized by agglomerates of $N \ge {10}^{6}$ constituent grains.
More recently, Grain Impact Analyser and Dust Accumulator (GIADA) and Optical, Spectroscopic, and Infrared Remote Imaging System (OSIRIS) onboard Rosetta have revealed that dust particles in the ${10}^{-8}$ to ${10}^{-7}\,\mathrm{kg}$ mass bin (i.e., $a_\mathrm{v} \sim 100$--$200\,\mathrm{\mu m}$, $N \sim {10}^{9}$--${10}^{10}$) dominate the coma of comet 67P/Churyumov-Gerasimenko \citep{rotundi-et-al2015}.
The COmetary Secondary Ion Mass Analyser (COSIMA) onboard Rosetta has collected fluffy agglomerates with $a_\mathrm{c} \ge 25\,\mathrm{\mu m}$ and $p>0.5$ from the comet \citep{schulz-et-al2015}.
The predominance of large agglomerates in the total cross sections is also true of debris disks around early-type stars such as $\beta$ Pic and Fomalhaut, because small particles of $a_\mathrm{v} < 10\,\mathrm{\mu m}$ are quickly blew out by stellar radiation pressure \citep[e.g.,][]{kimura-mann1999b,acke-et-al2012}.
In conclusion, it is of great importance to estimate light-scattering properties of large agglomerates consisting of  million or more grains of sumicrometer size.

It is worth emphasizing that available numerical techniques allow us to compute thermal emission from large ($N \ge {10}^{6}$) agglomerates of submicron constituent grains.
However, currently available numerical techniques do not allow us to properly simulate light scattering by such a large agglomerate consisting of submicron grains.
As a result, radiative transfer computations and SED fittings often rely on the Henyey-Greenstein phase function without proof of its applicability to light scattering by dust agglomerates.
Nevertheless, computing powers are, without doubt, increasing as the time goes by and numerical codes for calculation of light scattering by agglomerates are making steady progress.
In the end, we expect that our understanding of primitive dust particles in planetary systems will make great advances once light scattering by large agglomerates of more than ${10}^{6}$ submicron constituent grains becomes available.

As we have noted in Sect.\,\ref{intro}, the current review has focused on light-scattering and thermal-emission properties of primitive dust particles that were released from planetesimals in orbit around main-sequence stars.
Accordingly, we have not dealt with any subject of dust particles in protoplanetary disks, because the majority of the particles might be prestellar interstellar grains and thus not necessarily be incorporated into planetesimals.
We are, however, aware of the most recent findings that the majority of protoplanetary disks around weak-line T Tauri stars (WTTS) could be regarded as young debris disks \citep{hardy-et-al2015}.
If this identification is confirmed by future studies, then we have to admit that the current review is far from complete and any coming review on this subject cannot avoid additionally discussing light-scattering and thermal-emission properties of dust particles in young debris disks around WTTSs.
At the moment, we are confident that we have presented a new state-of-the-art review on light-scattering and thermal-emission properties of primitive dust particles in planetary systems.

\section{Summary}
\label{sec:8}

We have reviewed numerical approaches to deducing light-scattering and thermal-emission properties of primitive dust particles in planetary systems from astronomical observations. 
Currently available information on the particles indicates that they are often agglomerates of small constituent grains whose sizes are comparable to visible wavelength. 
If the particles are pristine, they are composed mainly of magnesium-rich silicates, ferrous metals and sulfides, and organic refractory materials. 
If dust particles were subjected to metamorphism, they could be composed of hydrous minerals, metals, and to a lesser extent organic refractory materials. 
These unique characteristics of primitive dust particles are associated with their formation and evolution around a star whose chemical composition is essentially solar. 
We have demonstrated that the development of light-scattering techniques has been offering powerful tools to make a thorough investigation of dust particles in various astrophysical environments. 
Among numerical techniques to solve light-scattering problems, the discrete dipole approximation, the T-matrix method, the generalized multiparticle Mie solution, and effective medium approximations are the most common ones for practical use in astronomy. 
The Monte Carlo method for radiative transfer could be applied to light scattering by an ensemble of dust particles in the environment where the optical depth is on the order of unity or larger. 
We have shown that numerical simulations of light scattering by and thermal emission from dust agglomerates provide new state-of-the-art knowledge of primitive dust particles in planetary systems, if combined with comprehensive collections of relevant results from not only astronomical observations, but also in-situ data analyses, laboratory sample analyses, laboratory analogue experiments, and theoretical studies on the formation and evolution of the particles.

\begin{acknowledgement}
We would like to thank anonymous reviewers for constructive comments and helpful suggestions that greatly improved the manuscript. We are grateful to Yu-lin Xu, Yasuhiko Okada, Ryo Tazaki, Sebastian Wolf, and Bruce T. Draine for useful information. 
We also express our gratitude to Alexander A. Kokhanovsky who encouraged us to write up a comprehensive review of this subject.
H. Kimura is thankful to Grants-in-Aid for Scientific Research (\#26400230, \#15K05273, \#23103004) from the Japan Society of the Promotion of Science.
\end{acknowledgement}
%
%

\end{document}